\def\year{2020}\relax
\documentclass[letterpaper]{article} 
\usepackage{aaai20}  
\usepackage{times}  
\usepackage{helvet} 
\usepackage{courier}  
\usepackage[hyphens]{url}  
\usepackage{graphicx} 
\usepackage{todonotes}
\usepackage{latexsym}
\usepackage{amsmath,amsfonts,mathrsfs,amssymb,bm,mathtools,mathabx}
\usepackage{multirow,mdframed}
\mdfsetup{skipabove=0pt,skipbelow=0pt}
\mdfdefinestyle{model}{%
	innertopmargin=0pt,
	innerbottommargin=0pt,
	innerleftmargin=0pt,
	innerrightmargin=0pt,
	skipbelow=0pt,%
	skipabove=0pt,
	splitbottomskip=0pt,
	splittopskip=0pt,
	leftmargin =0pt,%
  	rightmargin=0pt,
	splittopskip=0pt,
  	usetwoside=false,
}
\usepackage{float} 
\floatstyle{ruled}
\newfloat{model}{thp}{lop}
\floatname{model}{Model}

\usepackage{subcaption}

\usepackage{hyperref} 
\hypersetup{
colorlinks=true, 
breaklinks=true,
urlcolor= blue, 
linkcolor = magenta,
citecolor = cyan, 
}

\usepackage{booktabs}
\usepackage{xspace}
\usepackage[export]{adjustbox}
\usepackage{xcolor,colortbl}
\usepackage{subcaption,cleveref}
\usepackage[ruled,linesnumbered,noresetcount,vlined]{algorithm2e}
\SetKwFor{Repeat}{repeat}{:}{endw}

\def\thmspace{0.2em}

\DeclareMathOperator*{\argmin}{argmin}

\newcommand{\cO}{\mathcal{O}}
\newcommand{\bx}{\bm{x}}
\newcommand{\by}{\bm{y}}
\newcommand{\bw}{\bm{w}}
\newcommand{\blambda}{\bm{\lambda}}
\newcommand{\RR}{\mathbb{R}}

\newcommand{\citep}[1]{\citeauthor{#1} (\citeyear{#1})}
\let\oldnl\nl
\newcommand{\nonl}{\renewcommand{\nl}{\let\nl\oldnl}}
\newcommand{\cBox}{$\text{\rlap{$\checkmark$}}\Box$}

\newcommand{\B}{{\text{B}}}
\newcommand{\C}{{\text{C}}}
\newcommand{\CL}{{\text{CS}}}
\newcommand{\CLD}{{\text{CS}}^{\text{D}}}

\newcommand{\CLL}{{\text{CS}}^{\text{L}}}
\newcommand{\CD}{{\text{C}}^{\text{D}}}
\newcommand{\LFL}{{\text{LF}_S}}

\SetAlgoSkip{}
\SetAlgoInsideSkip{}

\nocopyright
\title{Predicting AC Optimal Power Flows: \\Combining Deep Learning and Lagrangian Dual Methods
}
\author{
Ferdinando Fioretto,\textsuperscript{\rm 1,2} 
Terrence W.K.~Mak,\textsuperscript{\rm 1} 
Pascal Van Hentenryck\textsuperscript{\rm 1}\\ 
\textsuperscript{\rm 1}Georgia Institute of Technology,
\textsuperscript{\rm 2}Syracuse University\\ 
ffiorett@syr.edu, wmak@gatech.edu, pvh@isye.gatech.edu
}

\begin{document}
\maketitle\sloppy\allowdisplaybreaks
\begin{abstract}
The Optimal Power Flow (OPF) problem is a fundamental building block
for the optimization of electrical power systems. It is nonlinear and
nonconvex and computes the generator setpoints for power and voltage,
given a set of load demands. It is often solved repeatedly under
various conditions, either in real-time or in large-scale
studies. This need is further exacerbated by the increasing
stochasticity of power systems due to renewable energy sources in
front and behind the meter. To address these challenges, this paper
presents a deep learning approach to the OPF. The learning model
exploits the information available in the similar states of the system
(which is commonly available in practical applications), as well as a
dual Lagrangian method to satisfy the physical and engineering
constraints present in the OPF.  The proposed model is evaluated on a
large collection of realistic medium-sized power systems. The
experimental results show that its predictions are highly accurate
with average errors as low as 0.2\%.  Additionally, the proposed
approach is shown to improve the accuracy of the widely adopted linear
DC approximation by at least two orders of magnitude.

\noindent
$\blacksquare$ \textbf{Note:} 
\textit{This paper is an extended version of \cite{Fioretto:dnnopf}}.

\end{abstract}


\section{Introduction}
The \emph{Optimal Power Flow} (OPF) problem determines the generator 
dispatch of minimal cost that meets the demands while satisfying 
the physical and engineering constraints of the power system 
\cite{OPF}.  The OPF (aka AC-OPF) is a non-convex non-linear 
optimization problem and the building bock of many applications, 
including security-constrained OPFs (Monticelli et al.~\citeyear{monticelli:87}), optimal 
transmission switching \cite{OTS}, capacitor placement 
\cite{baran:89}, expansion planning (Verma et al.~\citeyear{verma:16}), 
and security-constrained unit commitment \cite{wang:08}.

Typically, generation schedules are updated in intervals of 5
minutes \cite{Tong:11}, possibly using a solution to the OPF solved in
the previous step as a starting point. In recent years, the
integration of renewable energy in sub-transmission and distribution
systems has introduced significant stochasticity in front and behind
the meter, making load profiles much harder to predict and introducing
significant variations in load and generation. This uncertainty forces
system operators to adjust the generators setpoints with increasing
frequency in order to serve the power demand while ensuring stable
network operations. However, the resolution frequency to solve OPFs is
limited by their computational complexity. To address this issue,
system operators typically solve OPF approximations such as the linear
DC model (DC-OPF).  While these approximations are more efficient
computationally, their solution may be sub-optimal and induce
substantial economical losses, or they may fail to satisfy the
physical and engineering constraints.

Similar issues also arise in expansion planning and other 
configuration problems, where plans are evaluated by solving a massive number of multi-year Monte-Carlo simulations at 15-minute intervals \cite{pachenew,Highway50}. Additionally, the stochasticity introduced by renewable energy sources further increases the number of scenarios to consider. 
Therefore, modern approaches recur to the linear DC-OPF approximation and focus only on the scenarios considered most pertinent \cite{pachenew} at the expense of the
fidelity of the simulations.


To address these challenges, this paper studies how to approximate
OPFs using a Deep Neural Network (DNN) approach. The main goal of the
OPF is to find generator setpoints, i.e., the amount of real power and
the voltage magnitude for each generator. 
Approximating the OPF using DNNs can thus be seen as an empirical 
risk minimization problem. However, the resulting setpoints must 
also satisfy the physical and engineering constraints that regulate 
power flows, and these constraints introduce significant difficulties 
for machine learning-based approaches, as shown in 
\cite{ng2018statistical,deka:2019}.
To address these difficulties, this paper presents a DNN approach to
the OPF (OPF-DNN) that borrows ideas from Lagrangian duality and
models the learning task as the Lagrangian dual of the empirical risk
minimization problem under the OPF constraints. Note also that the
AC-OPF is an ideal application for machine learning, since it must be
solved almost continuously. Hence significant data is available to
train deep learning networks and improve them over time.

The contributions of this paper can be summarized as follows.  (1) It
proposes an approach (OPF-DNN) that uses a DNN to predict the
generator setpoints for the OPF; (2) It exploits the physical and
engineering constraints in a Lagrangian framework using violation
degrees; (3) It enhances the prediction accuracy by leveraging the
availability of a solution to a related OPF (e.g., the solution to a
closely related historical instances, which is almost always
available); (4) It recasts the OPF prediction as the Lagrangian dual
of the empirical risk minimization under constraints, using a
subgradient method to obtain a high-quality solution.

OPF-DNN is evaluated on realistic medium-sized power system
benchmarks: The computational results show significant improvements in
accuracy and efficiency compared to the ubiquitous DC model. In
particular, OPF-DNN provides accuracy improvements of up to two
orders of magnitude and efficiency speedups of several  orders
of magnitude. \emph{These
results may open new avenues for power system analyses and operations
under significant penetration of renewable energy}.

\def\lvec{\bm{\left(}}
\def\rvec{\bm{\right)}}
\newcommand{\vect}[1]{{\lvec{#1}\rvec}}

\section{Preliminaries}

The paper uses the following notations:  \emph{Variables} are denoted
by calligraph lowercase symbols, \emph{constants} by dotted symbols,
and \emph{vectors} by bold symbols. The hat notation $\hat{\bx}$ 
describes the prediction of a value $\bx$ and $\|\!\cdot\!\|$ denotes 
the $L2$-norm.  The power flow equations are expressed in terms of
complex \emph{powers} of the form $S \!=\! (p \!+\! jq)$, where $p$
and $q$ denote active and reactive powers,
\emph{admittance} of the form $Y \!=\! (g \!+\! jb)$,
where $g$ and $b$ denote the conductance and susceptance,
and \emph{voltages} of the form $V \!=\! (v \angle \theta)$, with
magnitude $v$ and phase angle $\theta$.

\begin{model}[!t]
	{\footnotesize
	\caption{AC Optimal Power Flow (AC-OPF)}
	\label{model:ac_opf}
	\vspace{-6pt}
	\begin{flalign}
		&{\cO}(\dot{\bm{p}}^d, \dot{\bm{q}}^d) = 
		\textstyle \bm{\argmin}_{\bm{p^g},\bm{v}}\;\;
		 \sum_{i \in {\cal N}} 	\text{cost}(p^g_i) && \label{ac_obj} \\
		&\mbox{\bf subject to:} \notag\\
		&\hspace{6pt}
		\dot{v}^{\min}_i \leq v_i \leq \dot{v}^{\max}_i 		
		&& \!\!\!\!\!\forall i \in {\cal N} 		\label{con:2a} \tag{2a}\\
		&\hspace{6pt}
		\text{ -- }\dot{\theta}^\Delta_{ij} \leq \theta_i \text{ -- } \theta_j  \leq \dot{\theta}^\Delta_{ij} 	
		&& \!\!\!\!\!\forall (ij) \in {\cal E}  	 \label{con:2b}\!\!\!\!\! \tag{$\bar{2b}$}\\
		&\hspace{6pt}
		\dot{p}^{g\min}_i \leq p^g_i \leq \dot{p}^{g\max}_i 	
		&& \!\!\!\!\!\forall i \in {\cal N} 		\label{con:3a} \tag{$\bar{3a}$}\\
		&\hspace{6pt}
		\dot{q}^{g\min}_i \leq q^g_i \leq \dot{q}^{g\max}_i 	
		&& \!\!\!\!\!\forall i \in {\cal N} 		\label{con:3b} \tag{3b}\\
		&\hspace{6pt}
		(p_{ij}^f)^2 + (q_{ij}^f)^2 \leq \dot{S}^{f\max}_{ij}			
		&& \!\!\!\!\!\forall (ij) \in {\cal E}	\label{con:4}  \tag{$\bar{4}$}\\
		&\hspace{6pt}
		p_{ij}^f \!=\! \dot{g}_{ij} v_i^2 \text{--}  
		v_i v_j (\dot{b}_{ij} \!\sin(\theta_i \text{--} \theta_j)
		+ \dot{g}_{ij} \!\cos(\theta_i \text{--} \theta_j)\!)	
		&& \!\!\!\!\!\forall (ij)\!\in\! {\cal E} 	\label{con:5a} \tag{$\bar{5a}$}\\
		&\hspace{6pt} 
		q_{ij}^f \!=\! \text{--} \dot{b}_{ij} v_i^2 \text{--}  v_i v_j (\dot{g}_{ij} \!\sin(\theta_i \text{--} \theta_j)
		\text{--} \dot{b}_{ij} \!\cos(\theta_i \text{--} \theta_j)\!)	
		&& \!\!\!\!\!\forall (ij)\!\in\! {\cal E}		\label{con:5b} \tag{5b}\\
		&\hspace{6pt}
		p^g_i \text{ -- } \dot{p}^d_i = \textstyle \sum_{(ij)\in {\cal E}} p_{ij}^f	
		&& \!\!\!\!\!\forall i\in {\cal N} 		\label{con:6a} \tag{$\bar{6a}$}\\
		&\hspace{6pt}
		q^g_i \text{ -- } \dot{q}^d_i = \textstyle 	\sum_{(ij)\in {\cal E}} q_{ij}^f	
		&& \!\!\!\!\!\forall i\in {\cal N} 		\label{con:6b} \tag{6b}\\
	&\textbf{output:}~~(\bm{p}^g, \bm{v}) \text{ -- The system operational parameters}
	\!\!\!\!\!\!\!\!\!\!\!\!\!\!\!\!\!\!\!\!
	\notag
	\end{flalign}
	\vspace{-12pt}
	}
\end{model}
\setcounter{equation}{6}

\subsection{Optimal Power Flow}

The \emph{Optimal Power Flow (OPF)} determines the least-cost
generator dispatch that meets the load (demand) in a power network. A power
network is viewed as a graph $({\cal N}, {\cal E})$ where the nodes
$\cal N$ represent the set of $n$ \emph{buses} and the edges $\cal E$
represent the set of $e$ \emph{transmission lines}. The OPF constraints
include physical and engineering constraints, which are captured in
the AC-OPF formulation of Model~\ref{model:ac_opf}.  The model uses
$\bm{p}^g$, and $\bm{p}^d$ to denote, respectively, the vectors of 
active power generation and load associated with each bus 
and $\bm{p}^f$
to describe the vector of active power flows associated with each
transmission line. Similar notations are used to denote the vectors of
reactive power $\bm{q}$.  Finally, the model uses $\bm{v}$ and
$\bm{\theta}$ to describe the vectors of voltage magnitude and angles
associated with each bus. The OPF takes as inputs the loads
$(\dot{\bm{p}}^d\!, \dot{\bm{q}}^d)$ and the admittance matrix
$\dot{\bm{Y}}$, with entries $\dot{g}_{ij}$ and $\dot{b}_{ij}$ for
each line $(ij) \!\in\!  {\cal E}$; It returns the active power
vector $\bm{p}$ of the generators, as well the voltage magnitude
$\bm{v}$ at the generator buses. The objective function \eqref{ac_obj}
captures the cost of the generator dispatch, and is typically expressed
as a quadratic function. Constraints \eqref{con:2a} and \eqref{con:2b}
restrict the voltage magnitudes and the phase angle
differences within their bounds.  Constraints \eqref{con:3a}
and \eqref{con:3b} enforce the generator active and
reactive output limits.  Constraints \eqref{con:4} enforce the line
flow limits.  Constraints \eqref{con:5a} and \eqref{con:5b} capture
\emph{Ohm's Law}. Finally, Constraint \eqref{con:6a}
and \eqref{con:6b} capture \emph{Kirchhoff's Current Law} enforcing
flow conservation.

\begin{model}[!t]
	{\small
	\caption{The Load Flow Model}
	\label{model:load_flow}
	\vspace{-6pt}
	\begin{flalign}
		\mbox{\bf minimize:}& \;\;
		\| \bm{p}^g - \hat{\bm{p}}^g \|^2 + \| \bm{v} - \hat{\bm{v}} \|^2 \label{load_flow_obj} \\
		\mbox{\bf subject to:} & \;\; \eqref{con:2a}-\eqref{con:6b} \notag
	\end{flalign}
	}
	\vspace{-12pt}
\end{model}

\paragraph{The DC Relaxation}
The DC model is a ubiquitous linear approximation to the OPF
\cite{Wood96}. It ignores reactive power and assumes that the voltage 
magnitudes are at their nominal values (1.0 in per unit notation).  
The model uses only the barred constraints in Model~\ref{model:ac_opf}.
Constrains \eqref{con:4} considers only the active flows and hence can
be trivially linearized and Constraints \eqref{con:5a}
becomes \mbox{$p_{ij}^f = -\dot{b}_{ij} (\theta_i - \theta_j)$}. The quadratic
objective is also replaced by a piecewise linear function.  Being an
approximation, a DC solution $\hat{\bm{p}}^g$ may not satisfy the AC
model constraints. As result, prior to being deployed, one typically
solves a \emph{load flow optimization}, described in
Model \ref{model:load_flow}. It is a least squares minimization 
problem that finds the closest AC-feasible solution to the approximated 
one.

\subsection{Deep Learning Models}

Supervised Deep Learning (SDL) can be viewed as the task of
approximating a complex non-linear mapping from labeled data.  Deep
Neural Networks (DNNs) are deep learning architectures composed of a
sequence of layers, each typically taking as inputs the results of the
previous layer \cite{lecun2015deep}. Feed-forward neural networks are
basic DNNs where the layers are fully connected and the function
connecting the layer is given by
$$
\bm{o} = \pi(\bm{W} \bm{x} + \bm{b}),
$$
where $\bx \!\in\! \RR^n$ and is the input vector, $\bm{o} \!\in\! \RR^m$ the output vector, $\bm{W} \!\in\! \RR^{m \times n}$ a matrix of weights, and
$\bm{b} \!\in\! \RR^m$ a bias vector. The function $\pi(\cdot)$ is
often non-linear (e.g., a rectified linear unit (ReLU)).

\section{OPF Learning Goals}

The goal of this paper is to learn the OPF mapping ${\cal
O}: \RR^{2n} \to \RR^{2n}$: Given the loads
$\vect{\bm{p}^d, \bm{q}^d}$, predict the setpoints
$\vect{\bm{p}^g, \bm{v}}$ of the generators, i.e., their active power
and the voltage magnitude at their buses. The input of the learning
task is a dataset ${\cal D} \!=\! \{(\bx_\ell,\by_\ell)\}_{\ell\!=\!1}^N$, where
$\bx_\ell \!\!=\!\! (\bm{p}^d, \bm{q}^d)$ and $\by_\ell \!=\!
(\bm{p}^g, \bm{v})$ represent the $\ell^{th}$ observation of load
demands and generator setpoints which satisfy $\by_\ell \!=\! {\cal
O}(\bx_\ell)$. The output is a function $\hat{\cal O}$ that ideally would be the result of the following optimization problem
\begin{flalign*}
\mbox{\bf minimize:} & \;\; \sum_{\ell=1}^N {\cal L}_o(\by_\ell,\hat{\cal O}(\bx_\ell)) \\ 
\mbox{\bf subject to:} & \;\; {\cal C}(\bx_\ell,\hat{\cal O}(\bx_\ell))
\end{flalign*}
\noindent
where the loss function is specified by
\begin{equation}
\label{basic_loss}
	{\cal L}_o(\by, \hat{\by}) = 
	\underbrace{\| \bm{p}^g - \hat{\bm{p}}^g \|^2}_{{\cal L}_p(\by, \hat{\by})}{\!} +
	\underbrace{\| \bm{v} - \hat{\bm{v}} \|^2}_{{\cal L}_v(\by, \hat{\by})}{\!}
\end{equation}
and ${\cal C}(\bx,\by)$ holds if there exist voltage angles
$\bm{\theta}$ and reactive power generated $\bm{q}^g$ that produce a
feasible solution to the OPF constraints with $\bx =
(\bm{p}^d, \bm{q}^d)$ and $\by = (\bm{p}^g, \bm{v})$.

One of the key difficulties of this learning task is the presence of
the complex nonlinear feasibility constraints in the OPF. The
approximation $\hat{\cal O}$ will typically not satisfy the problem
constraints. As a result, like in the case of the DC model discussed
earlier, the validation of the learning task uses a load flow
computation that, given a prediction $\hat{\by}\!=\!\hat{\cal O}(\bx_\ell)$, computes the \mbox{closest feasible generator setpoints}.


\section{Baseline Deep Learning Model}

\begin{figure*}[!t]
\centering\includegraphics[width=0.9\linewidth]{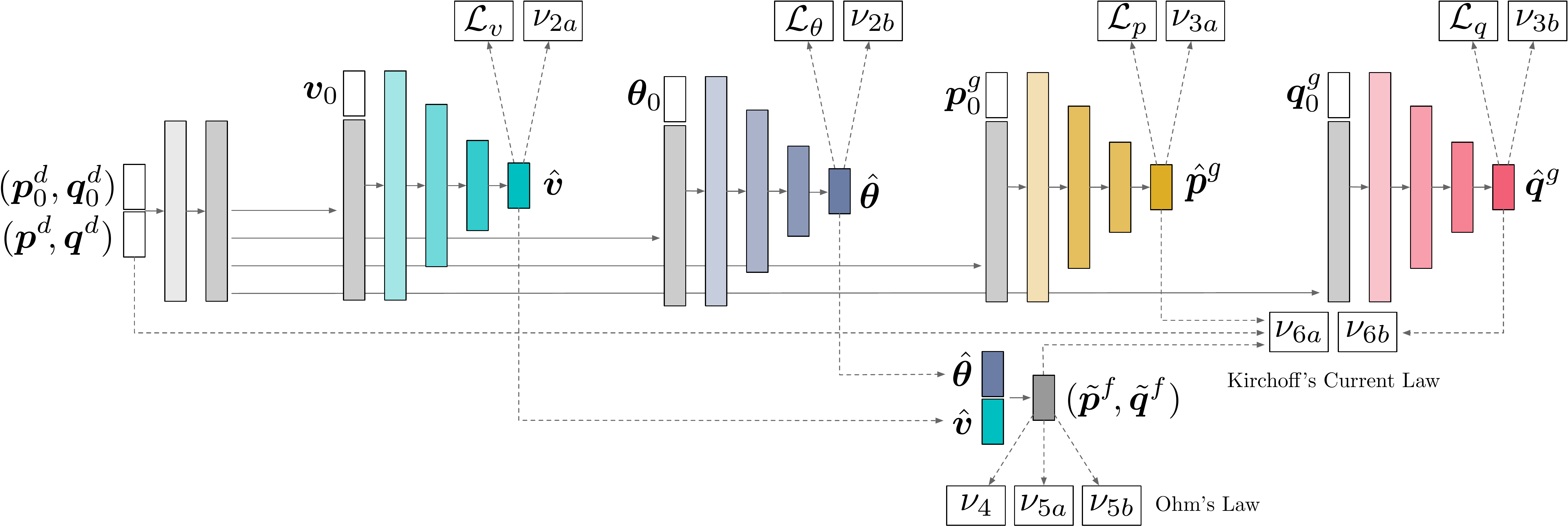}
\caption{\label{fig:dlopf} The OPF-DNN Model:
		Each layer is fully connected with ReLU activation. 
		White boxes correspond to input tensors, dark, colored, boxes 
		correspond to output layers. Loss components and violation 
		degrees are shown as white rectangles.}
\end{figure*}

The baseline model for this paper assumes that function $\hat{\cal O}$
is given by a feed-forward neural network, whose architecture is part
of the final network outlined in Figure \ref{fig:dlopf} and discussed
in detail later. While this baseline model is often accurate for many
regression problems, the experimental results show that it has low
fidelity for complex AC-OPF tasks. More precisely, a load flow
computation on the predictions of this baseline model to restore
feasibility produces generator setpoints with substantial errors.  The
rest of the paper shows how to improve the accuracy of the model by
exploiting the problem structure.

\section{Capturing the OPF Constraints}

To capture the OPF constraints, this paper uses a Lagrangian
relaxation approach based on constraint violations
\cite{Fontaine:14} used in generalized augmented Lagrangian
relaxation \cite{Hestenes:69}. The Lagrangian relaxation of an
optimization problem
\begin{flalign*}
\mbox{\bf minimize:} & \;\; f(\bx) \\
\mbox{\bf subject to:} & \;\; h(\bx) = 0\\ 
                       &\;\; g(\bx) \leq 0 
\end{flalign*}

\noindent
is given by
\begin{flalign*}
\mbox{\bf minimize:} & \;\; f(\bx) + \lambda_h h(\bx) + \lambda_g g(\bx)
\end{flalign*}
\noindent
where $\lambda_h$ and $\lambda_g \geq 0$ are the Lagrangian multipliers.
In contrast, the violation-based Lagrangian relaxation is 
\begin{flalign*}
\mbox{\bf minimize:} & \;\; f(\bx) + \lambda_h |h(\bx)| + \lambda_g \max(0,g(\bx))
\end{flalign*}
\noindent
with $\lambda_h,\lambda_g \geq 0$. In other words, the traditional
Lagrangian relaxation exploits the satisfiability degrees of
constraints, while the violation-based Lagrangian relaxation is
expressed in terms of violation degrees. 
The satisfiability degree of a constraint measures how well the
constraint is satisfied, with negative values representing the slack
and positive values representing violations, while the violation 
degree is always non-negative and represents how much the  constraint is 
violated.
More formally, the satisfiability degree of a constraint 
$c \!:\! \mathbb{R}^n \to {\it Bool}$
is a function $\sigma_c\!:\! \mathbb{R}^n \to \mathbb{R}$ such that
$\sigma_c(\bx) \leq 0 \equiv c(\bx)$. The violation degree of a
constraint $c\!:\! \mathbb{R}^n \to {\it Bool}$ is a function
$\nu_c\!:\! \mathbb{R}^n \to \mathbb{R^+}$ such that $\sigma_c(\bx) \equiv
0 \equiv c(\bx)$. For instance, for a linear constraints $c(\bx)$ of
type $A\bx \geq b$, the \emph{satisfiability degree} is defined as
\begin{equation*}
\sigma_c(\bx) \equiv \bm{b} - A\bx
\end{equation*}
and the \emph{violation degrees} for inequality and equality 
constraints are specified by
\begin{equation*}
\nu^{\geq}_c(\bx) = \max\left(0, \sigma_c(\bx)\right) 
\qquad
\nu^{=}_c(\bx) = \left| \sigma_c(\bx) \right|. 
\end{equation*}
\noindent
Although the resulting term is not differentiable (but admits
subgradients), computational experiments indicated that violation
degrees are more appropriate for predicting OPFs than satisfiability
degrees. Observe also that an augmented Lagrangian method uses both
the satisfiability and violation degrees in its objective.

\smallskip

To define the violation degrees of the AC-OPF constraints, the
baseline model needs to extended to predict the reactive power
dispatched $\bm{q}^g$ and the voltage angles $\bm{\theta}$ of the
power network.  Given the predicted values
$\hat{\bm{v}}, \hat{\bm{\theta}}, \hat{\bm{p}}^g,$ and $\hat{\bm{q}}^g$, the
satisfiability degree of the OPF constraints can be
expressed as:
This section extends the homonym section of the main paper by 
reporting the complete set of satisfiability and violation 
degrees of the OPF constraints. 

Given the predicted values $\hat{\bm{v}}, \hat{\bm{\theta}}, 
\hat{\bm{p}}^g,$ and $\hat{\bm{q}}^g$, the satisfiability degree of 
the OPF constraints are expressed as follows:
{\small
\begin{flalign*}
\sigma_{2a}^L(\hat{v}_i) &= 
    (\dot{v}^{\min}_i - \hat{v}_i) &\!\!\!\!\!\!
    \forall i \in {\cal N}\\
\sigma_{2a}^R(\hat{v}_i) &=  
    (\hat{v}_i - \dot{v}^{\max}_i) &\!\!\!\!\!\! \forall i \in {\cal N}\\
\sigma_{2b}^L(\hat{\theta}_{ij}) &=  
    ((\hat{\theta}_j - \hat{\theta}_i) - \dot{\theta}^{\Delta}_{ij}) &\!\!\!\!\!\! \forall (ij) \in {\cal E}\\
\sigma_{2b}^R(\hat{\theta}_{ij}) &= 
    ((\hat{\theta}_i - \hat{\theta}_j) -\dot{\theta}^{\Delta}_{ij}) &\!\!\!\!\!\! \forall (ij) \in {\cal E}\\
\sigma_{3a}^L(\hat{p_i}^g) &= 
    \dot{p}^{g\min}_i - \hat{p}{}^g_i &\!\!\!\!\!\! \forall i \in {\cal N}\\
\sigma_{3a}^R(\hat{p_i}^g) &=  
    \hat{p}^g_i - \dot{p}^{\max}_i &\!\!\!\!\!\! \forall i \in {\cal N}\\
\sigma_{3b}^L(\hat{q_i}^g) &= 
    \dot{q}^{g\min}_i - \hat{q}{}^g_i &\!\!\!\!\!\! \forall i \in {\cal N}\\
\sigma_{3b}^R(\hat{q_i}^g) &=  
    \hat{q}^g_i - \dot{q}^{\max}_i &\!\!\!\!\!\! \forall i \in {\cal N}\\
\sigma_4(\tilde{p}^f_{ij}, \tilde{q}^f_{ij}) &= 
    (\tilde{p}^f_{ij})^2 + (\tilde{q}^f_{ij})^2 - \dot{S}^{\max}_{ij} &\!\!\!\!\!\! \forall (ij) \in {\cal E}\\
\sigma_{5a}(\tilde{{p}}^f_{ij}, {p}^f_{ij}) &= 
    \tilde{p}^f_{ij} - p_{ij}^f &\!\!\!\!\!\! \forall (ij) \in {\cal E}\\
\sigma_{5b}(\tilde{{q}}^f_{ij}, {q}^f_{ij}) &= 
    \tilde{q}^f_{ij} - q^f_{ij} &\!\!\!\!\!\! \forall (ij) \in {\cal E}\\
\sigma_{6a}(\hat{p}^g_i, \dot{p}^d_i, \tilde{\bm{p}}^f) &= 
      \sum_{(ij)\in {\cal E}} \tilde{p}^f_{ij} - (\hat{p}^g_i \!-\! \dot{p}^d_i)
      &\!\!\!\!\!\! \forall i \in {\cal N}\\
\sigma_{6b}(\hat{q}^g_i, \dot{q}^d_i, \tilde{\bm{q}}^f) &= 
      \sum_{(ij)\in {\cal E}} \tilde{q}^f_{ij} - (\hat{q}^g_i \!-\! \dot{q}^d_i)
      &\!\!\!\!\!\! \forall i \in {\cal N}
\end{flalign*}
}
\noindent
where $\sigma_{2a}^L$ and $\sigma_{2a}^R$ correspond to Constraints 
\eqref{con:2a} and capture the distance of the predictions $\hat{v}_i$ 
from exceeding the voltage bounds. 
The functions $\sigma_{2b}^{L}$ and $\sigma_{2b}^{R}$ correspond to 
Constraints \eqref{con:2b} and express how much the difference between 
two voltage angles exceeds the bound. 
Similarly, $\sigma_{3a}^{L}$, $\sigma_{3a}^{R}$, and $\sigma_{3b}^{L}$, 
$\sigma_{3b}^{R}$, relate to Constraints \eqref{con:3a} and 
\eqref{con:3b}, respectively, and describe the distance of the 
predicted generator active and reactive dispatch from their bounds. 
Function $\sigma_4$ corresponds to Constraints \eqref{con:4} and 
captures the distance of the power flow on line $(ij)$ from its bound. 
Therein, $\tilde{p}^f_{ij}$ and $\tilde{q}^f_{ij}$ are, respectively,
 the active and reactive power flow for line $(ij) \in {\cal E}$. 
Notice that  $\tilde{p}^f_{ij}$ and $\tilde{q}^f_{ij}$ are not 
predicted directly, as an output of the DNN. Instead, they are 
computed using the predicted quantities  $\hat{v}_i, \hat{v}_j, 
\hat{\theta}_i$, and $\hat{\theta}_j$ according to 
Constraints  \eqref{con:5a} and \eqref{con:5b}.
The quantities ${p}^f_{ij}$ and ${q}^f_{ij}$ correspond 
 to the ground truths values. 
Functions $\sigma_{5a}$ and $\sigma_{5b}$ measure the deviation of 
the predicted flow (based on the other predicted quantities)
to the ground truth values
according to the Ohm's Law (Constraints (5a) and (5b)). 
Finally, the functions $\sigma_{6a}$ and $\sigma_{6b}$ relate to the 
Kirchhoff Current Law (Constraints \eqref{con:6a} and \eqref{con:6b}) 
and express the violation of flow conservation at a bus.

The violation degrees associated to the satisfiability degree 
above are defined as follows:
{\small
\begin{flalign*}
\nu_{2a}(\hat{\bm{v}}) &= 
    \frac{1}{n} \sum_{i \in {\cal N}}
    \left(\nu_c^{\geq}\big(\sigma_{2a}^L(\hat{v}_i)\big) + 
    \nu_c^{\geq}\big(\sigma_{2a}^R(\hat{v}_i)\big) \right)\\
\nu_{2b}(\hat{\bm{\theta}}) &= 
    \frac{1}{e} \sum_{(ij) \in {\cal E}}
    \left(\nu_c^{\geq}\big(\sigma_{2b}^L(\hat{\theta}_{ij})\big) + 
    \nu_c^{\geq}\big(\sigma_{2b}^R(\hat{\theta}_{ij})\big) \right)\\
\nu_{3a}(\hat{\bm{p}}) &= 
    \frac{1}{n} \sum_{i \in {\cal N}}
    \left(\nu_c^{\geq}\big(\sigma_{3a}^L(\hat{p}_i)\big) + 
    \nu_c^{\geq}\big(\sigma_{3a}^R(\hat{p}_i)\big) \right)\\
\nu_{3b}(\hat{\bm{q}}) &= 
    \frac{1}{n} \sum_{i \in {\cal N}}
    \left(\nu_c^{\geq}\big(\sigma_{3b}^L(\hat{q}_i)\big) + 
    \nu_c^{\geq}\big(\sigma_{3b}^R(\hat{q}_i)\big) \right)\\
\nu_{4}(\tilde{\bm{p}}^f\!, \tilde{\bm{q}}^f) &= 
    \frac{1}{e} \sum_{(ij) \in {\cal E}}
    \nu_c^{\geq}\big(\sigma_{4}(\tilde{p}_{ij}^f, \tilde{q}_{ij}^f)\big)\\
\nu_{5a}(\tilde{\bm{p}}^f\!, \bm{p}^f) &= 
    \frac{1}{e} \sum_{(ij) \in {\cal E}}
    \nu_c^{=}\big(\sigma_{5a}(\tilde{p}_{ij}^f, {p}_{ij}^f)\big)\\
\nu_{5b}(\tilde{\bm{q}}^f\!, \bm{q}^f) &= 
    \frac{1}{e} \sum_{(ij) \in {\cal E}}
    \nu_c^{=}\big(\sigma_{5a}(\tilde{q}_{ij}^f, {q}_{ij}^f)\big)\\
\nu_{6a}(\hat{\bm{p}}^g\!, \dot{\bm{p}}^d\!, \bm{p}^f) &= 
    \frac{1}{e} \sum_{(ij) \in {\cal E}}
    \nu_c^{=}\big(\sigma_{6a}(\hat{p}^g_i, \dot{p}^d_i, \tilde{\bm{p}}^f)\big)\\
\nu_{6b}(\hat{\bm{q}}^g\!, \dot{\bm{q}}^d\!, \bm{q}^f) &= 
    \frac{1}{e} \sum_{(ij) \in {\cal E}}
    \nu_c^{=}\big(\sigma_{6b}(\hat{q}^g_i, \dot{q}^d_i, \tilde{\bm{q}}^f)\big).
\end{flalign*}
}
where $n$ and $e$ denote the number of buses and transmission lines,
respectively. These functions capture the average deviation by which the
prediction violates the associated constraint. The violations degrees define 
penalties that will be used to enrich the DNN loss function to encourage their 
satisfaction.  
Prior describing the DNN objective, we introduce a further extension that 
exploits yet another aspect of the structure of the OPF.

\section{Exploiting Existing Solutions}

The solving of an OPF (or a load flow) rarely happens in a cold-start: OPFs are typically solved in the context of an existing operating point and/or with the availability of solutions to similar instances
(\emph{hot-start}). As a result, the learning task can exploit this existing configuration, which is called the \emph{hot-start state} in this
paper.  The hot-start state is a tuple
$\bm{s}_0=\vect{\bm{p}^d_0,\bm{q}^d_0,\bm{p}^g_0,
\bm{q}^g_0,\bm{v}_0,\bm{\theta}_0}$, describing the load, the generation, 
and the voltages that are solutions to a related OPF.
The learning can then use a new, enriched, training dataset, defined
as follows:
{\small
\begin{align*}
	{\cal D} \!=\! \Big\{ \!\!&\big(
	(\overbrace{\bm{p}_0^d, \bm{q}_0^d, \bm{p}^g_0, \bm{q}^g_0, \bm{v}_0, \bm{\theta}_0, \bm{p}^d, \bm{q}^d}^{\bm{x}_1})_1, 
    \overbrace{(\bm{p}^g, \bm{q}^g, \bm{v}, \bm{\theta}}^{\bm{y}_1})_1\big), 
	\ldots,\\
	&\big((\underbrace{\bm{p}_0^d, \bm{q}_0^d, \bm{p}^g_0, \bm{q}^g_0, \bm{v}_0, \bm{\theta}_0, \bm{p}^d, \bm{q}^d}_{\bm{x}_N})_N, 
    \underbrace{(\bm{p}^g, \bm{q}^g, \bm{v}, \bm{\theta}}_{\bm{y}_N})_N\big) 
	 	\Big\}.
\end{align*}
} The elements $\bx_\ell \!\in\! \RR^{8n}$ are vectors describing the
hot-start state $\bm{s}_0$ (e.g., the configuration in the previous timestep) 
and the current loads $(\bm{p}^d, \bm{q}^d)$.  The
elements $\by_\ell \!\in\! \RR^{4n}$ are vectors describing the optimal
generator and voltage settings for input data $\bx_\ell$.  The
collection of the elements $\{\bx_\ell\}_{\ell=1}^N$ is denoted by
${\cal X}$ and the elements $\{\by_\ell\}_{\ell=1}^N$ by ${\cal
Y}$. The goal remain that of learning a mapping $\hat{\cal O}$.  Note
that, despite some proximity of loads in subsequent states, the OPF
non linearities often cause severe variations in the operational
parameters outputs. Therefore, as confirmed by our experimental 
results, the learning mechanism cannot rely exclusively on the 
information encoded in the hot-start state.

\section{Objective}

It is now possible to define the final loss function used to train the
OPF-DNN. First, the loss is augmented to consider the predictions of
voltage phase angles and the reactive power of generators, since these
are required to compute the violation degrees associated with the
OPF constraints. The resulting loss function
${\cal L}_o(\by, \hat{\by})$ is:
\begin{flalign}
	\label{obj_advanced}
	\underbrace{\| \bm{v} - \hat{\bm{v}} \|}_{{\cal L}_v(\by, \hat{\by})}{\!}^2 + 
	\underbrace{\| \bm{\theta} - \hat{\bm{\theta}} \|}_{{\cal L}_\theta(\by, \hat{\by})}{\!}^2 +
	\underbrace{\| \bm{p}^g - \hat{\bm{p}}^g \|}_{{\cal L}_p(\by, \hat{\by})}{\!}^2 +
	\underbrace{\| \bm{q}^g - \hat{\bm{q}}^g \|}_{{\cal L}_q(\by, \hat{\by})}{\!}^2.
\end{flalign}
It minimizes the mean squared error between the optimal voltage 
and generator settings $\bm{y}$ and the predicted ones $\hat{\bm{y}}$.

Moreover, the objective function includes the Lagrangian relaxation 
based on the OPF physical and engineering constraints violation degrees. 
Given the set ${\cal C}$ of OPF constraints, the associated loss is 
captured by the expression
\begin{flalign*}
{\cal L}_c(\bx, \hat{\by}) &= \sum_{c \in {\cal C}} \lambda_c \nu_c(\bx,\hat{\by}).
\end{flalign*}
The model loss function sums these two terms, i.e.,
\[
{\cal L}(\bx,\by,\hat{\by}) = {\cal L}_o(\by, \hat{\by}) + {\cal L}_c(\bx, \hat{\by}).
\]

\section{The Network Architecture}

The network architecture is outlined in Figure \ref{fig:dlopf}.  The
input layers on the left process the tensor of loads
$(\bm{p}^d_0, \bm{q}^d_0)$ of the hot-start state $\bm{s}_0$ and
the input loads $(\bm{p}^d, \bm{q}^d)$.  The network has 4 basic
units, each following a decoder-encoder structure and composed by a
number of fully connected layers with ReLU activations. Each
subnetwork predicts a target variable: voltage magnitudes
$\hat{\bm{v}}$, phase angles $\hat{\bm{\theta}}$, active power
generations $\hat{\bm{p}}^g$, and reactive power generations
$\hat{\bm{q}}^g$.  Each sub-network takes as input the corresponding
tensor in the hot-start state $\bm{s}_0$ (e.g., the sub-network 
responsible for predicting the voltage magnitude $\hat{\bm{v}}$ 
takes as input $\bm{v}_0$), as well as the last hidden layer of its 
input subnetwork, that processes the load tensors.

The predictions for the voltage magnitude $\hat{\bm{v}}$ and angle
$\hat{\bm{\theta}}$ are used to compute the load flows
$(\tilde{\bm{p}}^f, \tilde{\bm{q}}^f\!)$, as illustrated on the bottom
of the Figure. The components of the losses are highlighted in the
white boxes and a full description of the network architecture is
provided in the Appendix.

\section{Lagrangian Duality}

Let $\hat{\cal O}[\bw]$ be the resulting OPF-DNN with weights $\bw$
and let ${\cal L}[\blambda]$ be the loss function parametrized by the
Lagrangian multipliers $\blambda = \{\lambda_c\}_{c \in {\cal C}}$.
The training aims at finding the weights $\bw$ that minimize the loss
function for a given set of Lagrangian multipliers, i.e., it computes
\[
{\it LR}(\blambda) = \min_{\bw} {\cal L}[\blambda](\bx,\by,\hat{\cal O}[\bw](\bx)).
\]
It remains to determine appropriate Lagrangian multipliers. This paper proposes the use of Lagrangian duality to
obtain the optimal Lagrangian multipliers when training the OPF-DNN, i.e., it solves
\[
{\it LD} = \max_{\blambda} {\it LR}(\blambda).
\]
The Lagrangian dual is solved through a subgradient method that 
computes a sequence of multipliers
$\blambda^1,\ldots,\blambda^k,\ldots$ by solving a sequence of
trainings ${\it LR}(\blambda^0),\ldots,{\it
LR}(\blambda^{k-1}),\ldots$ and adjusting the multipliers using the
violations, i.e.,
\begin{align}
\bm{w}^{k+1} &= \argmin_{\bm{w}} {\cal L}[\blambda^k](\bx,\by,\hat{\cal O}[\bw^k](\bx)) \label{eq:L1} \tag{L1}\\
\bm{\lambda}^{k+1} &= \vect{\lambda^k_c + \rho\,\nu_c(\bx,\hat{\cal O}[\bw^{k+1}](\bx)) \;|\; c\in {\cal C}}. \label{eq:L2} \tag{L2}
\end{align}
\noindent
In the implementation, step \eqref{eq:L1} is approximated using a
Stochastic Gradient Descent (SGD) method. Importantly, this step does not
recomputes the training from scratch but uses a hot start for the
weights $\bw$.

The overall training scheme is presented in
Algorithm \ref{alg:learning}.  It takes as input the training dataset
$({\cal X}, {\cal Y})$, the optimizer step size $\alpha > 0$ and the
Lagrangian step size $\rho > 0$.  The Lagrangian multipliers are
initialized in line \ref{line:1}. The training is performed for a
fixed number of epochs, and each epoch optimizes the weights using a 
minibatch of size $b$. 
After predicting the voltage and generation power quantities
(line \ref{line:4}), the objective and constraint losses are computed
(lines \ref{line:6} and \ref{line:7}).  The latter uses
the Lagrangian multipliers $\bm{\lambda}^k$ associated with current
epoch $k$. The model weights are updated in line \ref{line:8}.
Finally, after each epoch, the Lagrangian multipliers are updated
following step \eqref{eq:L2} described above (lines \ref{line:9}
and \ref{line:10}).

\begin{algorithm}[!t]
  \caption{Learning Step}
  \label{alg:learning}
  \setcounter{AlgoLine}{0}
  \SetKwInOut{Input}{input}

  \Input{$({\cal X}, {\cal Y}):$ Training data\\
  		 $\alpha, \rho:$ Optimizer and Lagrangian step sizes, reps.\!\!\!\!\!\!\!\!\!\!}
  \label{line:1}
  $\blambda^0 \gets 0 \;\; \forall c \in {\cal C}$\\
  \For{epoch $k = 0, 1, \ldots$} { 
  \label{line:2}
  	\ForEach{$(\bx, \by) \!\gets\! \mbox{minibatch}({\cal X}, {\cal Y})$ of size $b$}{
	  \label{line:3}
	  	$\hat{\by} \gets \hat{\cal O}[\bw](\bx)$\\
  		\label{line:4}
	  	${\cal L}_o(\hat{\by}, \by) \gets \frac{1}{b}
	  	 	\sum_{\ell \in [b]} 
	  	 	{\cal L}_v(\by_\ell, \hat{\by}_\ell) + 
	  	 	{\cal L}_\theta(\by_\ell, \hat{\by}_\ell)+$\!\!\!\!\\
	  	 	\nonl
	  	 	$\hspace*{87pt}
	  	 	{\cal L}_p(\by_\ell, \hat{\by}_\ell) + 
	  	 	{\cal L}_q(\by_\ell, \hat{\by}_\ell)$\\
	  	\label{line:6}
	  	${\cal L}_c(\bx,\hat{\by}) \gets \frac{1}{b} 
	  	 	\sum_{\ell \in [b]} 
	  	 	\sum_{c \in {\cal C}} \lambda_c^k \nu_c(\bx_\ell,\hat{\by}_\ell) $\\
	  	\label{line:7}
	  	$\omega \gets \omega - \alpha \nabla_{\omega} 
	  		({\cal L}_o(\hat{\by}, \by) 
	  		+ {\cal L}_c(\bx,\hat{\by}))\!\!\!\!$
	  	\label{line:8}
	}
	\ForEach{$c \in {\cal C}$} {
  	\label{line:9}
  		$\lambda^{k+1}_c \gets \lambda^k + \rho \nu_c(\bx,\hat{\by})$ 
  		\label{line:10}
  	}
  }
\end{algorithm}


\begin{table*}[!t]
\centering
\resizebox{0.65\linewidth}{!}
{
  \begin{tabular}{l|rrrrr |rr rr rr}
	\toprule
  	\multicolumn{6}{l}{}  &
  	\multicolumn{2}{c}{$\Delta_{1\%}~p^d$} &
  	\multicolumn{2}{c}{$\Delta_{2\%}~p^d$} &
  	\multicolumn{2}{c}{$\Delta_{3\%}~p^d$} \\
    \cmidrule(r){7-8}
    \cmidrule(r){9-10}
    \cmidrule(r){11-12}
  	\multicolumn{1}{l}{\textbf{Test Case}} &  $|{\cal N}|$ & $|{\cal E}|$ & $l$ & $g$ &
  	\multicolumn{1}{r}{$|({\cal X}, {\cal Y})|$}
   	&$(\%)$ & MW &$(\%)$ & MW &$(\%)$ & MW\\
  	\midrule
	\textbf{14\_ieee     }& 14& 40&  11 & 2  &	 395806 &2.05& 5.3  &2.59& 6.7   &3.15 & 8.2	\\
	\textbf{30\_ieee     }& 30& 82&  21 & 2  &	 273506 &2.47& 7.0  &2.94& 8.3   &3.36 & 9.5	\\
	\textbf{39\_epri     }& 39& 92&  21 & 10 &	 287390 &2.49&156.3 &2.94& 183.9 &3.42 & 213.9	\\
	\textbf{57\_ieee     }& 57& 160& 42 & 4  & 	 269140 &2.65&33.2  &3.19& 39.9  &3.67 & 45.9	\\
	\textbf{73\_ieee\_rts}& 73& 240& 51 & 73 & 	 373142 &2.72&233.2 &3.28& 281.2 &3.80 & 324.9	\\
	\textbf{89\_pegase   }& 89& 420& 35 & 12 & 	 338132 &2.50&204.0 &3.06& 250.1 &3.53 & 288.0	\\
	\textbf{118\_ieee    }& 118& 372& 99 & 19 &  395806 &3.03&128.6 &3.50& 148.8 &3.98 & 169.1	\\
	\textbf{162\_ieee\_dtc}& 162& 568& 113& 12 & 237812 &3.10&296.5 &3.54& 337.9 &4.04 & 385.9	\\
	\textbf{189\_edin    }& 189& 412& 41 & 35 &  69342  &2.85&39.1  &3.27& 44.8  &3.72 & 50.9	\\
	\textbf{300\_ieee    }& 300& 822& 201& 57 & 235732  &3.25&775.9 &3.78& 902.8 &4.22 & 1007.0	\\
	\bottomrule
	\end{tabular}
}
\caption{The Power Networks Adopted as Benchmarks.}
\label{tbl:dataset} 
\end{table*}

\section{Experiments}

This section evaluates the predictive accuracy of OPF-DNN and compares 
it to the AC model and its linear DC approximation. 
It also analyzes various design decisions in detail.

\paragraph{Data sets}
The experiments examine the proposed models on a variety of mid-sized 
power networks from the NESTA library \cite{Coffrin14Nesta}. 
The ground truth data are constructed as follows: For each network,
different benchmarks are generated by altering the amount of nominal
load $\bx \!=\! (\bm{p}^d, \bm{q}^d)$ within a range of $\pm 20\%$. 
The loads are thus sampled from the distributions $\bx'=(\bm{p}^d{}',
\bm{q}^d{}') \sim \text{Uniform}(0.8\bx, 1.2\bx)$. 
Notice that the resulting benchmarks have load demands that vary by 
a factor of up to $20\%$ of their nominal values: Many of them become 
congested and significantly harder computationally than their original
counterparts.  A network value that constitutes a dataset entry
$(\bm{x}', \bm{y}')$ is a feasible OPF solution obtained by solving
the AC-OPF problem \mbox{detailed in Model~\ref{model:ac_opf}}.

When the learning step exploits an existing hot-start state $\bm{s}_0$, 
the training test cases have the property that the total active loads
$\| \bm{p}^d_0 \|_1$ in $\bm{s}_0$ are within 1, 2, and 3\% of the total
active loads $\| \bm{p}^d \|_1$.  Note that, while the aggregated
loads follow this restriction, the individual loads may have 
greater variations. Those are illustrated in Table \ref{tbl:dataset}
for the 1\% ($\Delta_{1\%} p^d$), 2\% ($\Delta_{2\%} p^d$) and 
3\% ($\Delta_{3\%} p^d$) cases, where the average variations are 
expressed both in percentage of the total load and in absolute values (MWs). 
As can be seen, the variations are significant.  
The table also describes the dataset sizes, including the number of buses 
$|{\cal N}|$ and transmission lines $|{\cal E}|$ of the networks. The column 
titled $l$ and $g$ denote, respectively, the number of load and generator 
buses of the networks. 
The data are normalized using the per unit (pu) system so that 
all quantities are close to 1. 
The experiments use a $80/20$ train-test split and report results on
the test set.

\paragraph{Settings}
The experiments examine the OPF-DNN models whose features are
summarized in Table \ref{tbl:models}.  ${\cal M}_\B$ refers to the
baseline model: It minimizes the loss function ${\cal L}_o$ described
in Equation \eqref{basic_loss}.  ${\cal M}_\C$ exploits the problem
constraints and minimizes the loss: ${\cal L}_o + \sum_{c \in {\cal
C}} \lambda_c \nu_c$, with ${\cal L}_o$ defined in
Equation \eqref{obj_advanced} and all $\lambda_c$ set to $1$.  
The suffix $S$ is used for the models that exploit a hot-start state,
and $D$ is used for the model that exploit the Lagrangian dual scheme.  
In particular, ${\cal M}_\CD$ extends ${\cal M}_\C$ by learning the 
Lagrangian multipliers $\lambda_c$ using the Lagrangian dual scheme 
described in Algorithm \ref{alg:learning}.
${\cal M}_\CL$ uses the same loss function as ${\cal M}_\C$, 
but it adopts the architecture outlined in Figure \ref{fig:dlopf}.
${\cal M}_\CLL$ sets the Lagrangian weights as trainable parameters 
and learns them during the training cycle. 
Finally, ${\cal M}_\CLD$ extends ${\cal M}_\CL$ by learning the 
Lagrangian multipliers $\lambda_c$ using the Lagrangian dual scheme 
(see Algorithm \ref{alg:learning}).  
The latter model is also denoted with OPF-DNN in the paper.  
The details of the models architectures and loss functions are provided
in the appendix. 
All the models that exploit a hot-start state are trained over datasets
using states differing by at most $1\%$.  
The section also reports a comparison of the DNN-OPF model trained over 
hot-start state datasets using states differing by at most $1$, $2$, and 
$3\%$.

\begin{table}[!tb]
\centering
\resizebox{\linewidth}{!}
{
  \begin{tabular}{l|cccccc}
  \toprule
    \multicolumn{1}{l}{\textbf{Model}} 
    & {${\cal M}_\B$} & {${\cal M}_\C$} & {${\cal M}_\CD$} & {${\cal M}_\CL$} & {${\cal M}_\CLL$} & {${\cal M}_\CLD$}  \\
    \midrule
    Exploit \textbf{C}onstraints             
    	& $\Box$ & \cBox  & \cBox  &\cBox   &\cBox   & \cBox \\
    Exploit hot-start \textbf{S}tate          
    	& $\Box$ & $\Box$ & $\Box$ &\cBox   &\cBox   & \cBox \\
    Train \textbf{L}agrangian weights 
    	& $\Box$ & $\Box$ & $\Box$ & $\Box$ &\cBox   & $\Box$ \\    
    Lagrangian \textbf{D}ual update  
    	& $\Box$ & $\Box$ & \cBox & $\Box$ & $\Box$ & \cBox \\
    \bottomrule
  \end{tabular}
}
\caption{The DNN Models Adopted.}
\label{tbl:models} 
\end{table}

The models were implemented using the Julia package PowerModels.jl
\cite{Coffrin:18} with the nonlinear solver IPOPT \cite{wachter06on} 
for solving the nonlinear AC model and its the DC approximation. 
The DDN models were implemented using PyTorch \cite{paszke:17} with 
Python 3.0. 
The training was performed using NVidia Tesla V100 GPUs and and 2GHz 
Intel Cores. The AC and DC-OPF models were solved using the same CPU
cores. Training each network requires less than 2GB of RAM.  
The training uses the Adam optimizer with learning rate ($\alpha\!=\!0.001$)
and $\beta$ values $(0.9, 0.999)$ and was performed for $80$
epochs using batch sizes $b=64$. 
Finally, the Lagrangian step size $\rho$ is set to $0.01$.

\subsection{Prediction Errors}

\def\b{\textbf}
\begin{table*}[t]
\centering
\small
\resizebox{0.9\linewidth}{!}
  {
  \begin{tabular}{llrrrrr|llrrrrr}
  \toprule
  \textbf{Test case} & \textbf{Model} & $\hat{\bm{p}}^g$ & $\hat{\bm{q}}^g$ & $\hat{\bm{v}}$& $\hat{\bm{\theta}}$& $\tilde{\bm{p}}^f$  &
  \textbf{Test case} & \textbf{Model} & $\hat{\bm{p}}^g$ & $\hat{\bm{q}}^g$ & $\hat{\bm{v}}$& $\hat{\bm{\theta}}$& $\tilde{\bm{p}}^f$  \\
  \midrule
  \multirow{5}{*}{\textbf{14\_ieee}}   
     &     ${\cal M}_\B$  &5.7820 &11.004 &0.7310 &1.4050 &1.9070 &\multirow{5}{*}{\textbf{89\_pegase}}&   ${\cal M}_\B$  &0.2516&0.2250&90.689&37.176&3133.4\\
     &     ${\cal M}_\C$  &6.1396 &11.315 &1.2790 &1.4100 &0.4640 &&                                       ${\cal M}_\C$  &0.3589&0.2320&77.295&7.9760&42.962\\
     &     ${\cal M}_\CD$  &5.5698 &7.1120 &6.0682 &0.0500 &0.4970 &&                                       ${\cal M}_\CD$  &0.3549&0.3361&2.8380&17.921&24.529\\
     &     ${\cal M}_\CL$ &0.2756 &0.6980 &0.1180 &0.1480 &0.1050 &&                                       ${\cal M}_\CL$ &0.1074&0.0860&9.4168&0.8240&6.4130\\
     &     ${\cal M}_\CLL$&0.2703 &0.7450 &0.1860 &0.0760 &0.1690 &&                                       ${\cal M}_\CLL$&0.1014&0.0830&10.199&0.9120&9.3560\\
     &     ${\cal M}_\CLD$&\b{0.0234}&\b{0.0470} &\b{0.0050} &\b{0.0070} &\b{0.0530} &&                    ${\cal M}_\CLD$&\b{0.0797}&\b{0.0770}&\b{0.0862}&\b{0.0530}&\b{5.0160}\\
    \hline 
    \multirow{5}{*}{\textbf{30\_ieee}}
     &    ${\cal M}_\B$  &3.3465&2.0270&14.699&4.3400&27.213&\multirow{5}{*}{\textbf{118\_ieee}}&    ${\cal M}_\B$  &0.2150&2.9910&7.1520&4.2600&38.863\\
     &    ${\cal M}_\C$  &3.1289&1.3380&2.7346&1.5930&1.6820&&                                       ${\cal M}_\C$  &0.1810&3.2570&6.9150&4.6520&6.4730\\
     &    ${\cal M}_\CD$  &3.1230&1.1096&0.1596&0.2590&2.3000&&                                       ${\cal M}_\CD$  &0.1787&1.0840&10.002&0.2160&2.8100\\
     &    ${\cal M}_\CL$ &0.3052&0.1104&0.3130&0.0580&0.2030&&                                       ${\cal M}_\CL$ &0.0380&0.6900&0.1170&1.2750&0.6640\\
     &    ${\cal M}_\CLL$&0.2900&0.3200&0.3120&0.0600&0.1600&&                                       ${\cal M}_\CLL$&0.0380&0.6870&0.1380&1.2750&0.6100\\
     &    ${\cal M}_\CLD$&\b{0.0055}&\b{0.0320}&\b{0.0070}&\b{0.0041}&\b{0.0620}&&                   ${\cal M}_\CLD$&\b{0.0340}&\b{0.6180}&\b{0.0290}&\b{0.2070}&\b{0.4550}\\
    \hline
    \multirow{5}{*}{\textbf{39\_epri}}     
     &    ${\cal M}_\B$   &0.2299&1.2600&98.726&58.135&202.67&\multirow{5}{*}{\textbf{162\_ieee}}& ${\cal M}_\B$ &0.2310&1.2070&9.1810&5.4800&82.076\\
     &    ${\cal M}_\C$   &0.2180&1.2790&17.104&2.5940&80.064&&                                    ${\cal M}_\C$ &0.2820&1.6120&7.1210&5.3620&14.706\\
     &    ${\cal M}_\CD$   &0.2216&1.2610&2.7346&4.0730&41.395&&                                    ${\cal M}_\CD$ &0.2772&0.9873&6.8359&1.1950&15.456\\
     &    ${\cal M}_\CL$  &0.0559&0.1080&2.2350&0.2880&1.8360&&                                    ${\cal M}_\CL$ &0.0750&0.3760&0.1760&0.3720&0.7520\\
     &    ${\cal M}_\CLL$ &0.0533&0.1440&2.2010&0.2040&2.2000&&                                    ${\cal M}_\CLL$&0.0750&0.3690&0.1750&0.3950&0.6750\\
     &    ${\cal M}_\CLD$ &\b{0.0024}&\b{0.0720}&\b{0.0280}&\b{0.0100}&\b{1.2660}&&               ${\cal M}_\CLD$&\b{0.0710}&\b{0.2440}&\b{0.0770}&\b{0.3660}&\b{0.4920}\\
    \hline
    \multirow{5}{*}{\textbf{57\_ieee}}    
    &    ${\cal M}_\B$ &2.3255&1.6380&5.1002&1.5680&14.386&\multirow{5}{*}{\textbf{189\_edin}}&     ${\cal M}_\B$ &0.4979&0.1160&42.295&5.2970&4371.1\\
    &    ${\cal M}_\C$ &2.2658&1.4850&3.5402&2.0890&2.8850&&                                        ${\cal M}_\C$ &0.5748&0.0890&18.577&3.9640&24.918\\
    &    ${\cal M}_\CD$ &2.2708&1.5138&9.5861&0.0680&1.6170 &&                                        ${\cal M}_\CD$ &0.4081&0.0711&7.3091&3.2220&15.774\\
    &    ${\cal M}_\CL$ &0.1308&0.3320&0.2150&0.0430&0.2410&&                                        ${\cal M}_\CL$ &0.1178&0.0190&1.9913&0.7040&3.8470\\
    &    ${\cal M}_\CLL$&0.1340&0.3300&0.2110&0.0360&0.2280&&                                        ${\cal M}_\CLL$&0.1178&0.0180&2.4300&0.4960&3.5810\\
    &    ${\cal M}_\CLD$&\b{0.0170}&\b{0.0231}&\b{0.0150}&\b{0.0080}&\b{0.1520}&&                    ${\cal M}_\CLD$&\b{0.0907}&\b{0.0110}&\b{0.0982}&\b{0.3330}&\b{1.6520}\\
    \hline
    \multirow{5}{*}{\textbf{73\_ieee}}    
    &    ${\cal M}_\B$  &0.2184&0.0380&18.414&5.0550&106.08&  \multirow{5}{*}{\textbf{300\_ieee}} &  ${\cal M}_\B$ &0.0838&0.0900&28.025&12.137&125.47\\
    &    ${\cal M}_\C$  &0.0783&0.0360&2.8074&1.2500&7.8630&&                                        ${\cal M}_\C$ &0.0914&0.0860&14.727&7.7450&34.133\\
    &    ${\cal M}_\CD$  &0.0775&0.5302&2.7038&0.3880&6.2980&&                                        ${\cal M}_\CD$ &0.0529&0.0491&11.096&7.3830&27.554\\
    &    ${\cal M}_\CL$ &0.0061&0.0160&0.2192&0.1190&0.4890&&                                        ${\cal M}_\CL$ &0.0174&0.0240&3.1130&7.2330&26.905\\
    &    ${\cal M}_\CLL$&0.0063&0.0150&0.3156&0.1260&0.4160&&                                        ${\cal M}_\CLL$&0.0139&0.0240&0.2180&4.6480&2.0180\\
    &    ${\cal M}_\CLD$&\b{0.0050}&\b{0.0101}&\b{0.0235}&\b{0.1180}&\b{0.3300}&&                    ${\cal M}_\CLD$&\b{0.0126}&\b{0.0190}&\b{0.0610}&\b{2.5670}&\b{1.1360}\\
  \bottomrule
  \end{tabular}
  }
  \caption{Prediction Errors (\%).}
  \label{tbl:prediction_errors} 
\end{table*}

This section first analyzes the prediction error of the DNN models.
Table \ref{tbl:prediction_errors} reports the average L1 distance
between the predicted generator active $\hat{\bm{p}}^g$ and 
reactive $\hat{\bm{q}}^g$ power, 
voltage magnitude $\hat{\bm{v}}$ and angles $\hat{\bm{\theta}}$ and the
original quantities. It also reports the errors of the predicted flows
$\tilde{\bm{p}}^f$ (which use the generator power and voltage
predictions) and are important to assess the fidelity of the
predictions.  The distances are reported in percentage: 
$\frac{\|\hat{\bx} - \bx\|_1}{\|\bx\|_1} \!\times\! 100$,
for quantity $\bx$, and best results are highlighted in bold.  
For completeness, the results report an extended version of model 
${\cal M}_\B$, that allow us to predict quantities $\bm{\theta}$ 
and $\bm{q}^d$. The latter were obtained by extending ${\cal M}_\B$ 
network using two additional layers, Out-$\bm{\theta}$ and Out-$\bm{q}^g$ 
for, respectively, predicting the voltage angles and the reactive generator 
power, analogously to those in model ${\cal M}_\C$. Additionally, its loss 
function was extended as: 
\begin{align*}
  {\cal L}_o(\bm{y}, \hat{\bm{y}}) &= 
     \| \bm{v} - \hat{\bm{v}}\|^2 + \| \bm{\theta} - \hat{\bm{\theta}}\|^2 \\
  &+ \| \bm{p}^g - \hat{\bm{p}}^g\| ^2 + \| \bm{q}^g - \hat{\bm{q}}^g\| ^2.
\end{align*}
The prediction errors for quantities $\bm{p}^g$ and $\bm{v}$ did not 
degrade in this extended version with respect to those predicted by 
the simple ${\cal M}_\B$ network.

A clear trend appears: The prediction errors decrease with the
increasing of the model complexity. In particular, model ${\cal M}_\C$, 
which exploits the problem constraints, predicts much better
voltage quantities and power flows than ${\cal M}_\B$.  
The use of the Lagrangian Duals, in model $\cal{M}_\CL$, further improve 
the predictions, especially those associated to the voltage magnitude and 
angles and power flows. 
${\cal M}_\CL$, which exploits the problem constraints and a hot-start state, 
improves ${\cal M}_\C$ predictions by one order of magnitude in most of the 
cases. {\em Finally, the use of the Lagrangian dual to find the best weights
(${\cal M}_\CLD$) further improves ${\cal M}_\CL$ predictions by up to
an additional order of magnitude.}

\begin{figure*}[!h]
\centering
\includegraphics[width=0.5\linewidth]{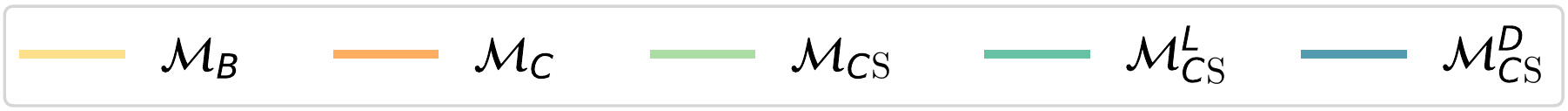}\\
\footnotesize{NESTA case 14\_ieee}\\
\includegraphics[width=0.24\linewidth]{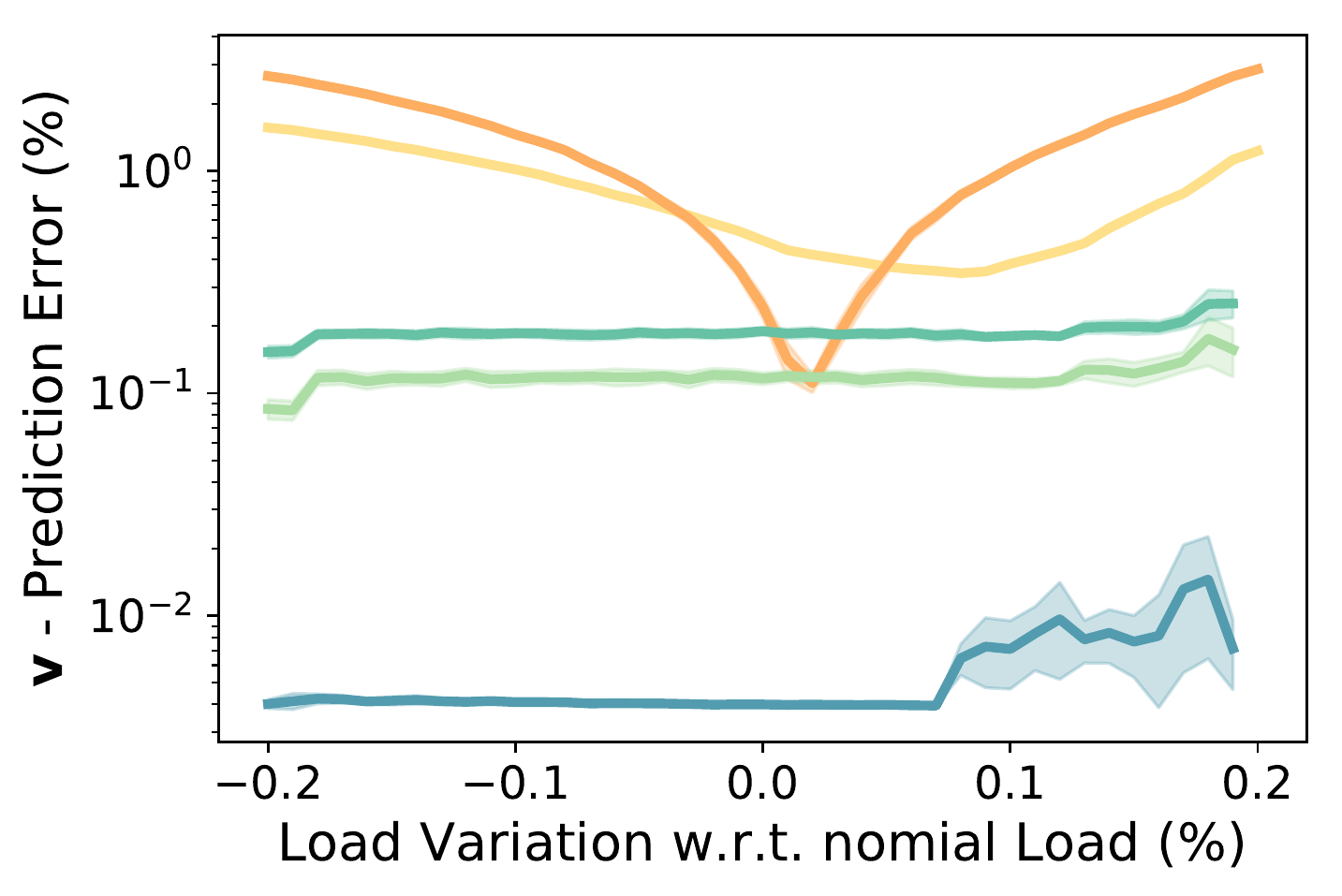}
\includegraphics[width=0.24\linewidth]{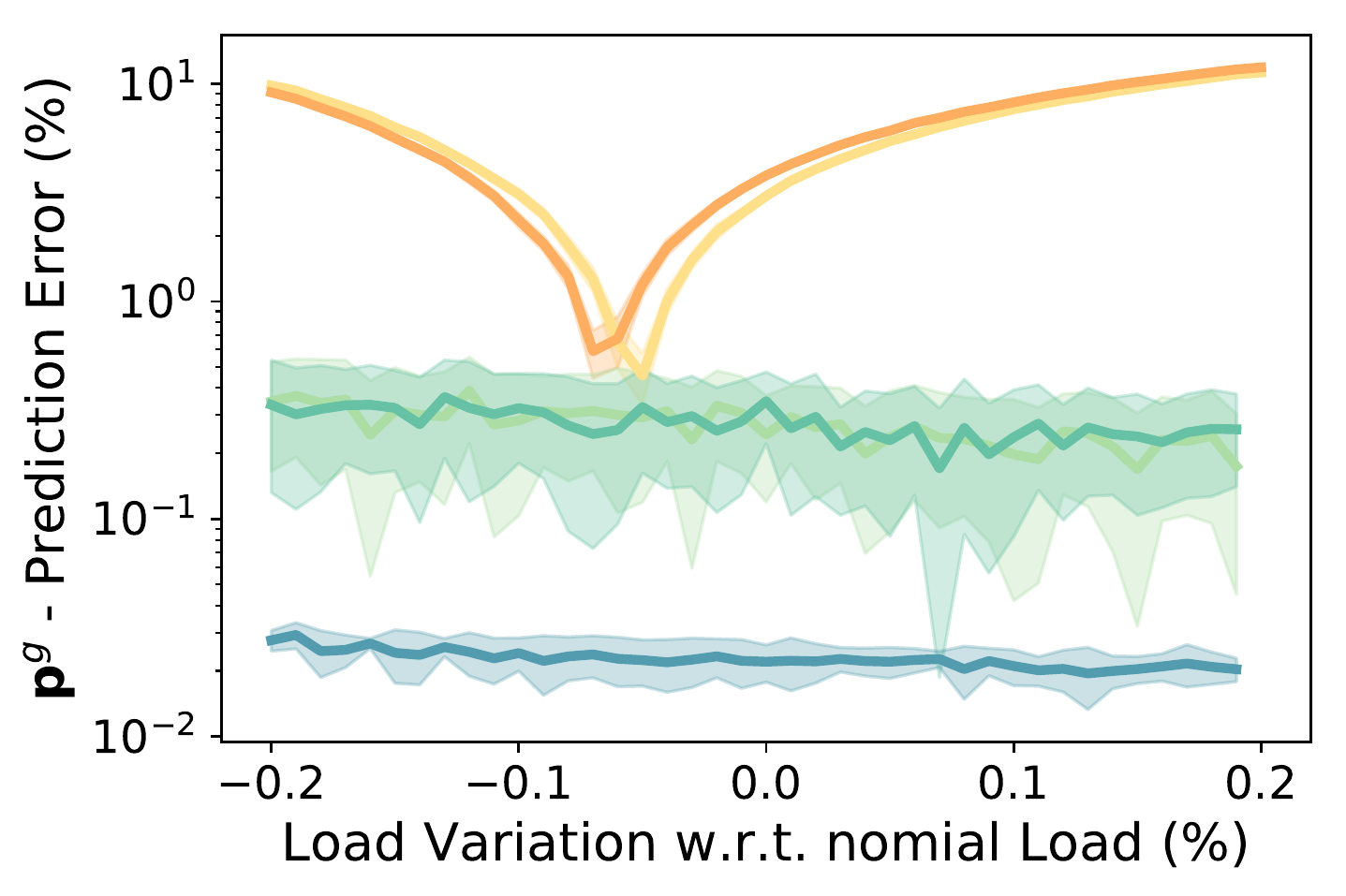}
\includegraphics[width=0.24\linewidth]{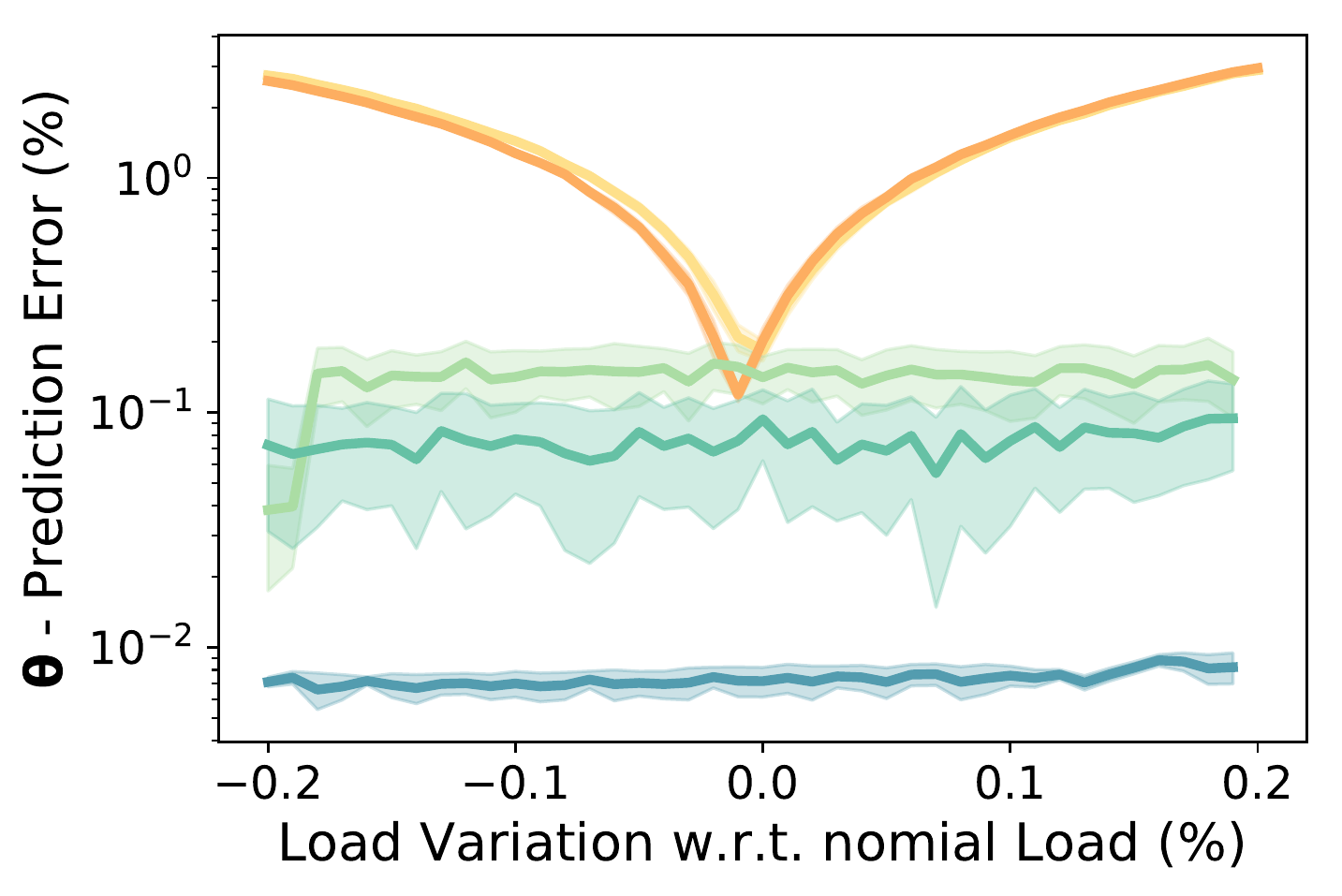}
\includegraphics[width=0.24\linewidth]{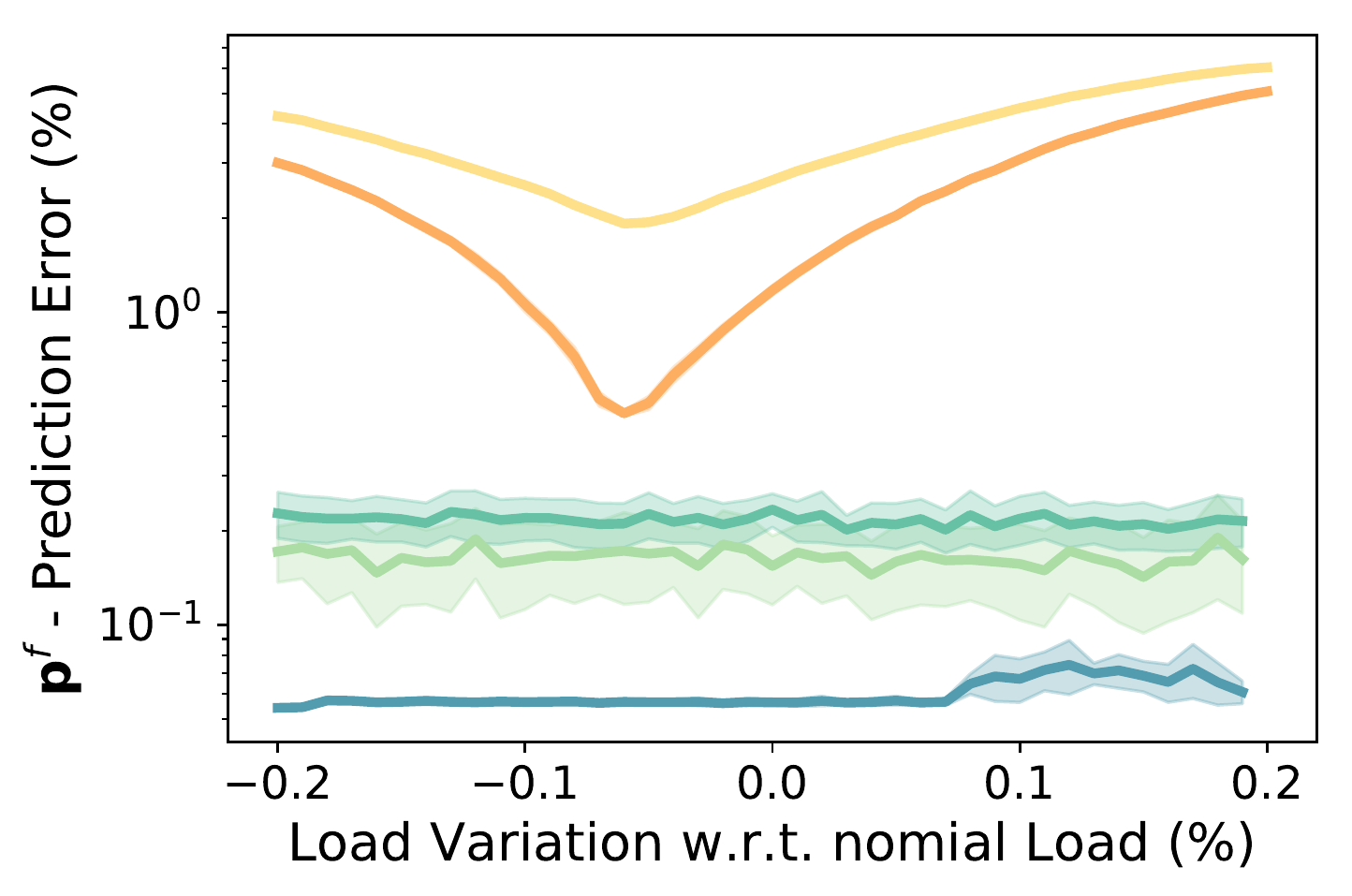}\\
\footnotesize{NESTA case 30\_ieee}\\
\includegraphics[width=0.24\linewidth]{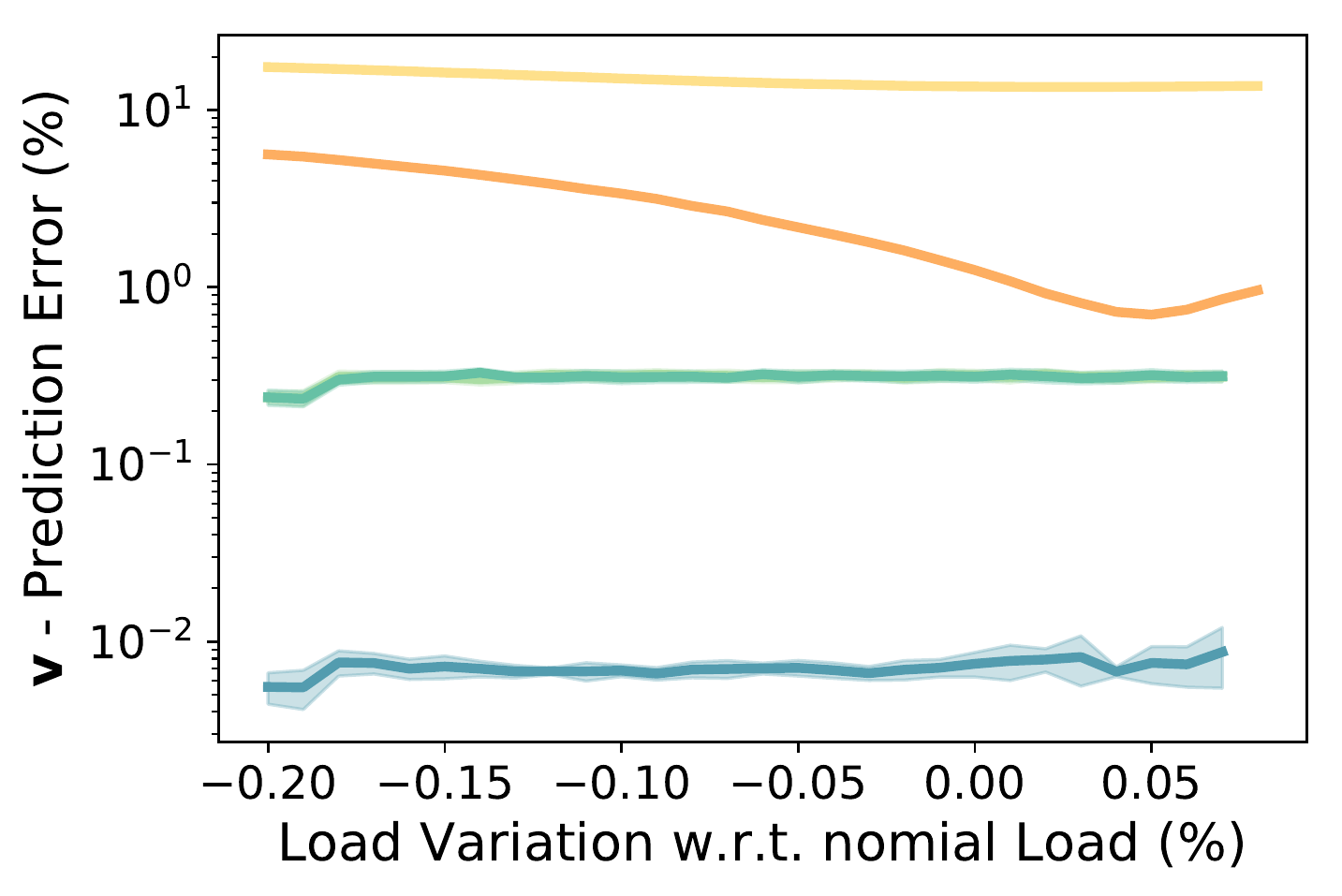}
\includegraphics[width=0.24\linewidth]{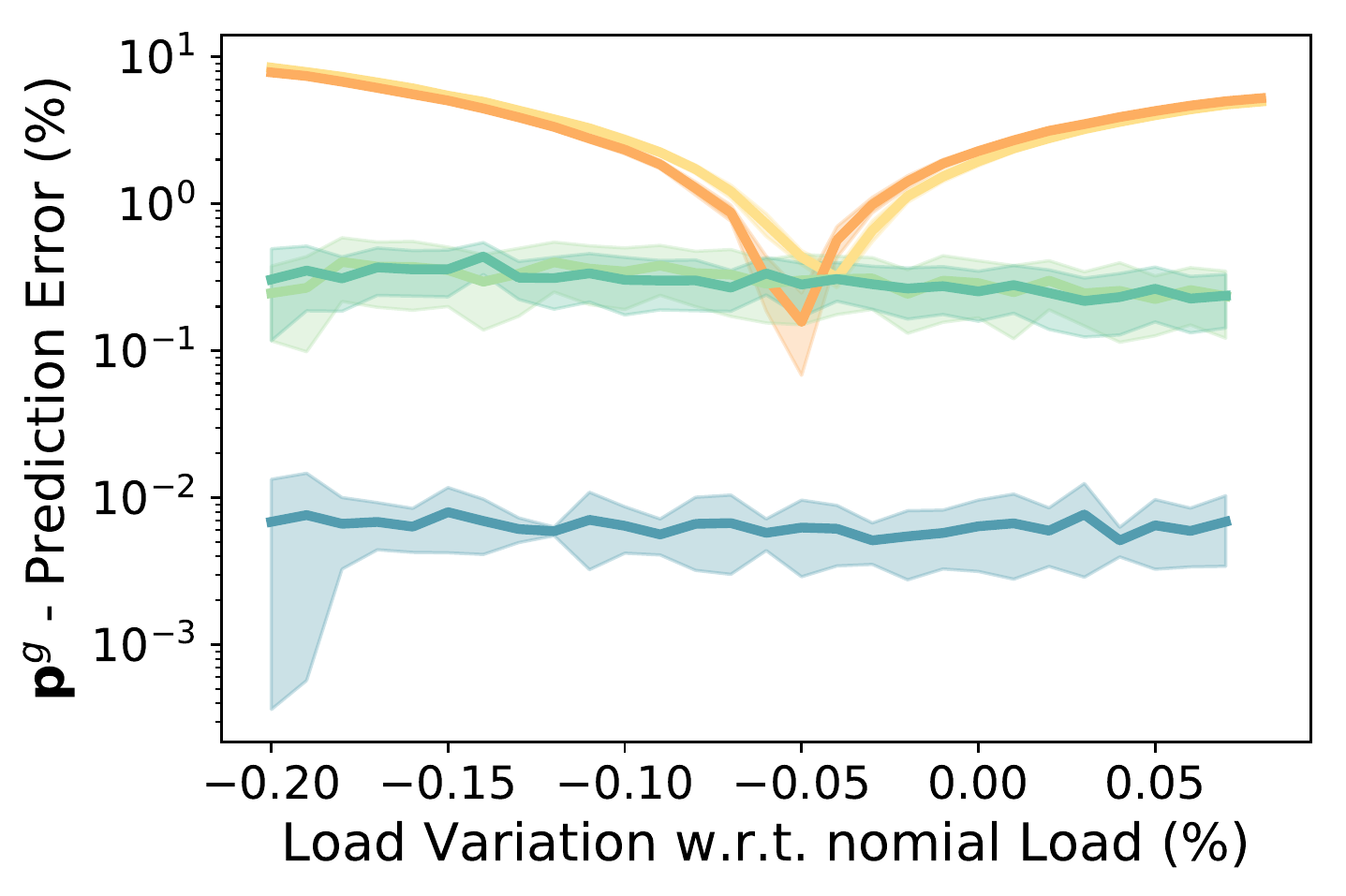}
\includegraphics[width=0.24\linewidth]{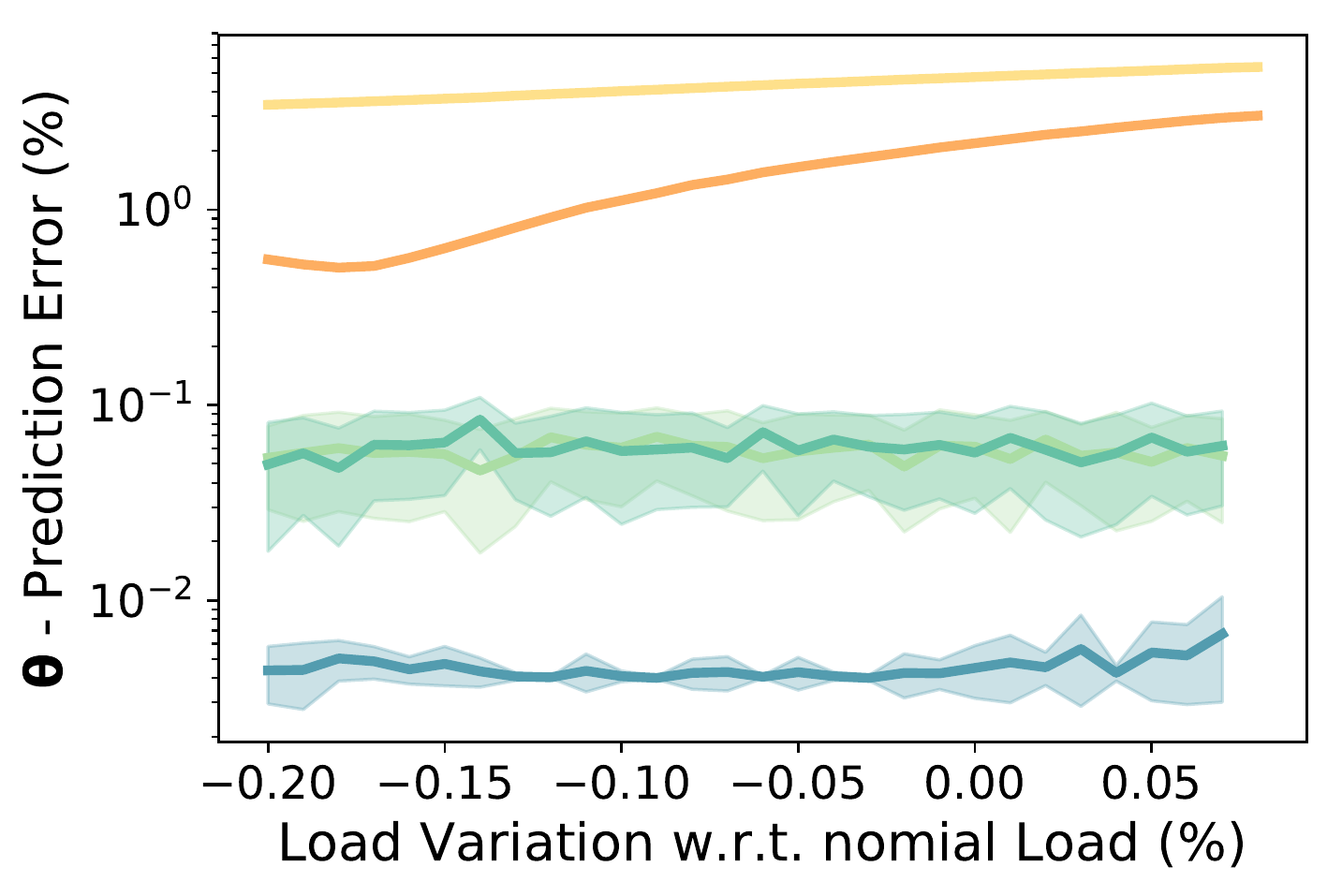}
\includegraphics[width=0.24\linewidth]{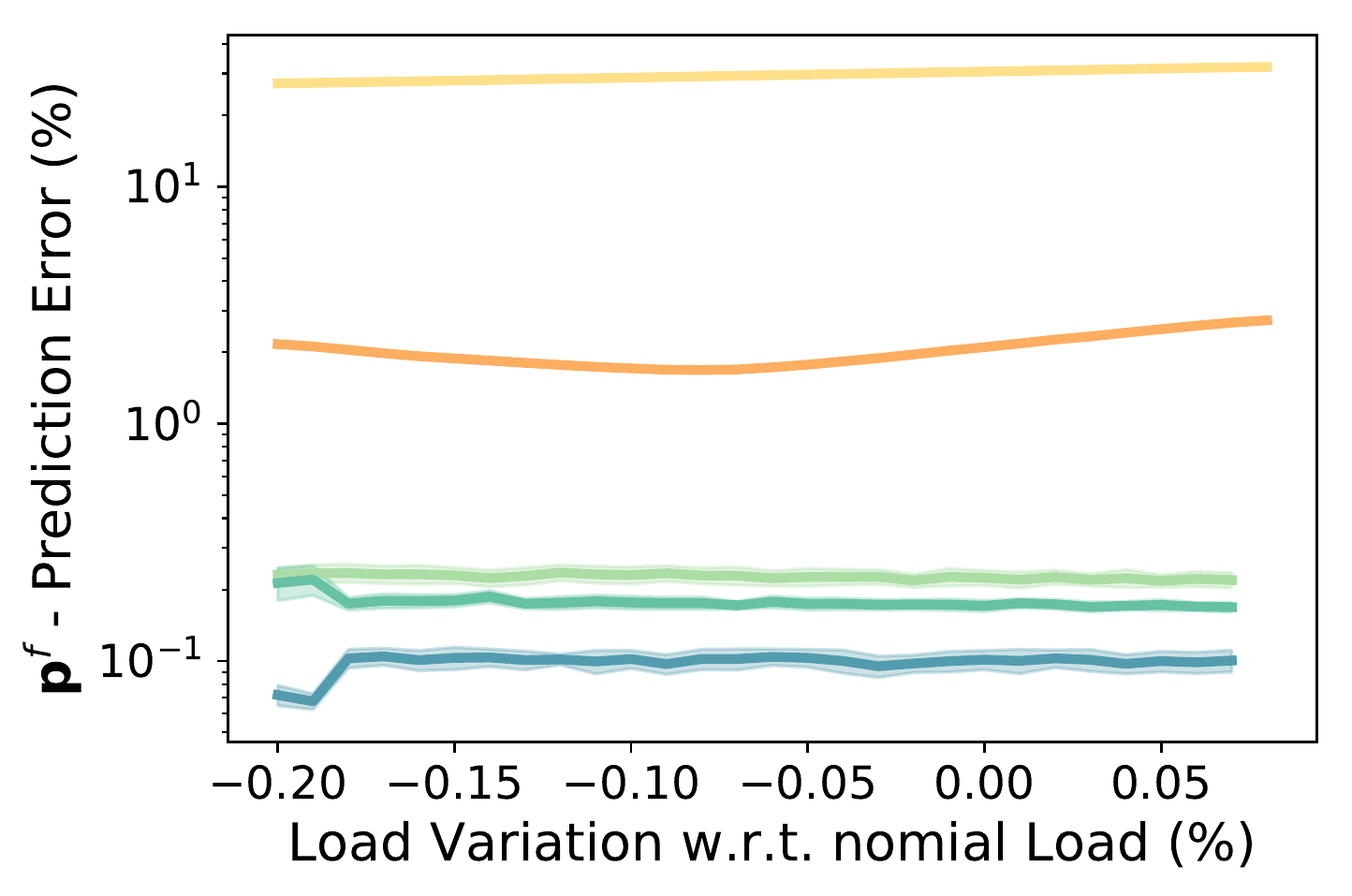}\\
\footnotesize{NESTA case 39\_epri}\\
\includegraphics[width=0.24\linewidth]{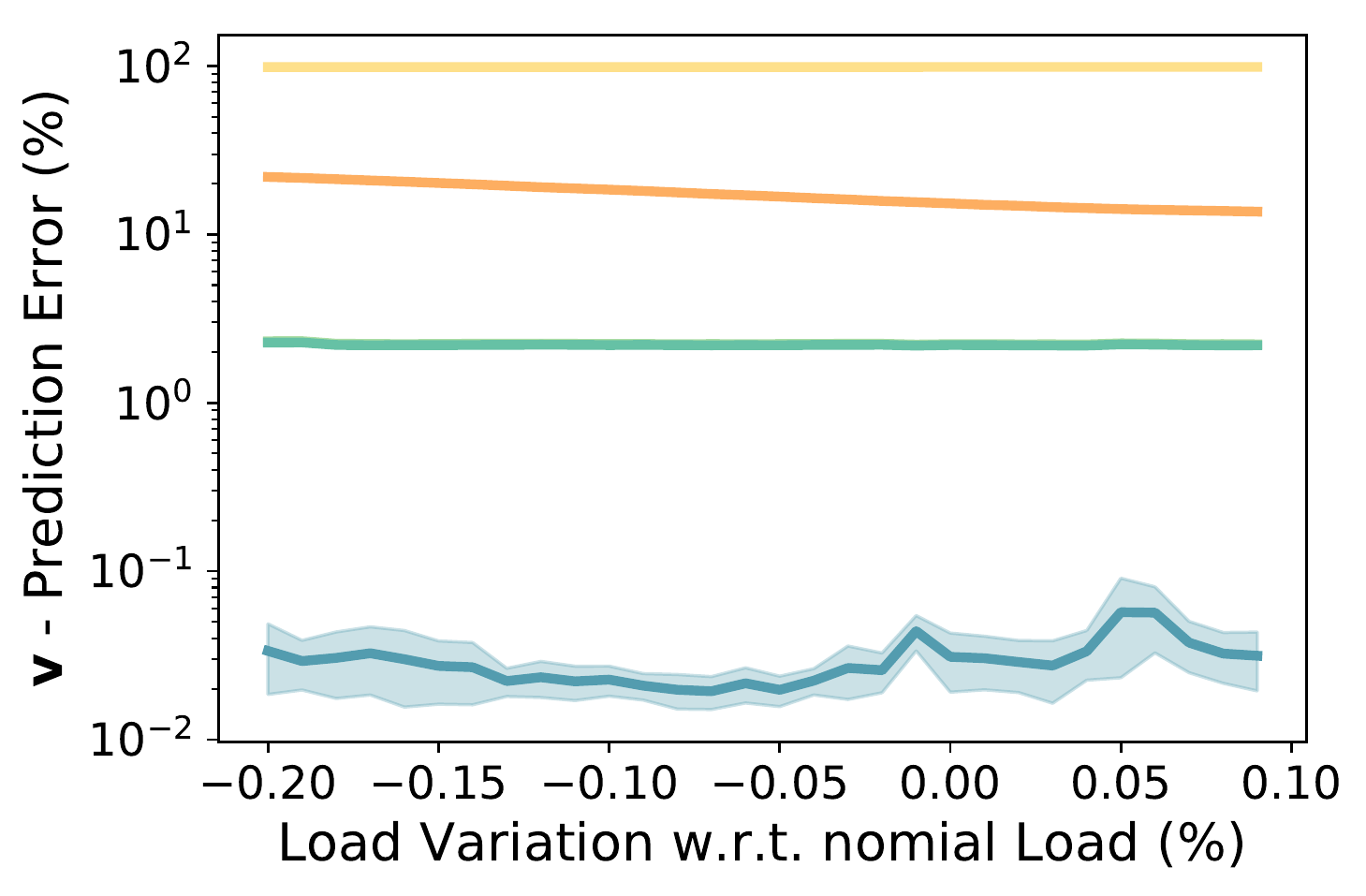}
\includegraphics[width=0.24\linewidth]{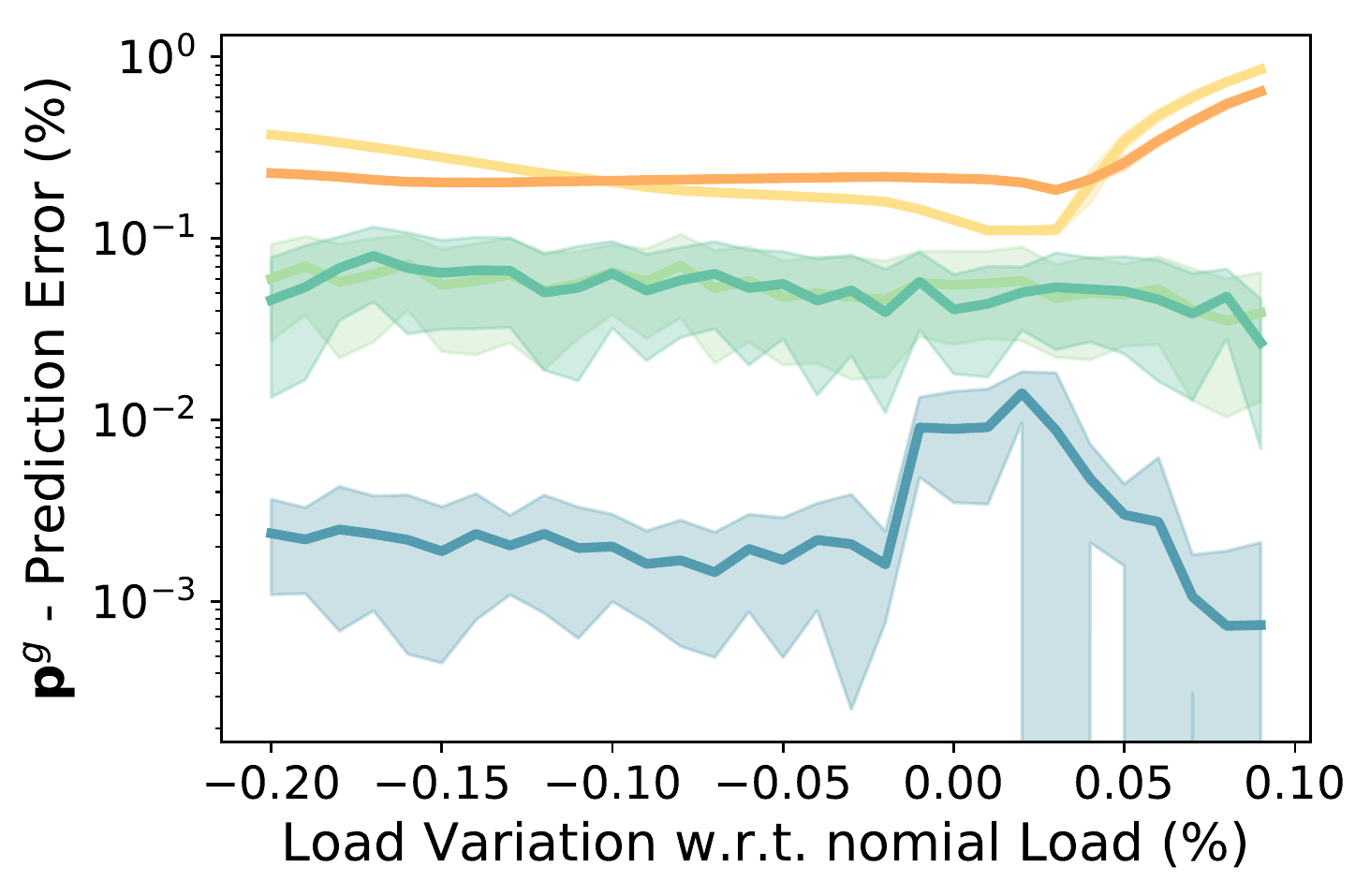}
\includegraphics[width=0.24\linewidth]{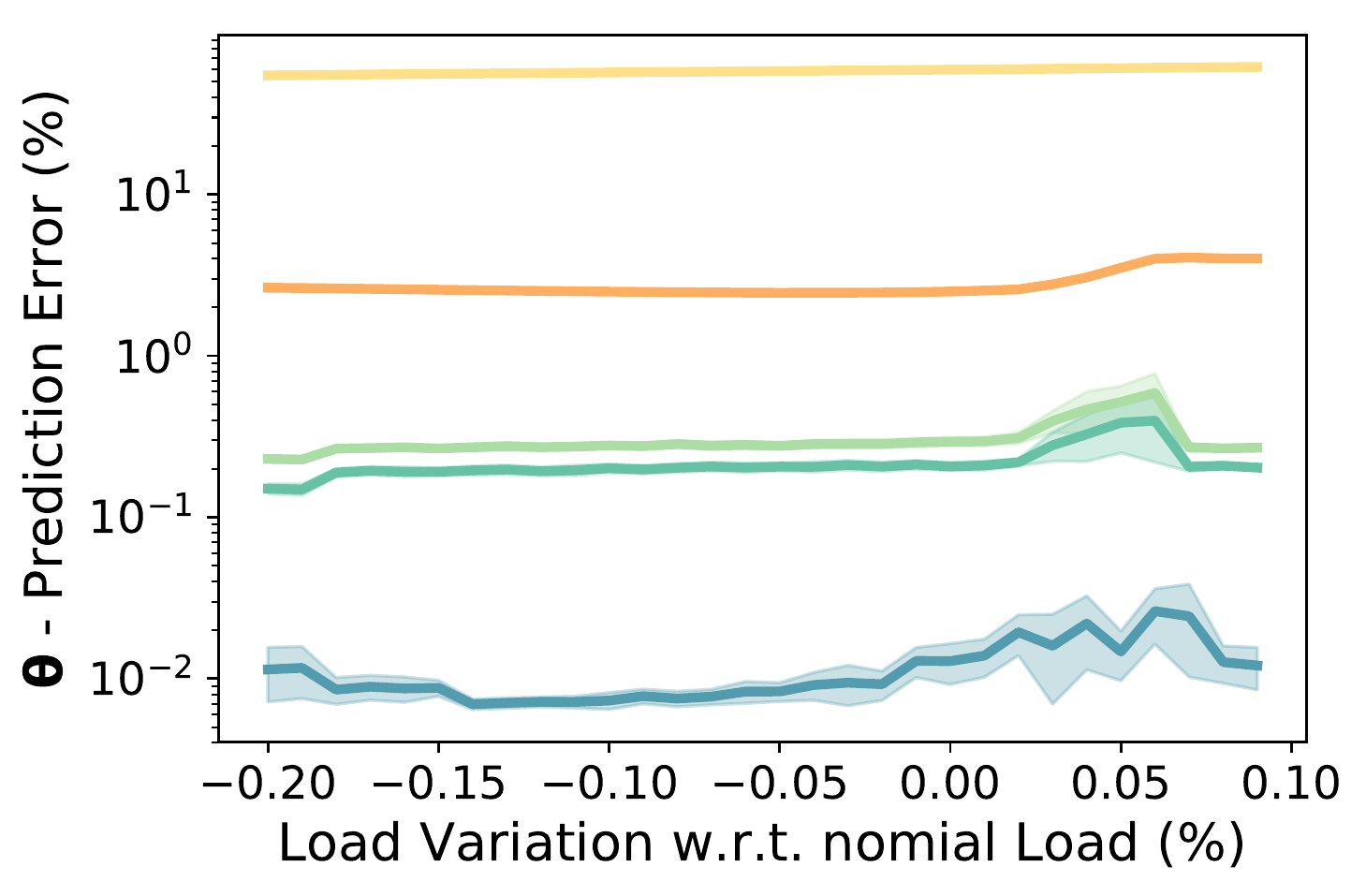}
\includegraphics[width=0.24\linewidth]{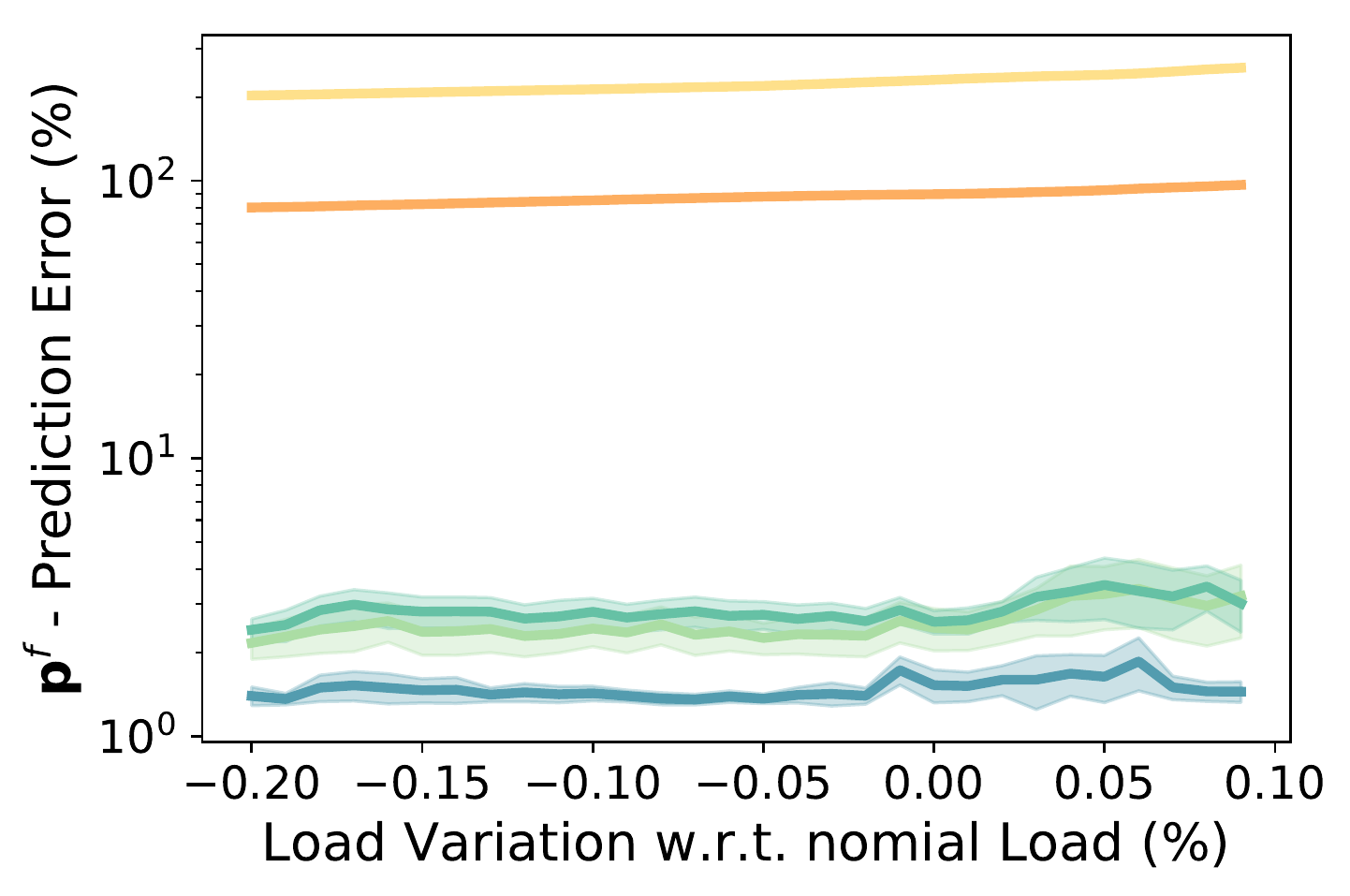}\\
\footnotesize{NESTA case 57\_ieee}\\
\includegraphics[width=0.24\linewidth]{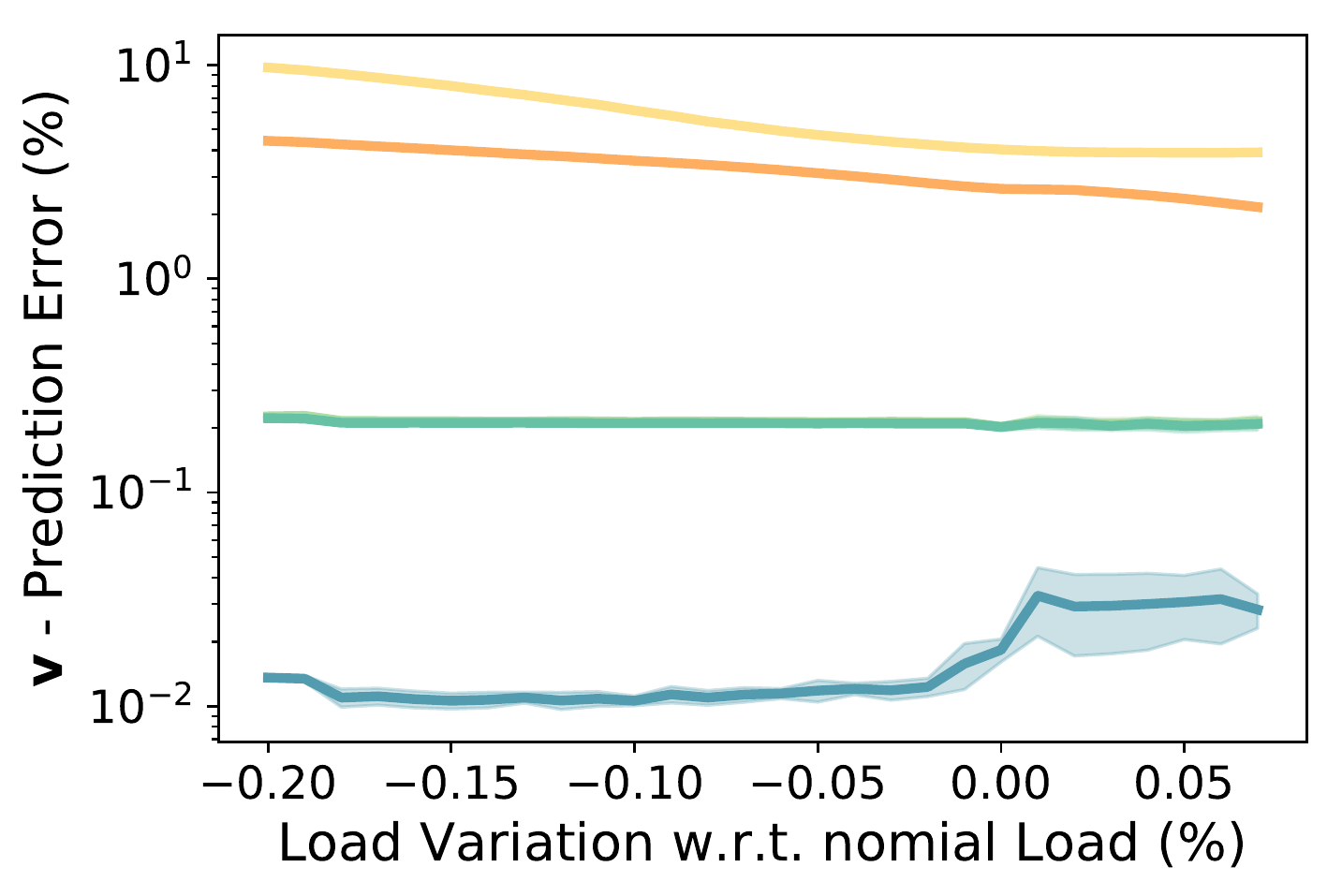}
\includegraphics[width=0.24\linewidth]{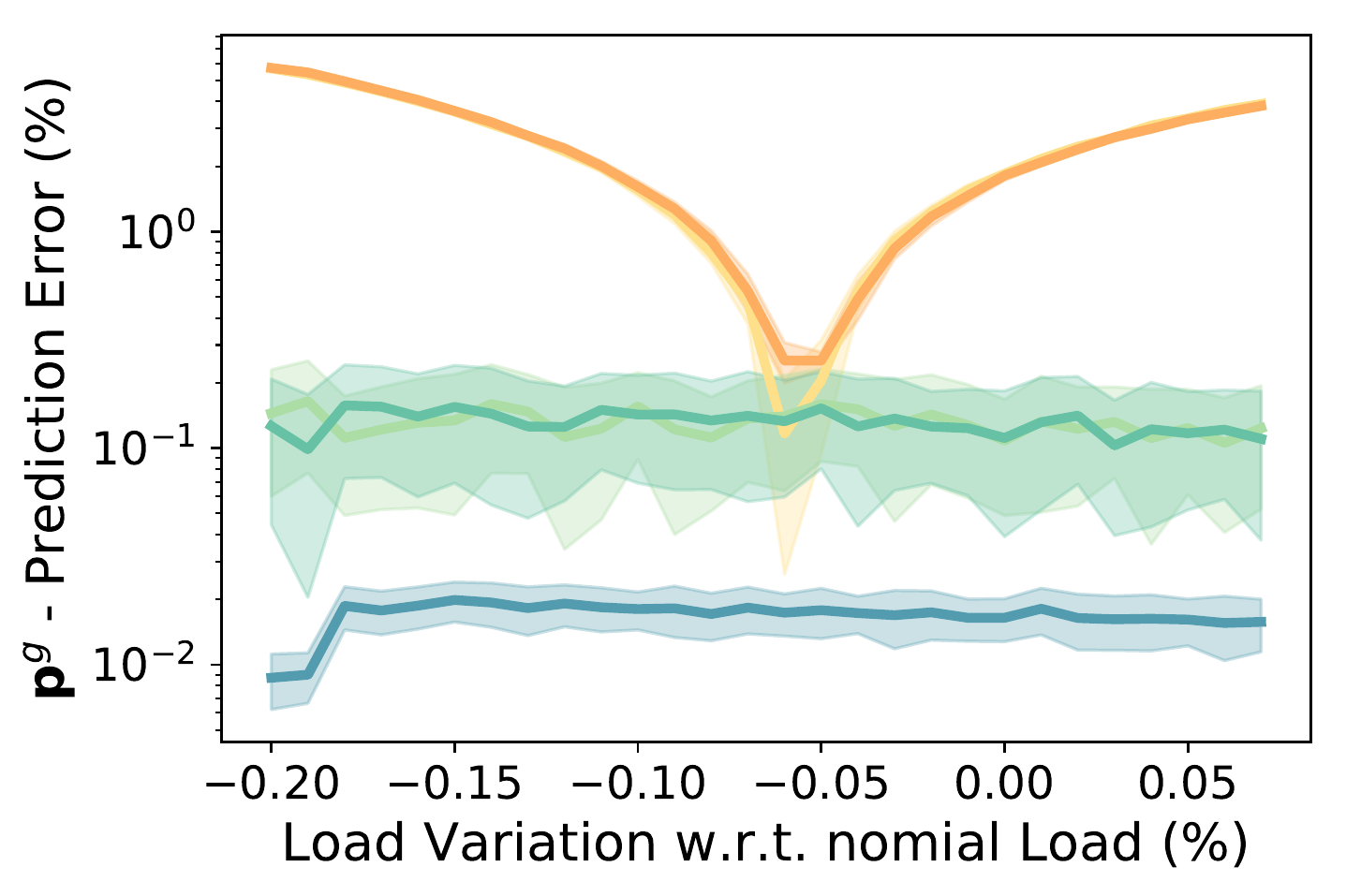}
\includegraphics[width=0.24\linewidth]{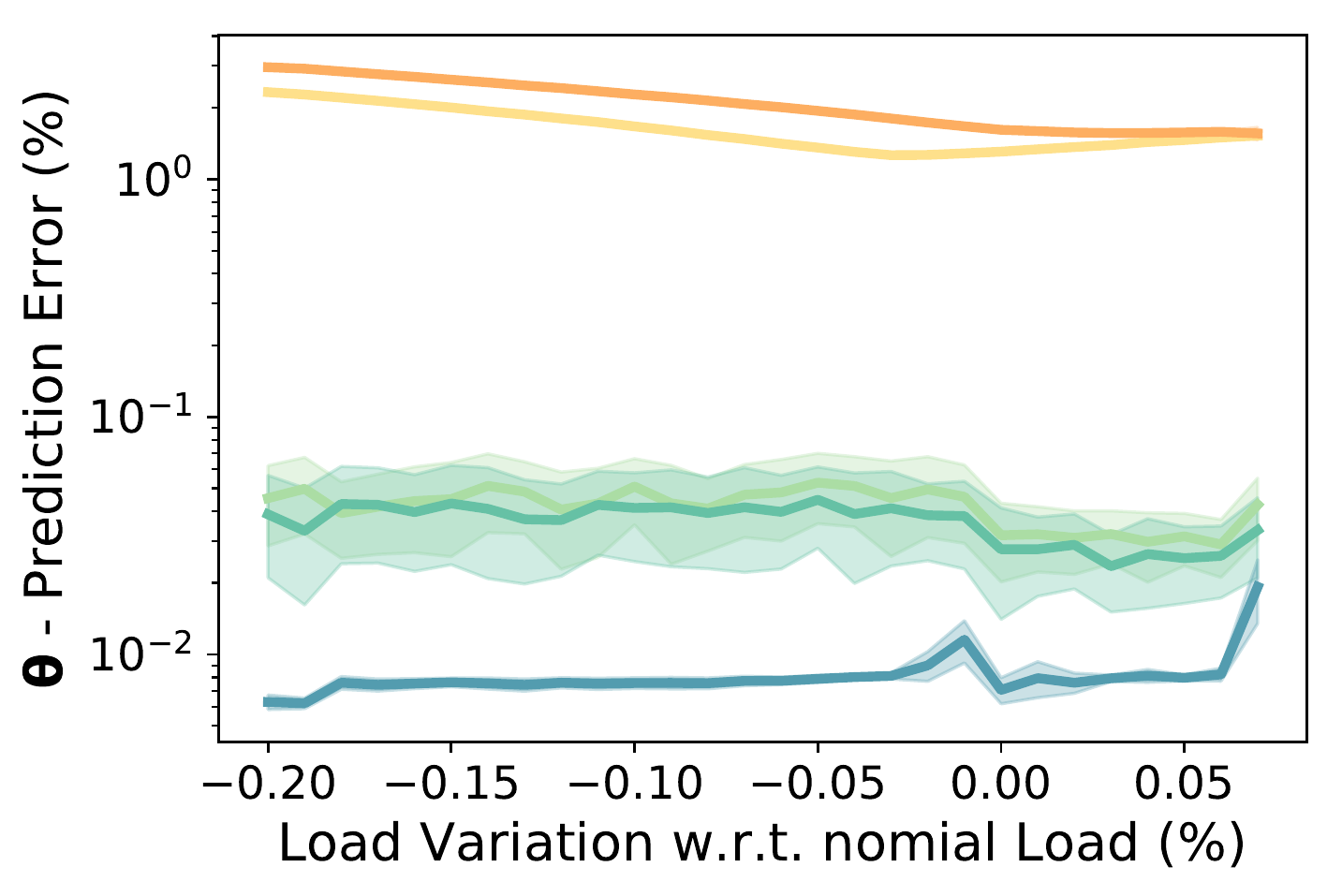}
\includegraphics[width=0.24\linewidth]{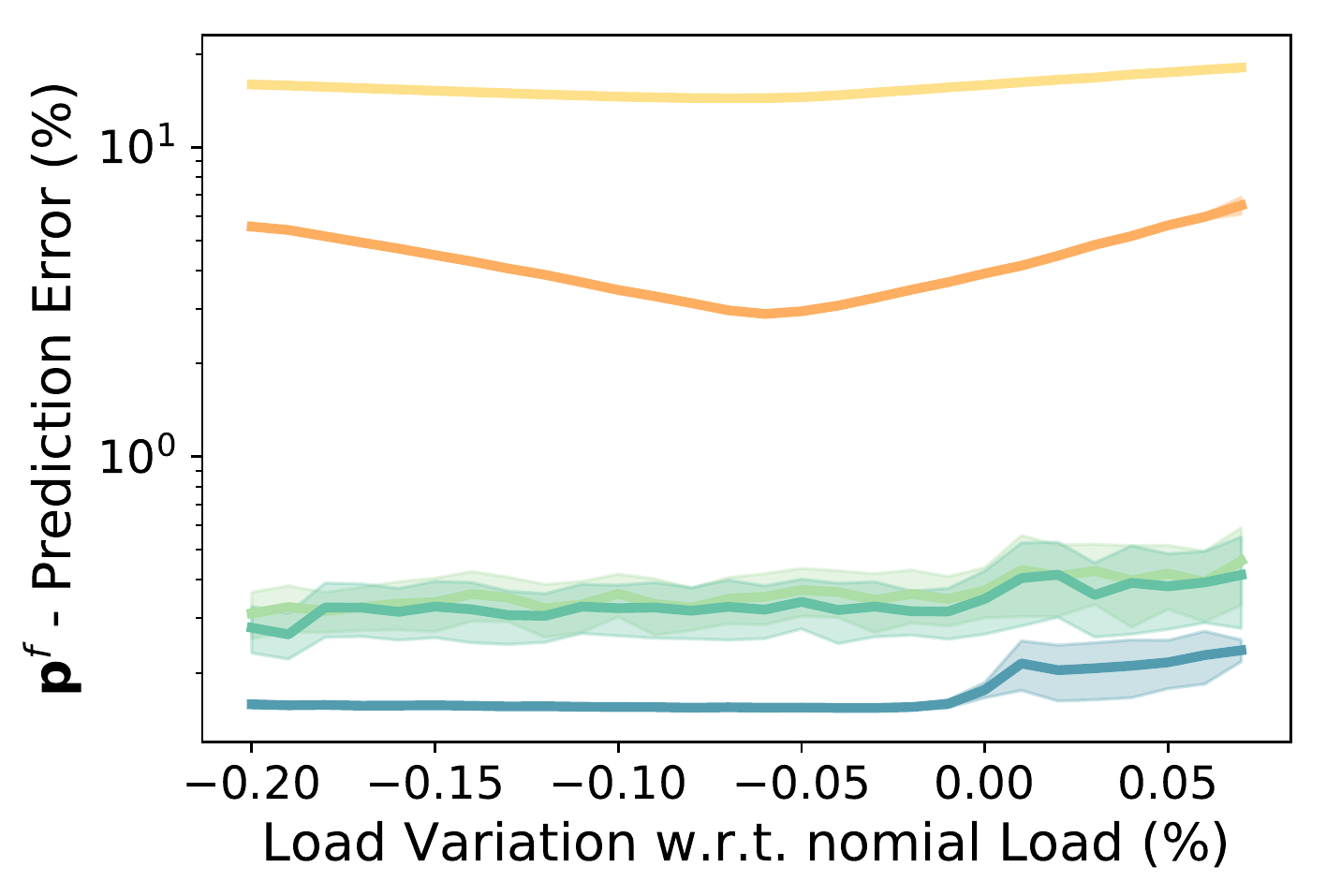}
\\
\footnotesize{NESTA case 73\_ieee\_rts}\\
\includegraphics[width=0.24\linewidth]{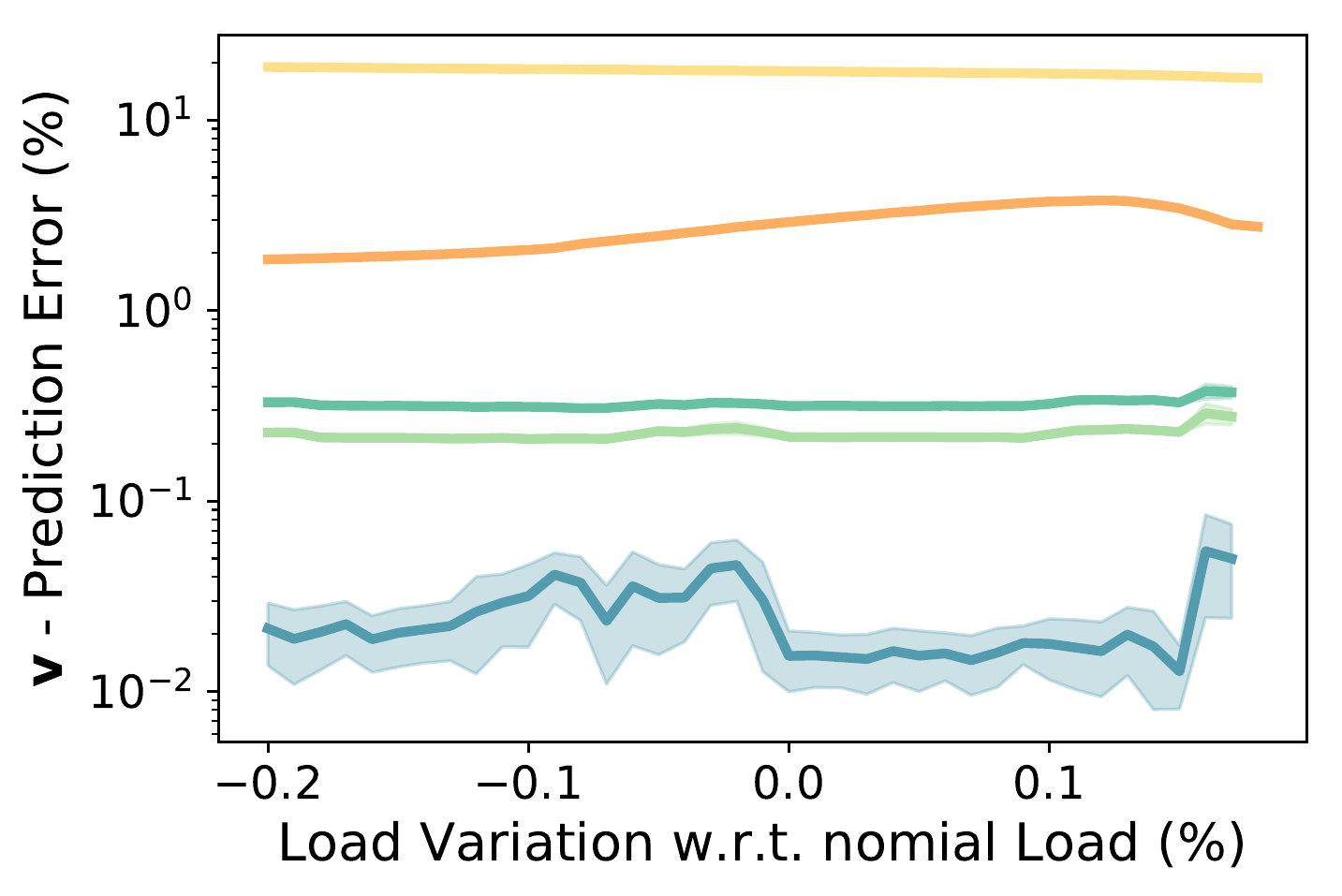}
\includegraphics[width=0.24\linewidth]{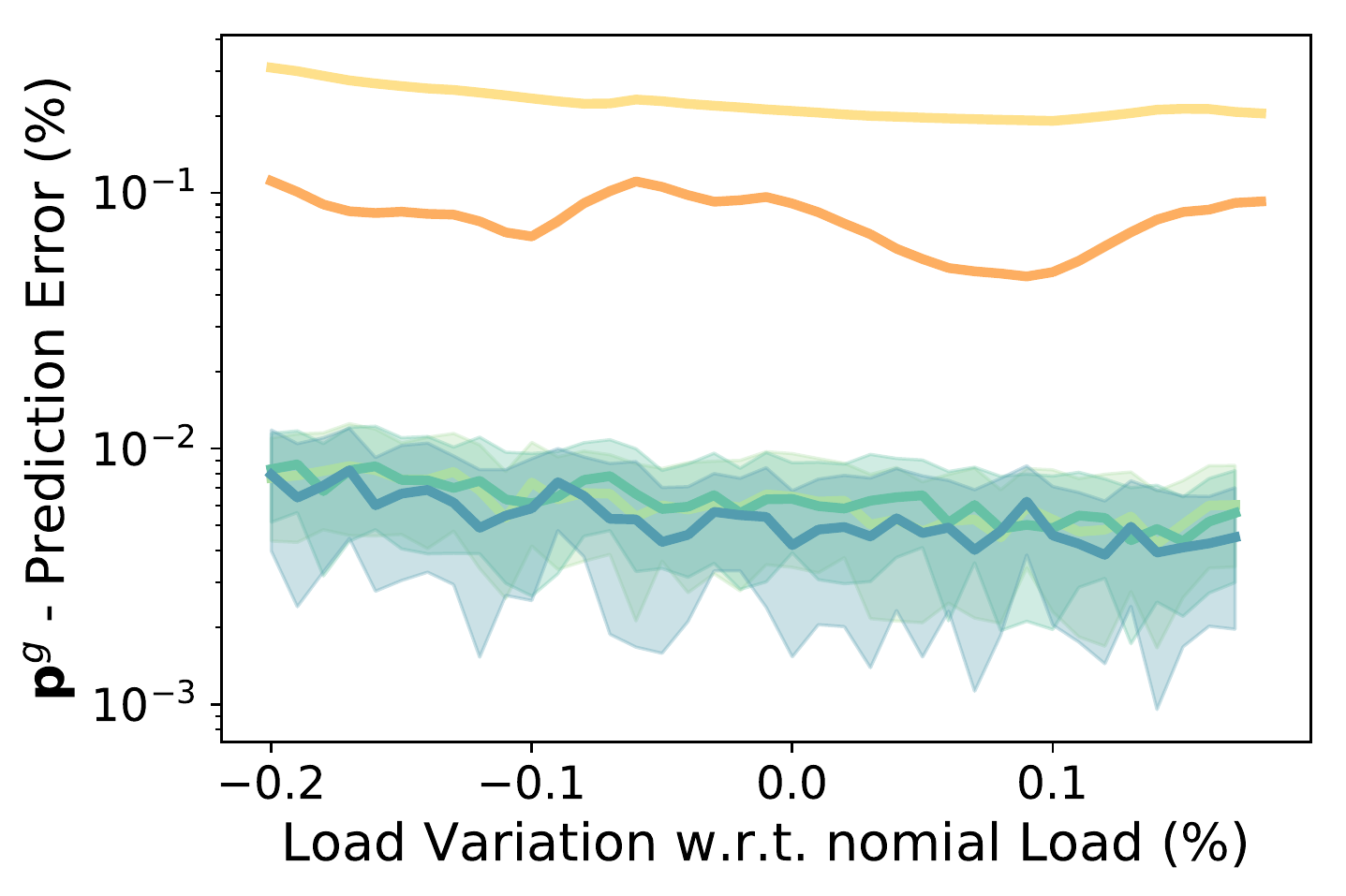}
\includegraphics[width=0.24\linewidth]{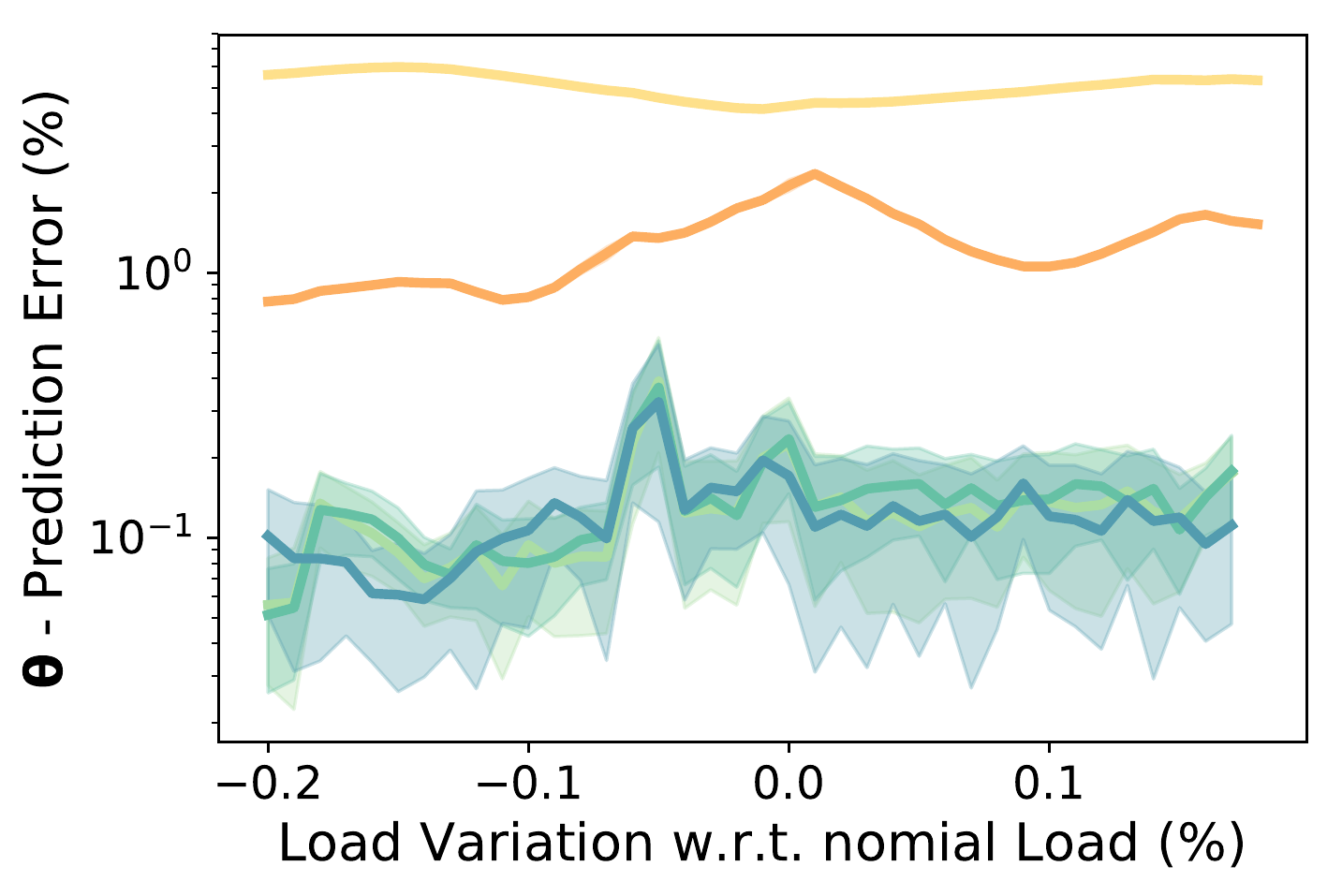}
\includegraphics[width=0.24\linewidth]{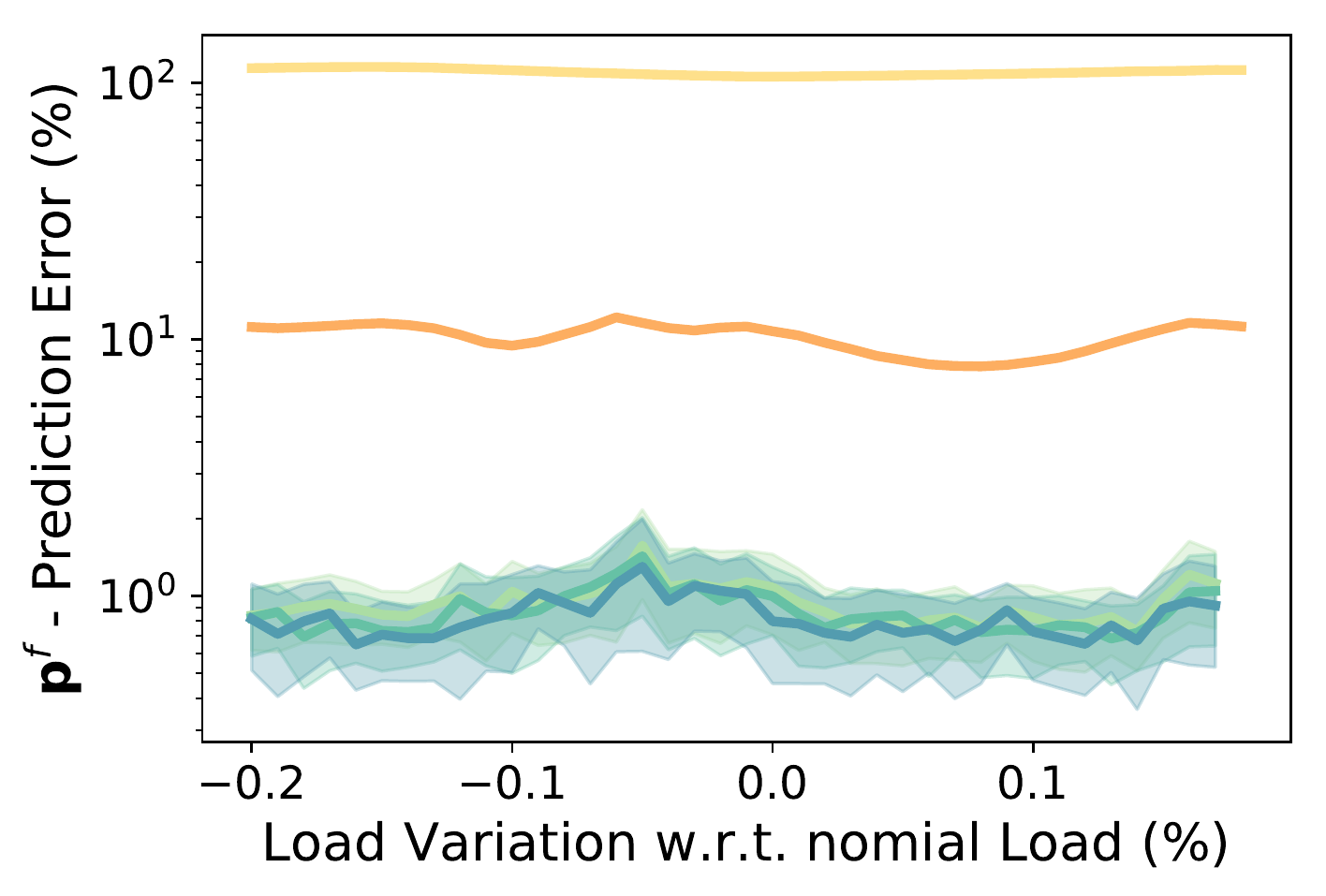}

\caption{Prediction Errors (\%) at the varying of the nominal network loads.}
\label{fig:prediction_errors1}
\end{figure*}
\begin{figure*}[!h]
\centering
\includegraphics[width=0.5\linewidth]{legend_5} 
\footnotesize{NESTA case 89\_pegase}\\
\centering
\includegraphics[width=0.24\linewidth]{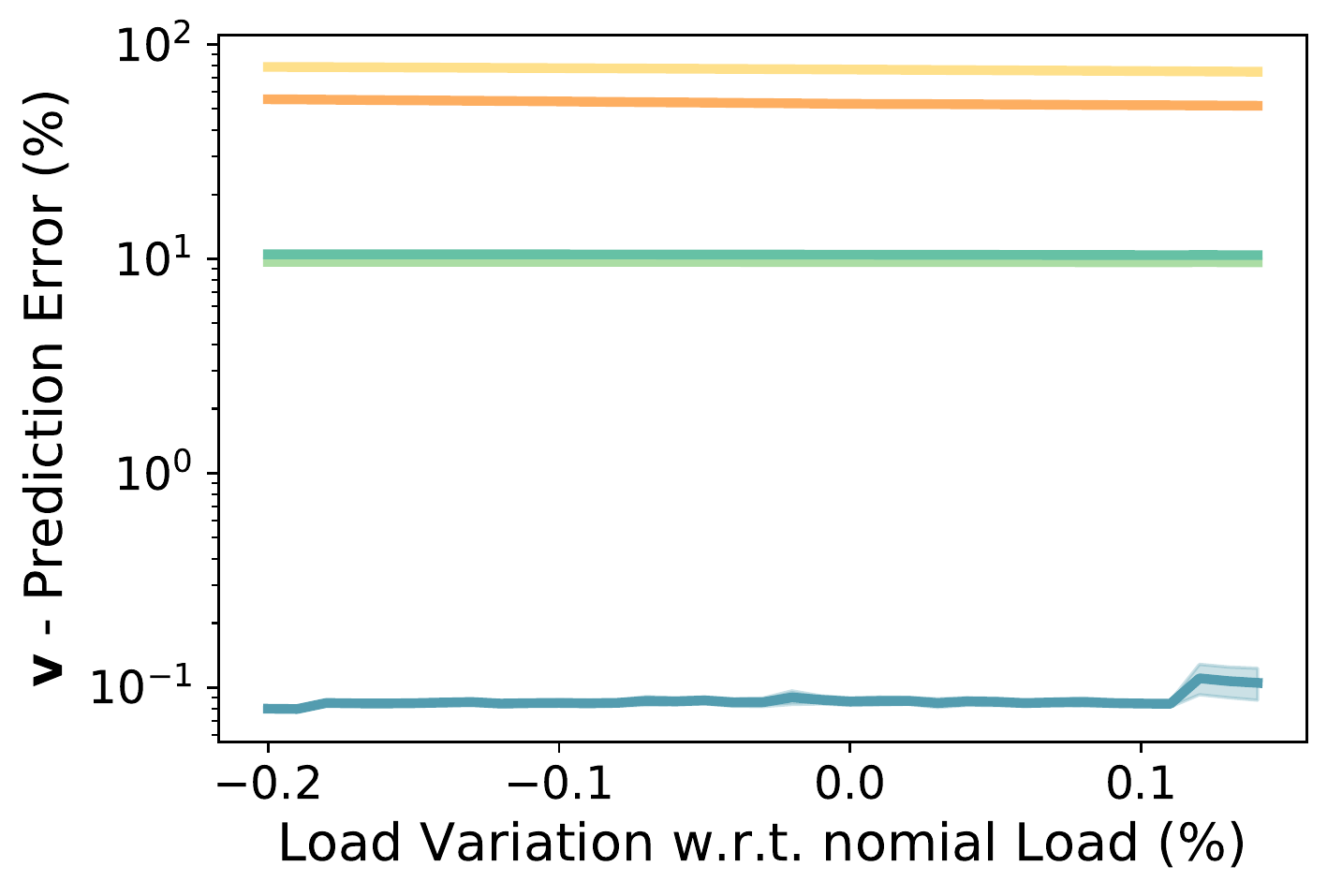}
\includegraphics[width=0.24\linewidth]{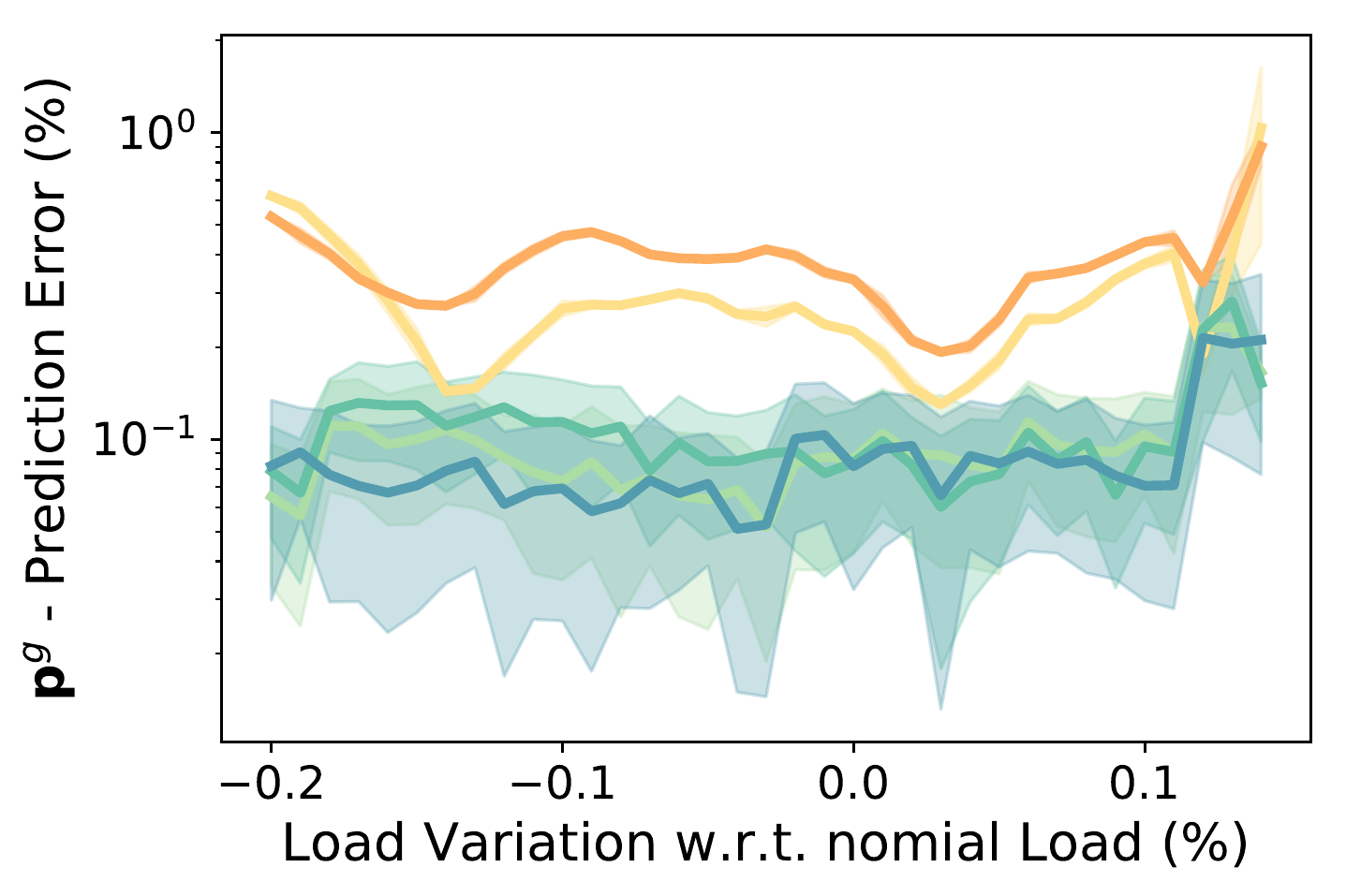}
\includegraphics[width=0.24\linewidth]{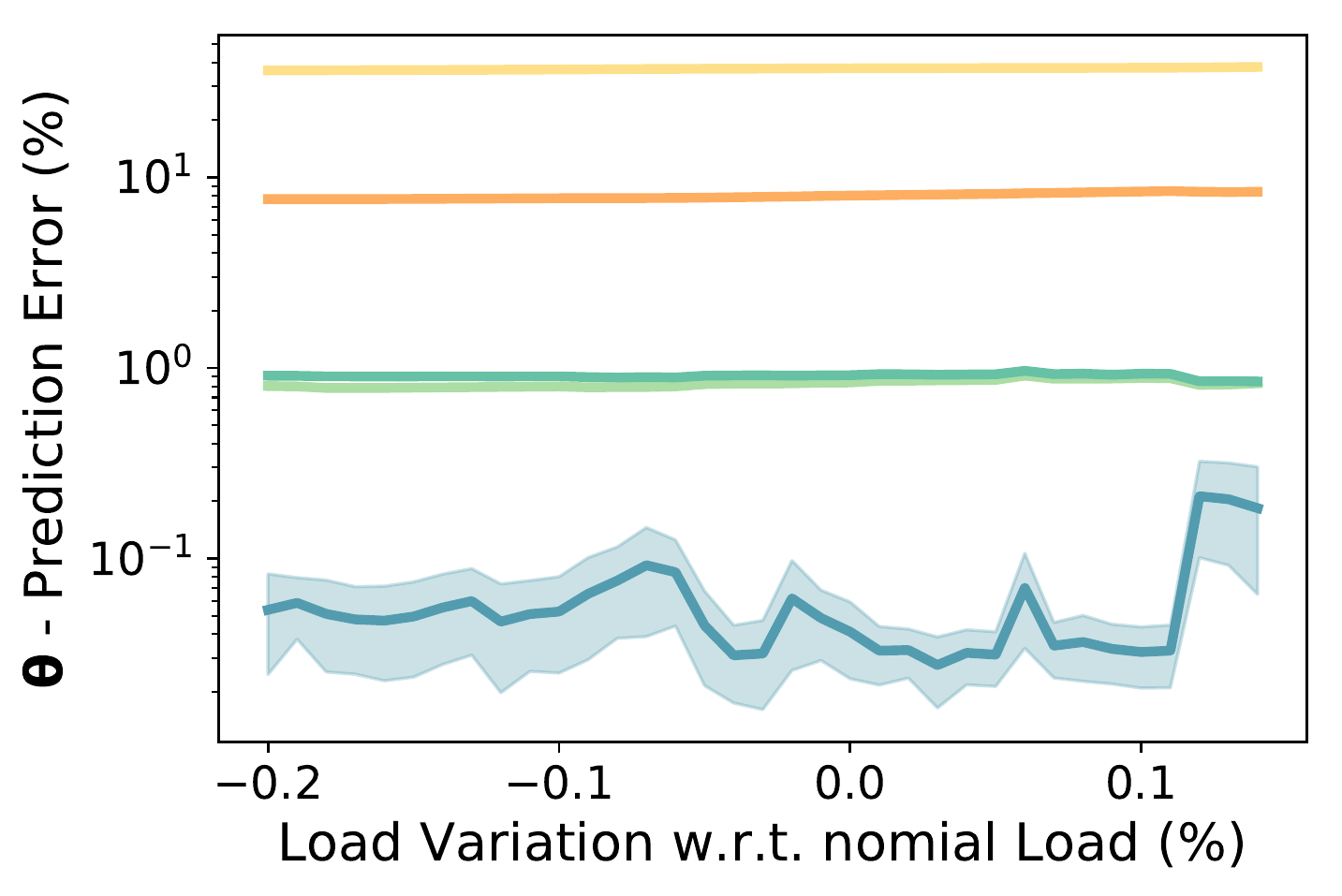}
\includegraphics[width=0.24\linewidth]{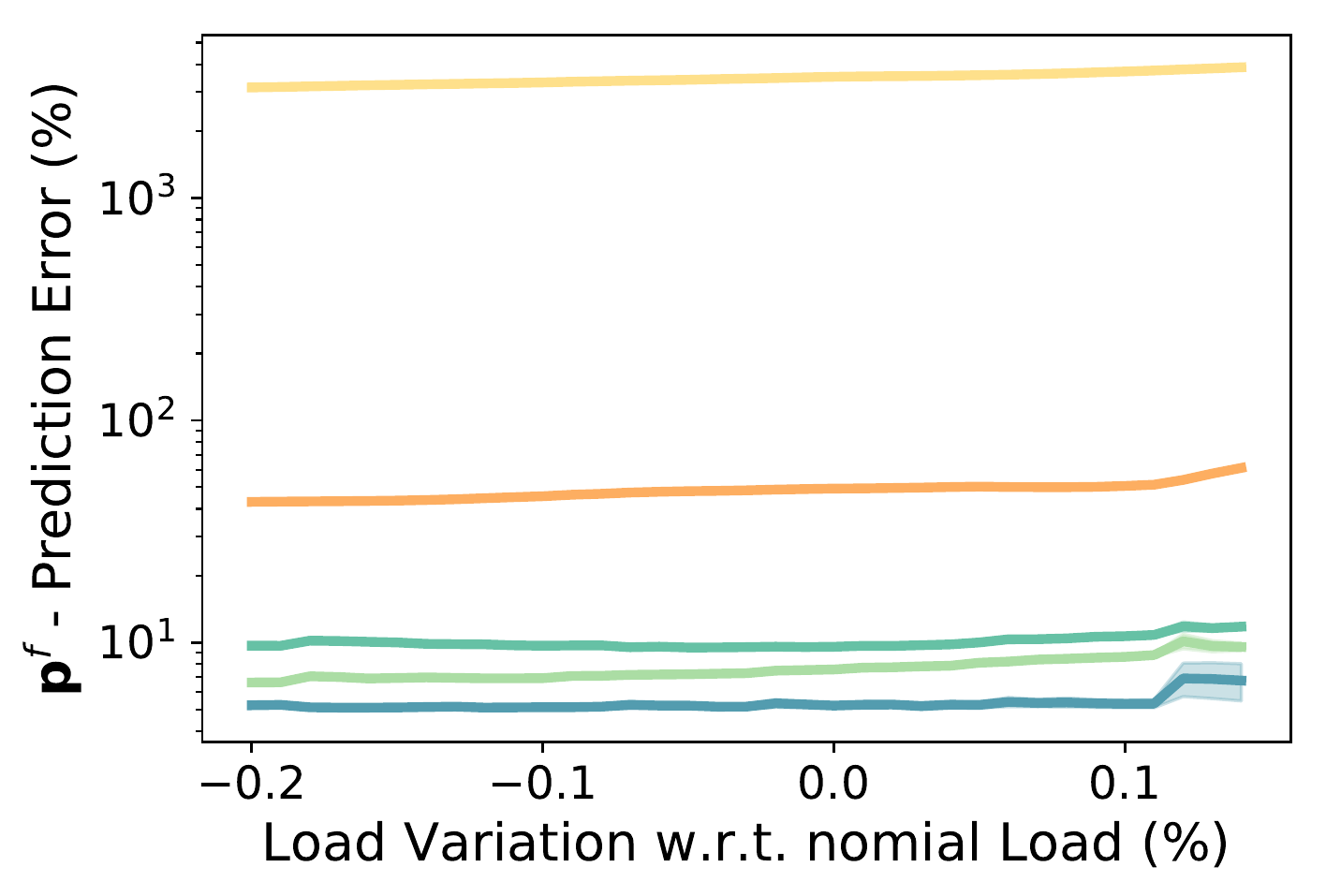}

\footnotesize{NESTA case 118\_ieee}\\
\centering
\includegraphics[width=0.24\linewidth]{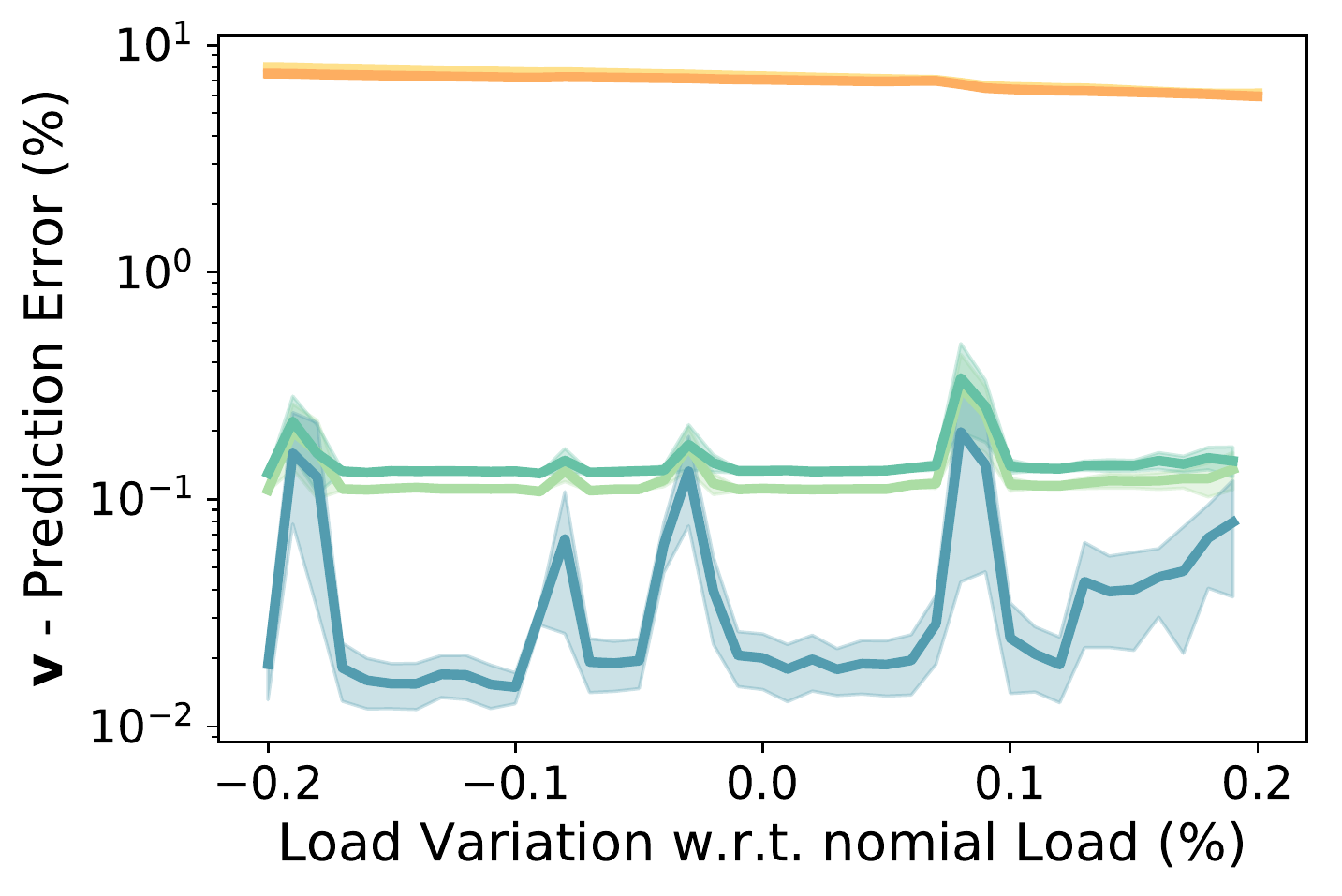}
\includegraphics[width=0.24\linewidth]{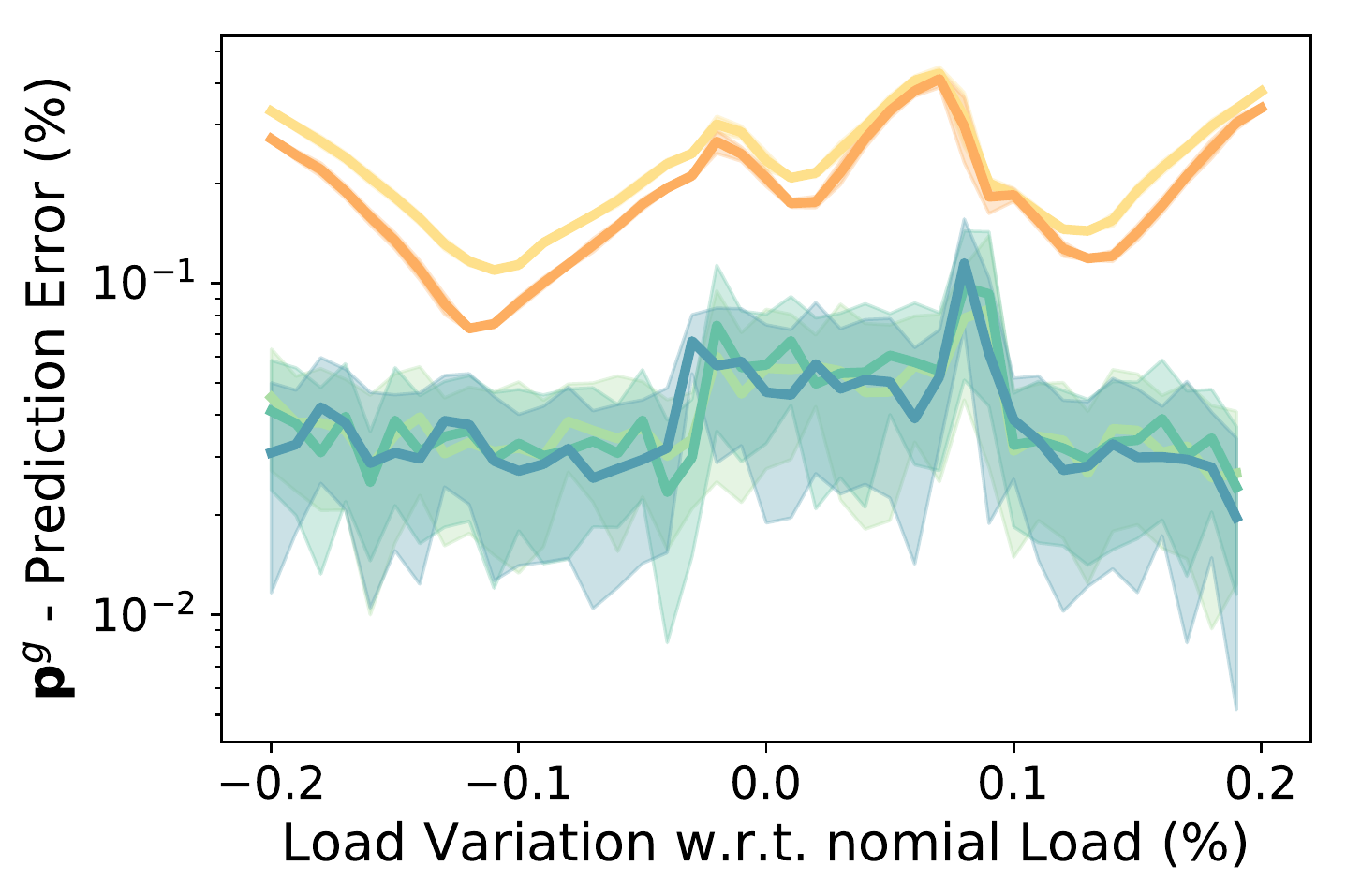}
\includegraphics[width=0.24\linewidth]{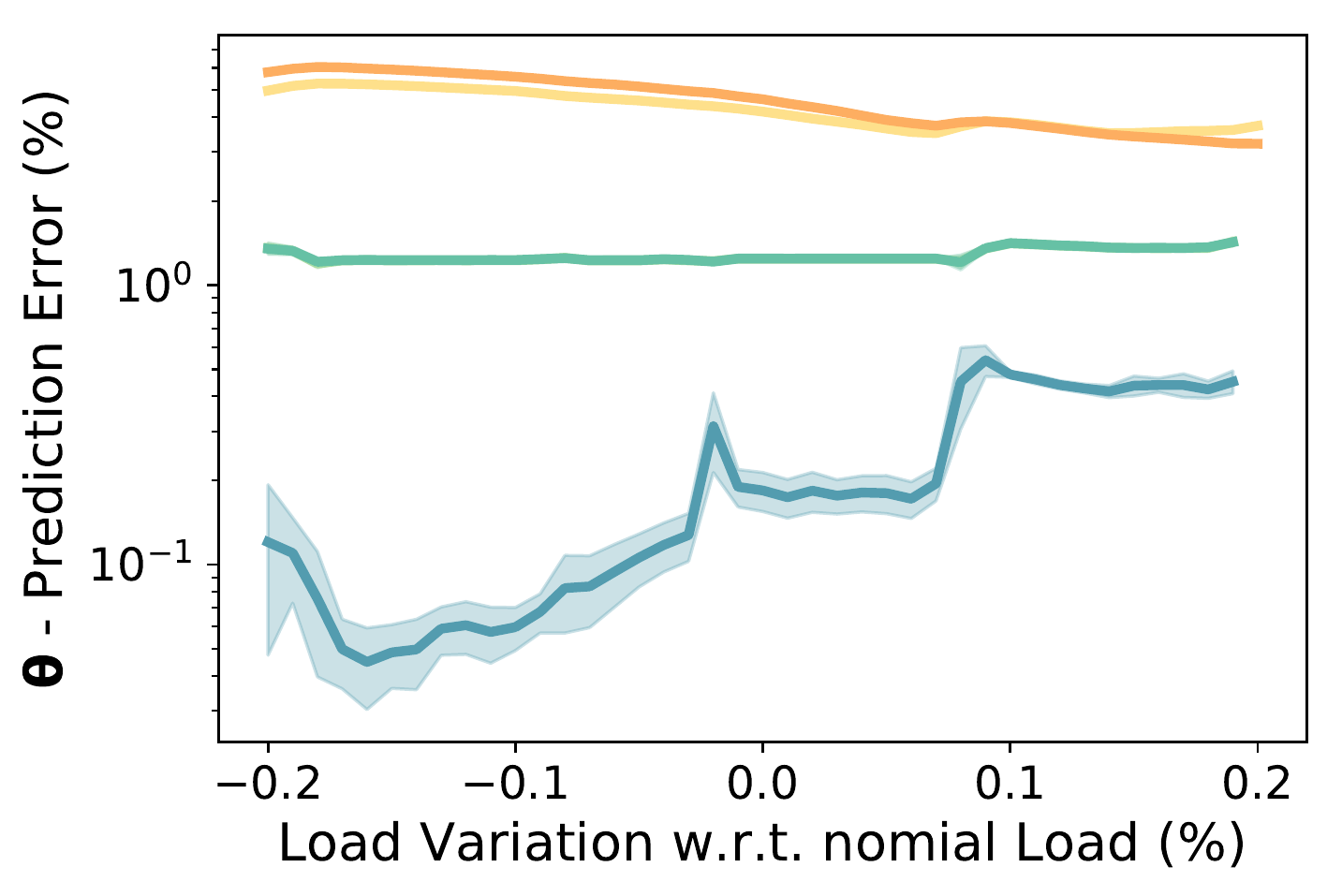}
\includegraphics[width=0.24\linewidth]{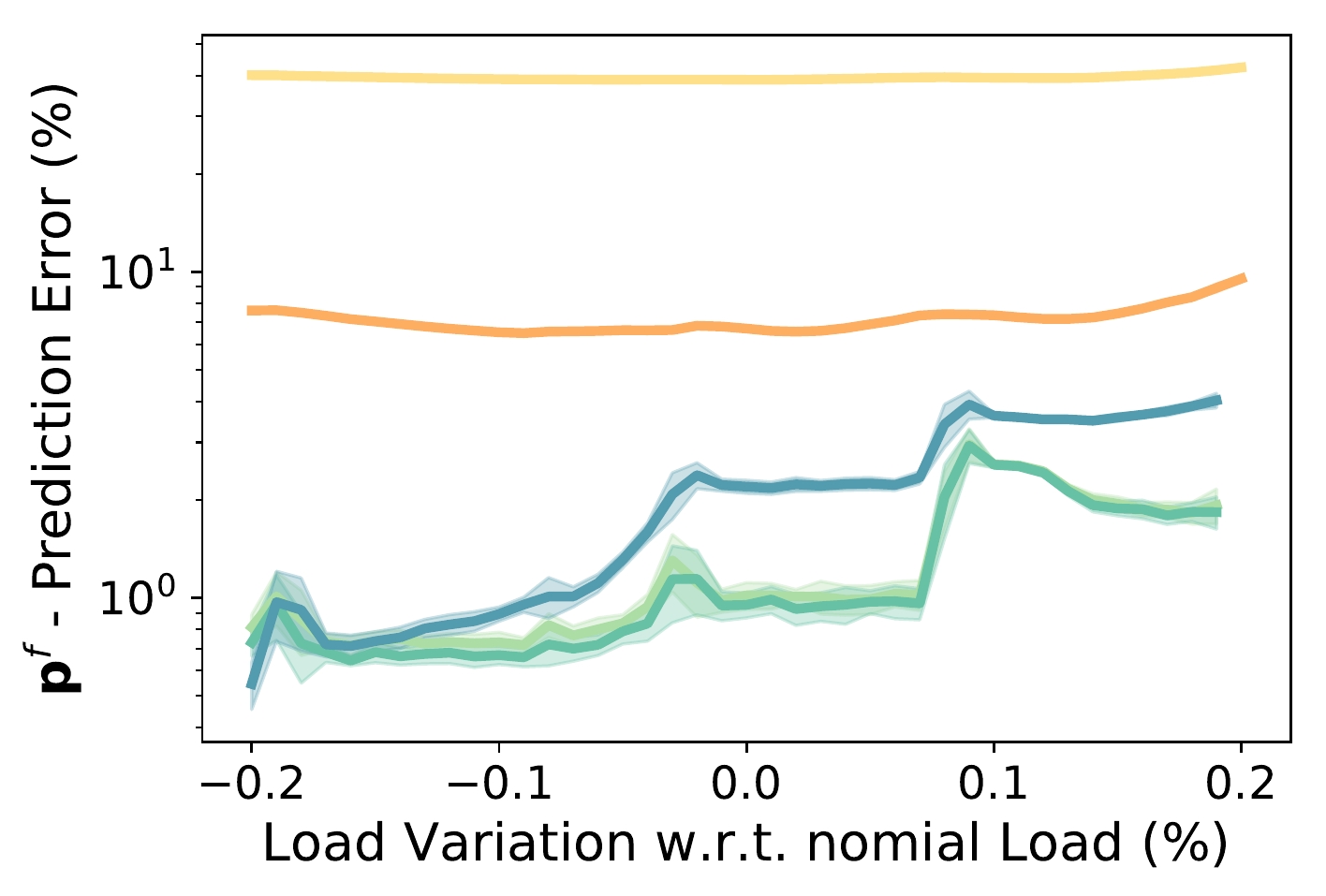}

\footnotesize{NESTA case 162\_ieee\_dtc}\\
\centering
\includegraphics[width=0.24\linewidth]{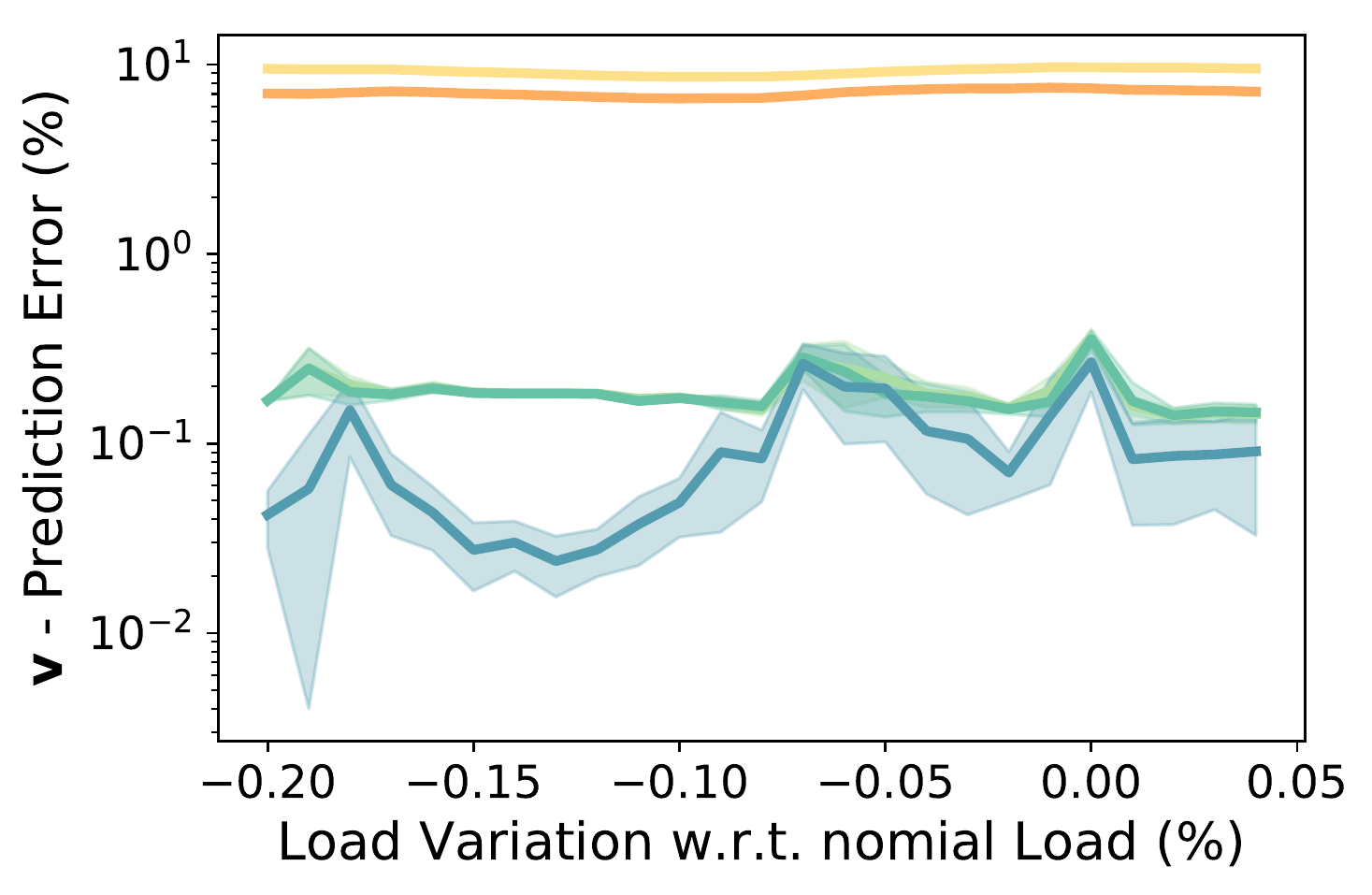}
\includegraphics[width=0.24\linewidth]{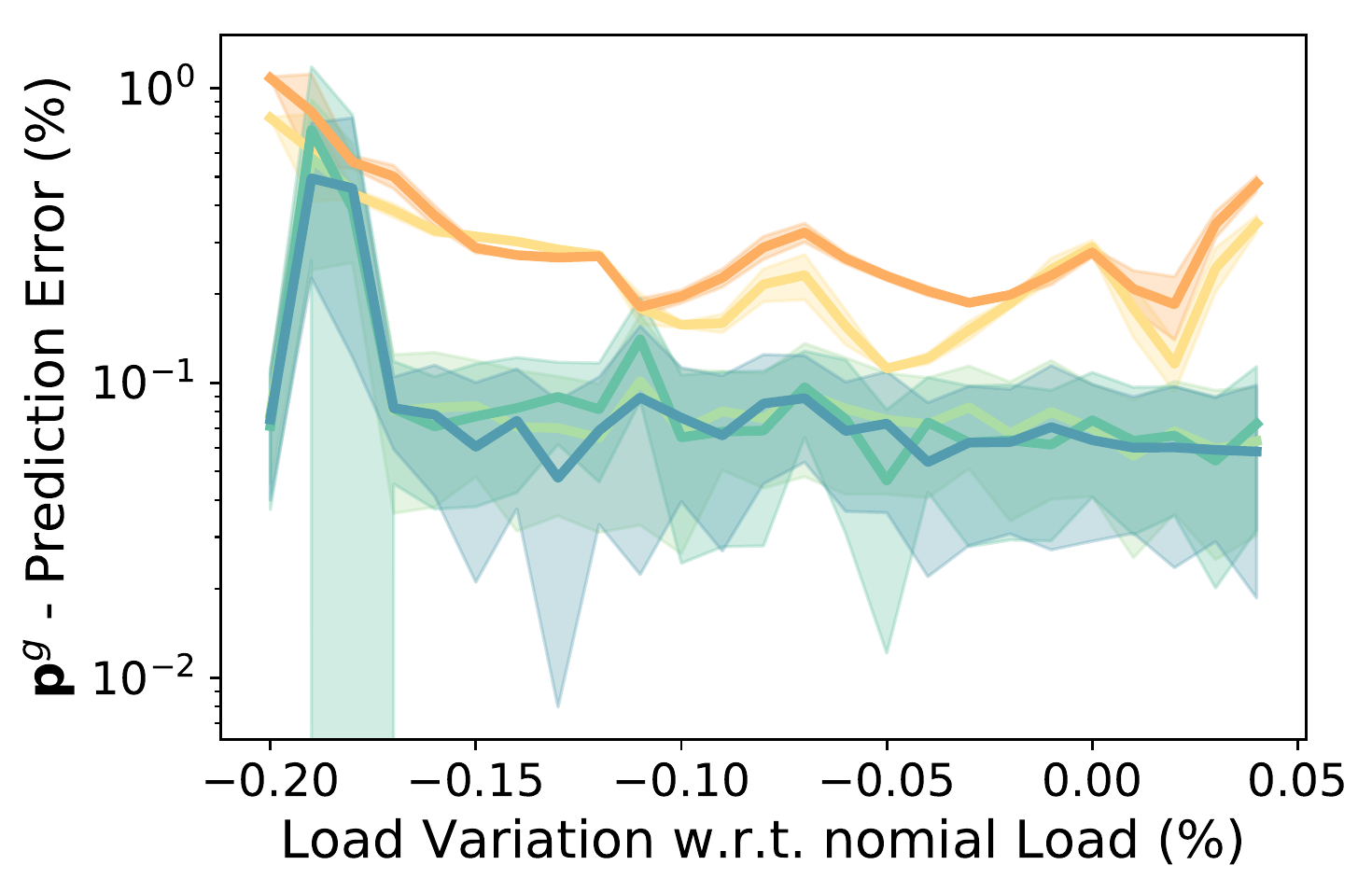}
\includegraphics[width=0.24\linewidth]{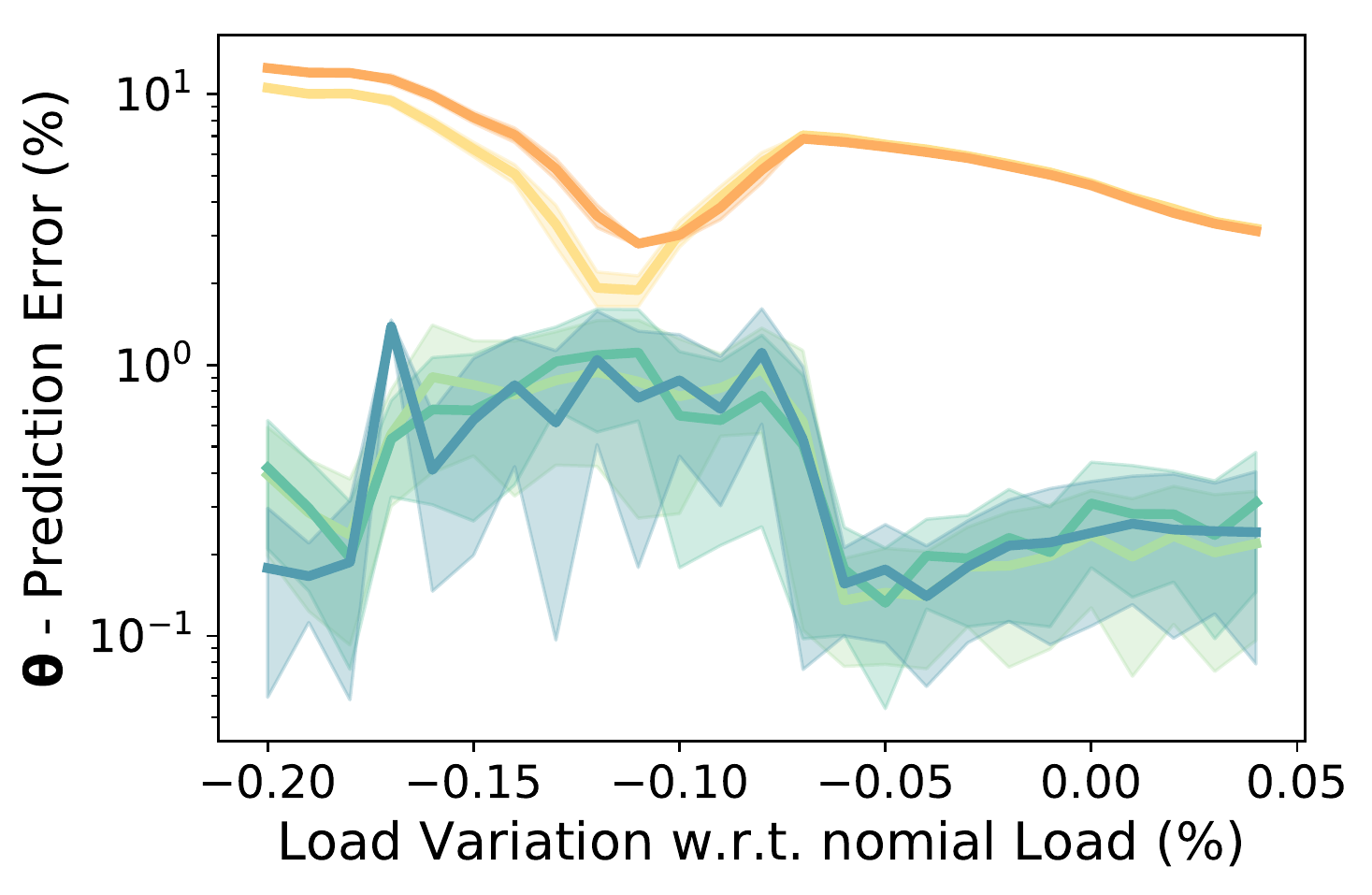}
\includegraphics[width=0.24\linewidth]{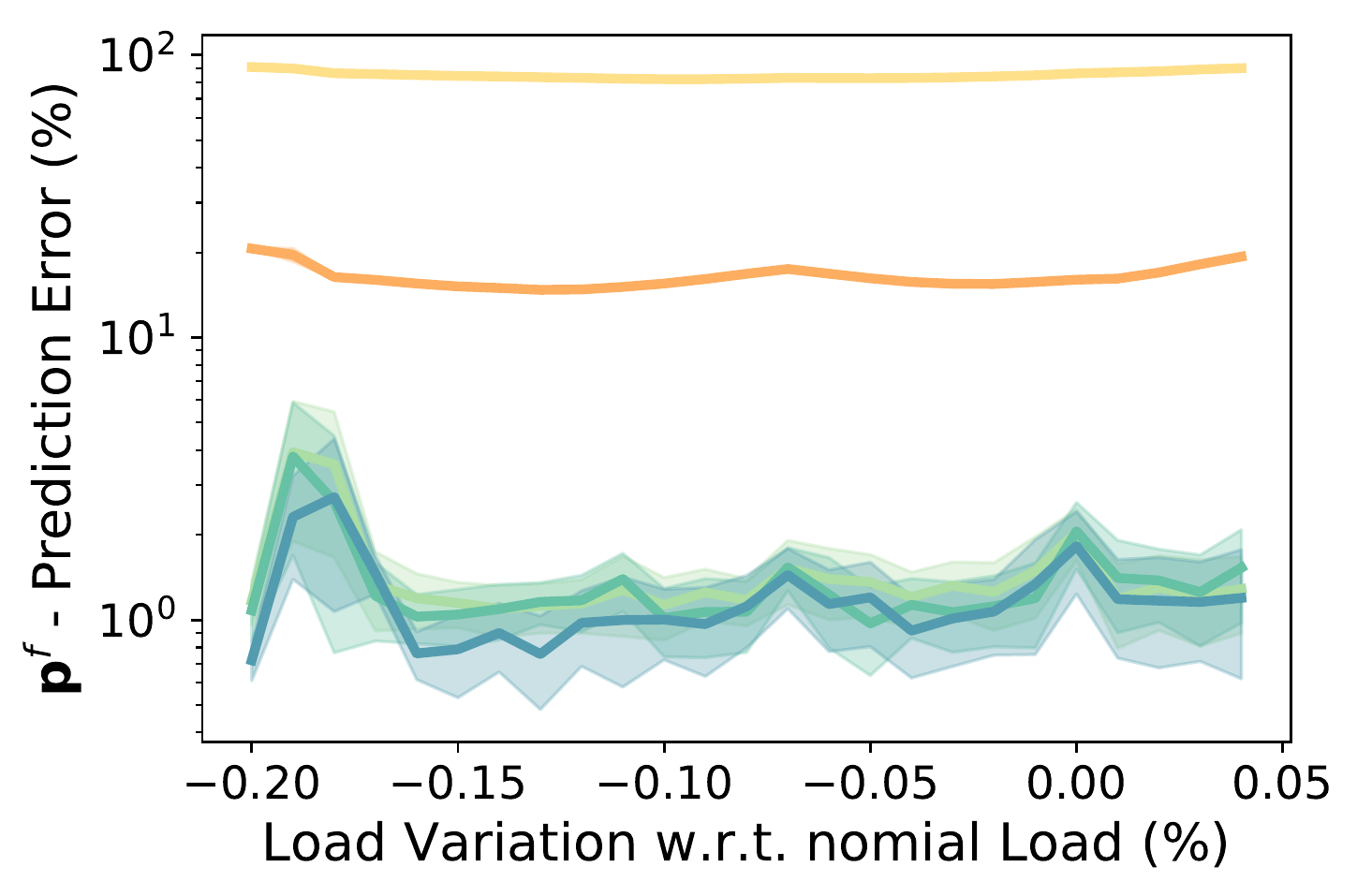}

\footnotesize{NESTA case 189\_edin}\\
\centering
\includegraphics[width=0.24\linewidth]{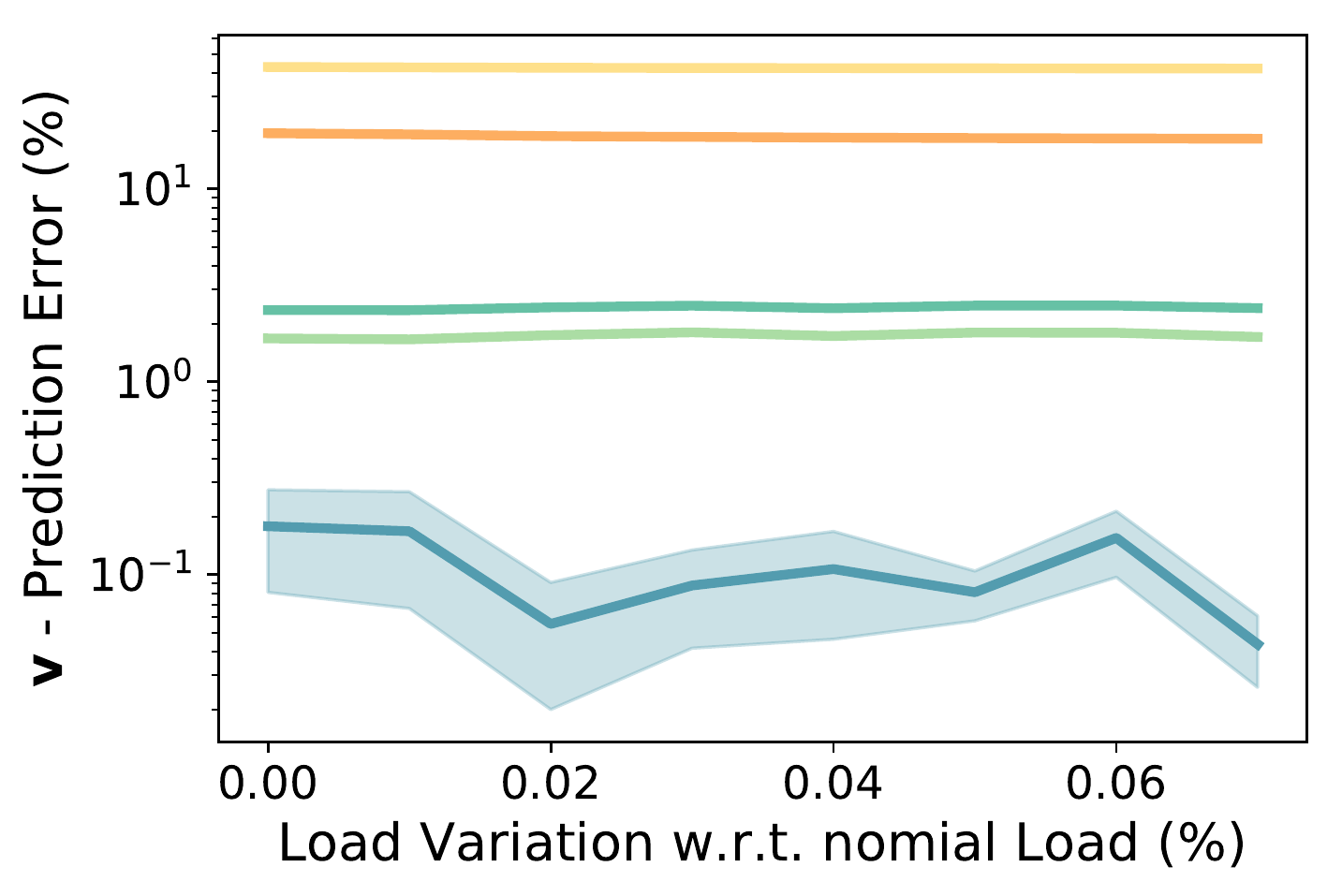} 
\includegraphics[width=0.24\linewidth]{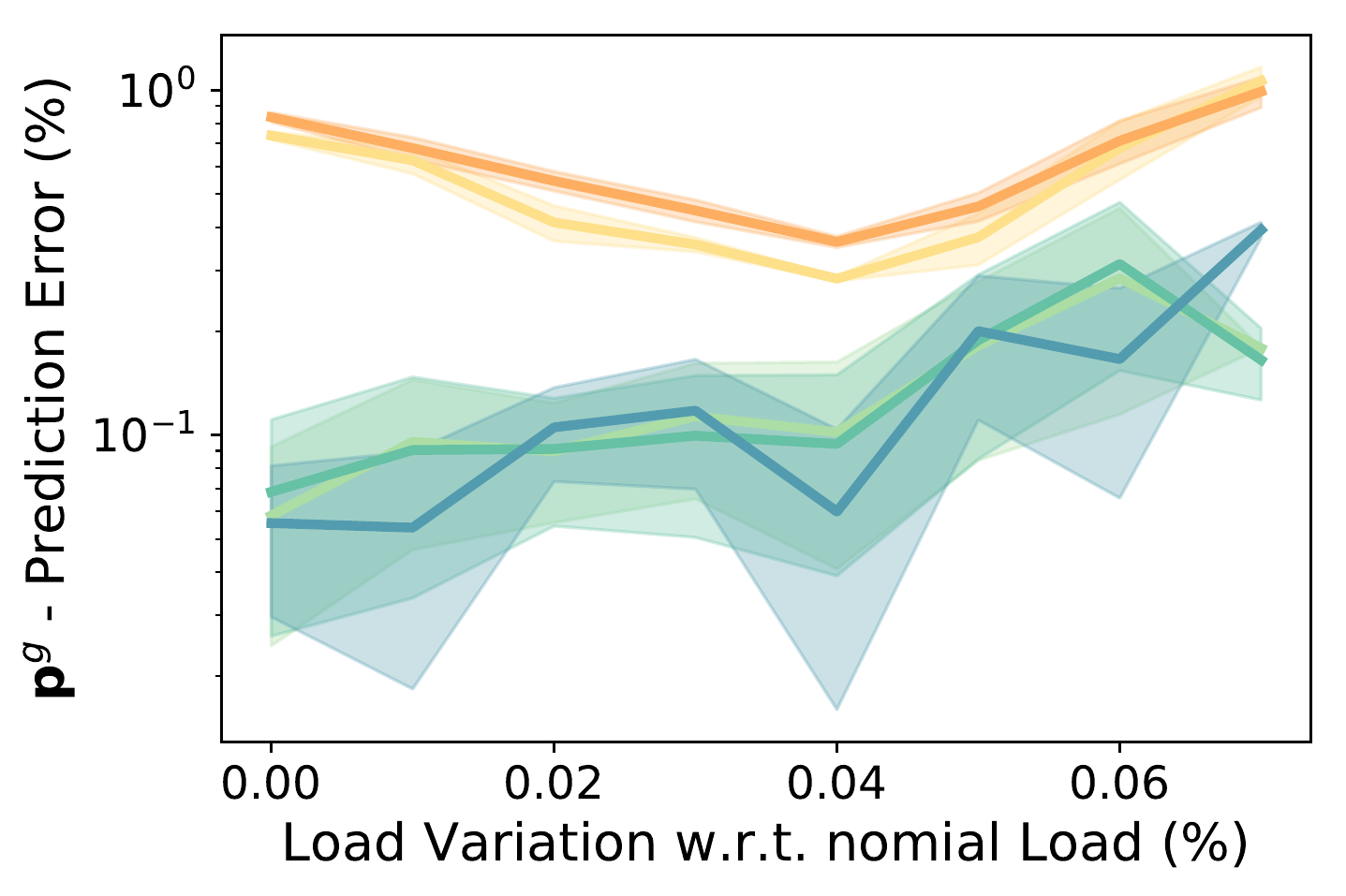}
\includegraphics[width=0.24\linewidth]{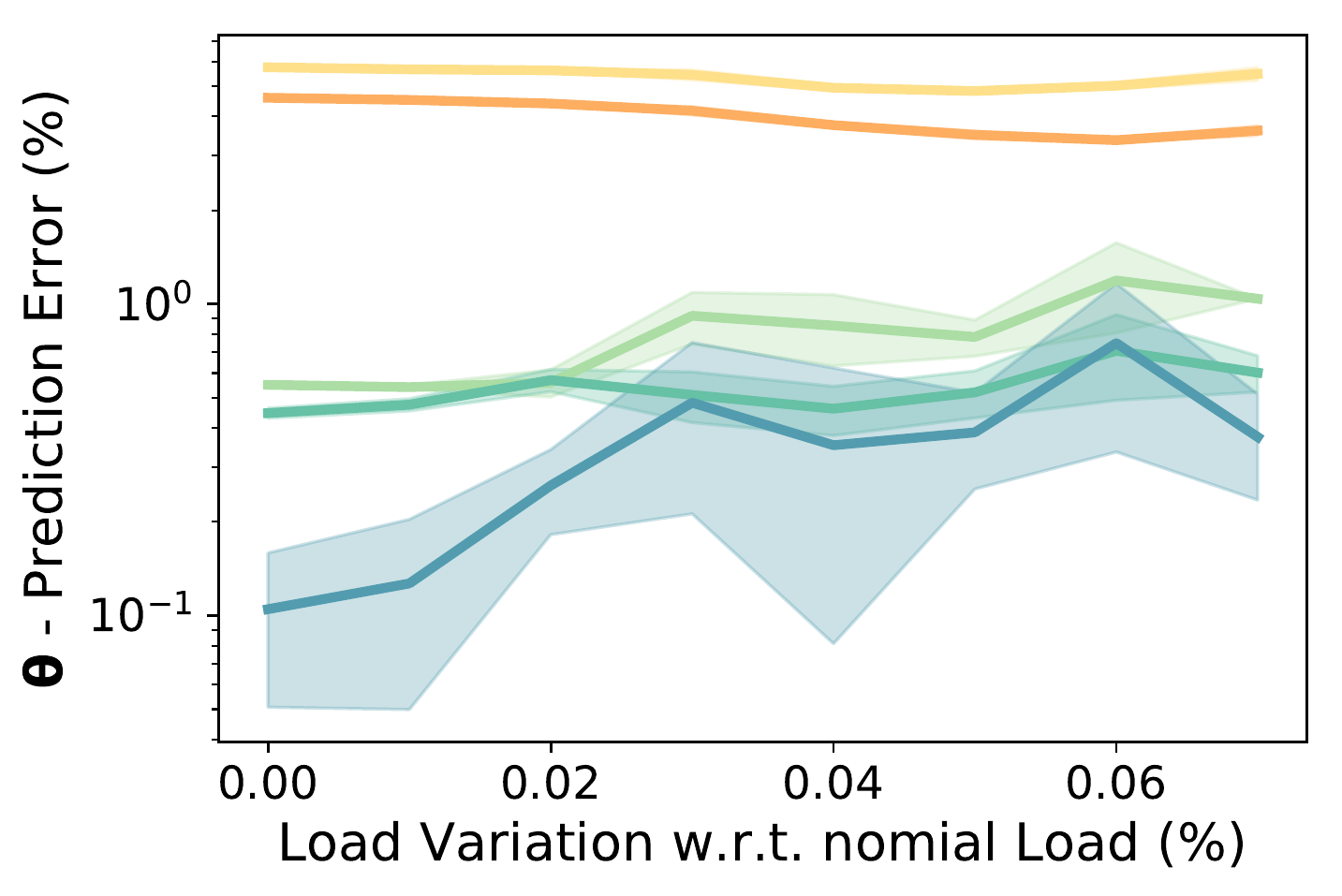}
\includegraphics[width=0.24\linewidth]{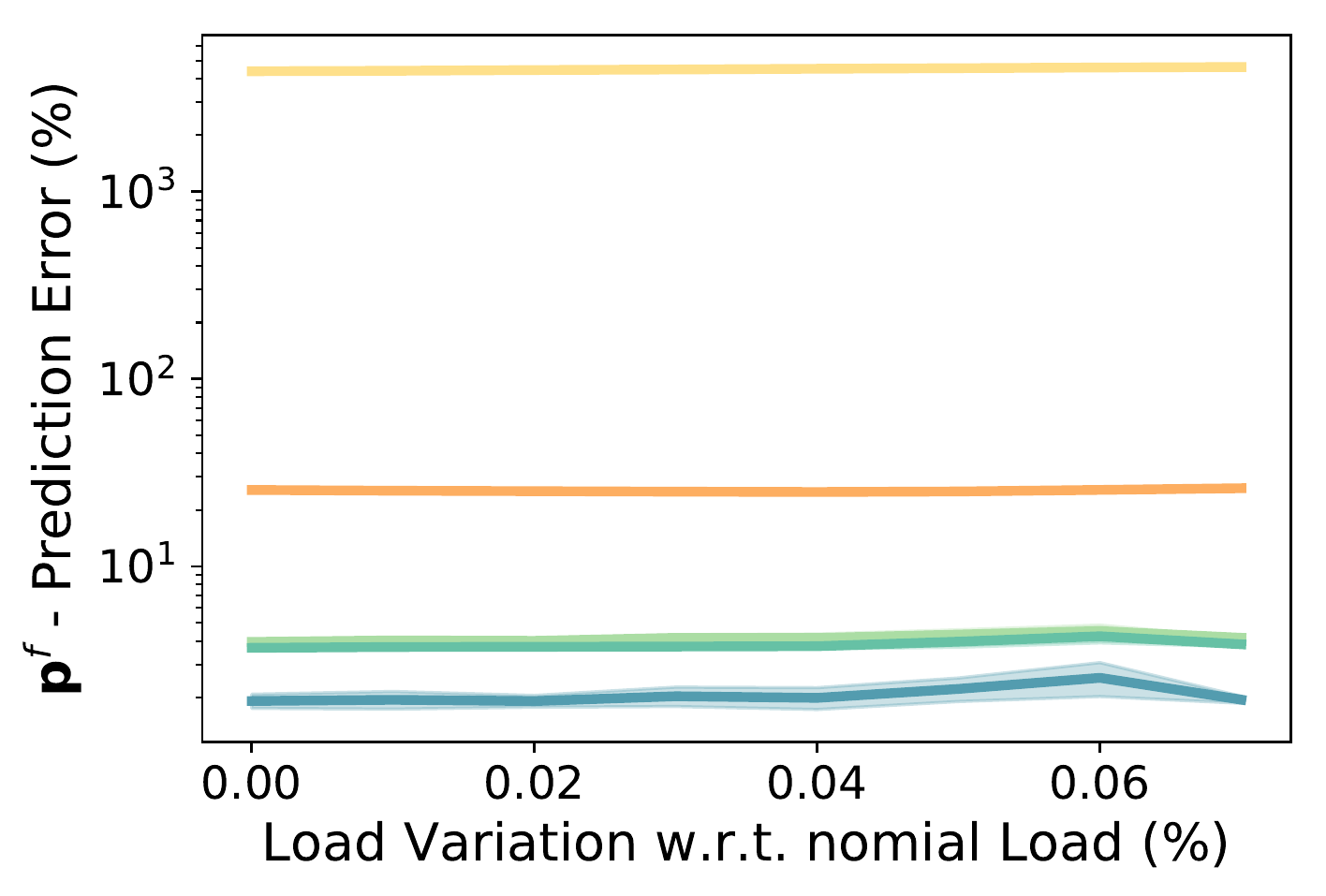}

\footnotesize{NESTA case 300\_ieee}\\
\centering
\includegraphics[width=0.24\linewidth]{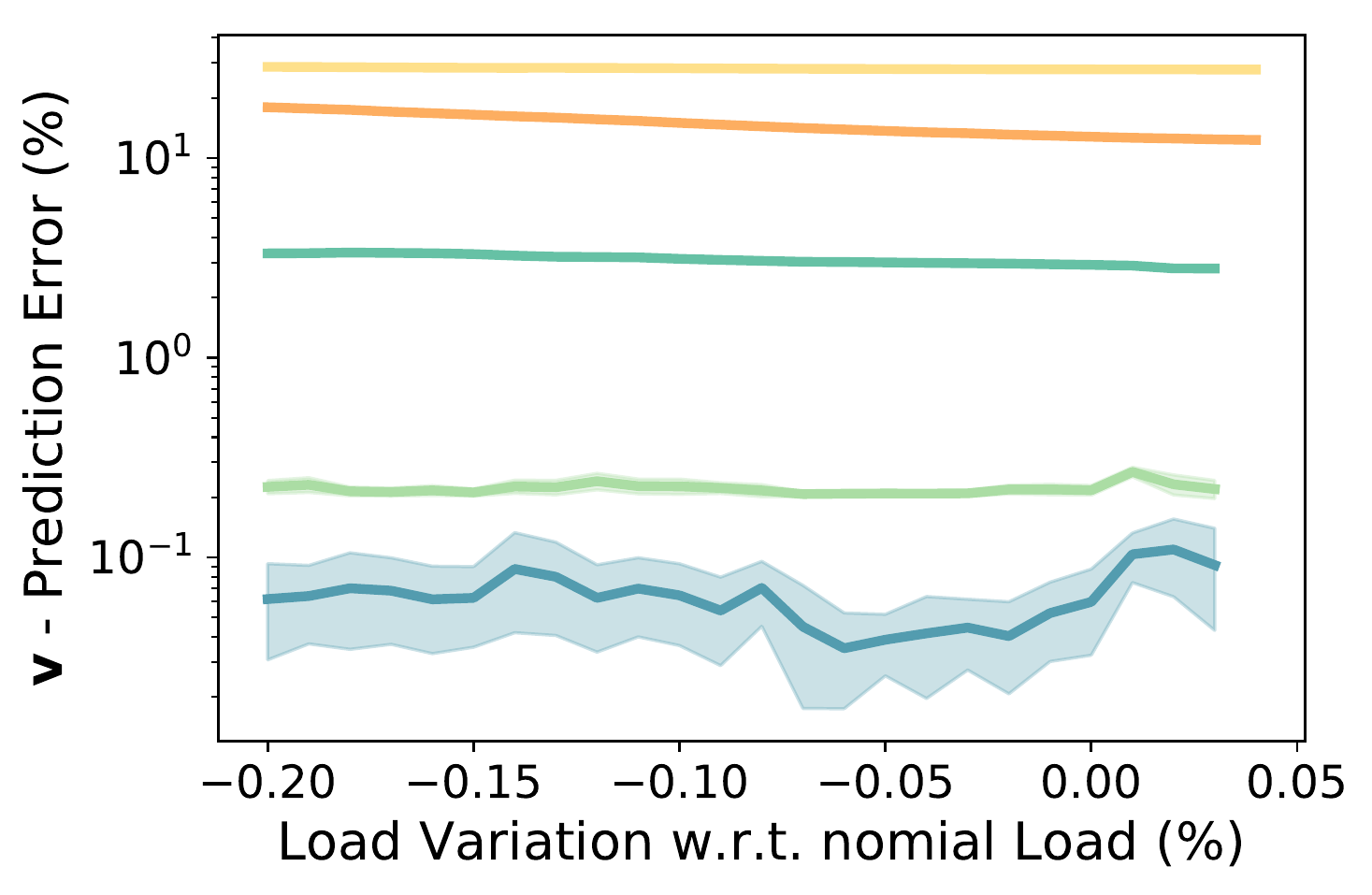}
\includegraphics[width=0.24\linewidth]{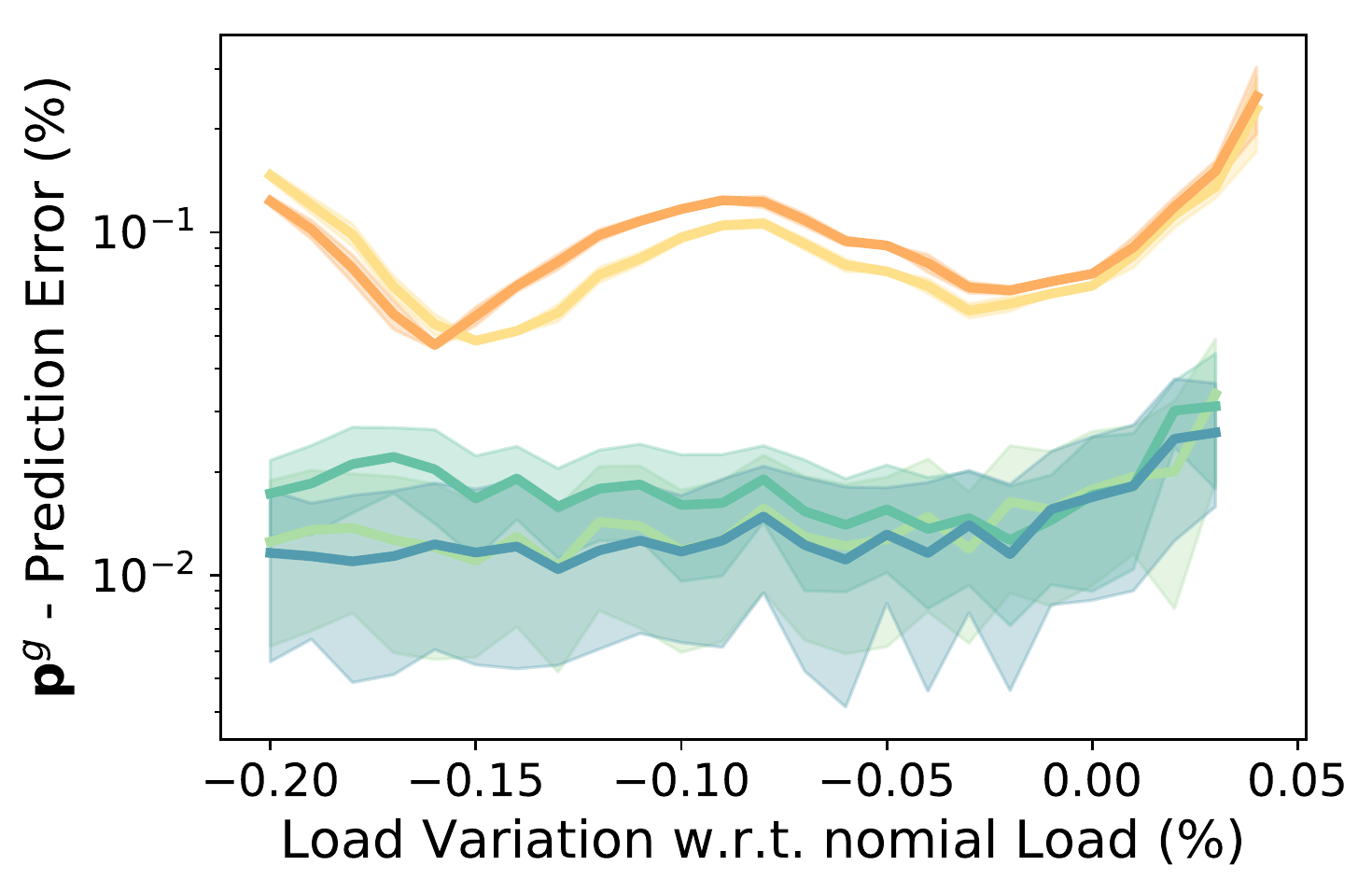}
\includegraphics[width=0.24\linewidth]{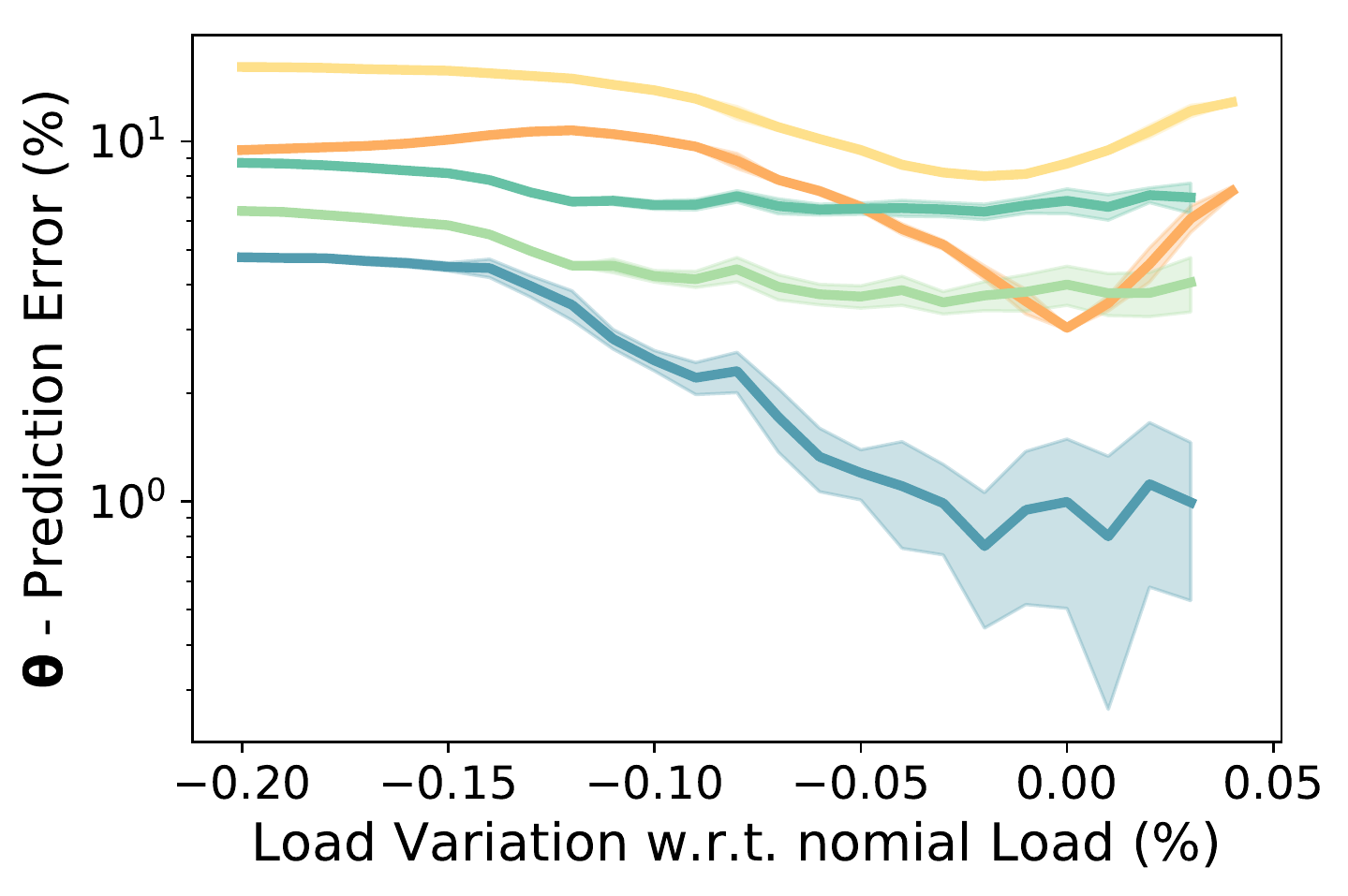}
\includegraphics[width=0.24\linewidth]{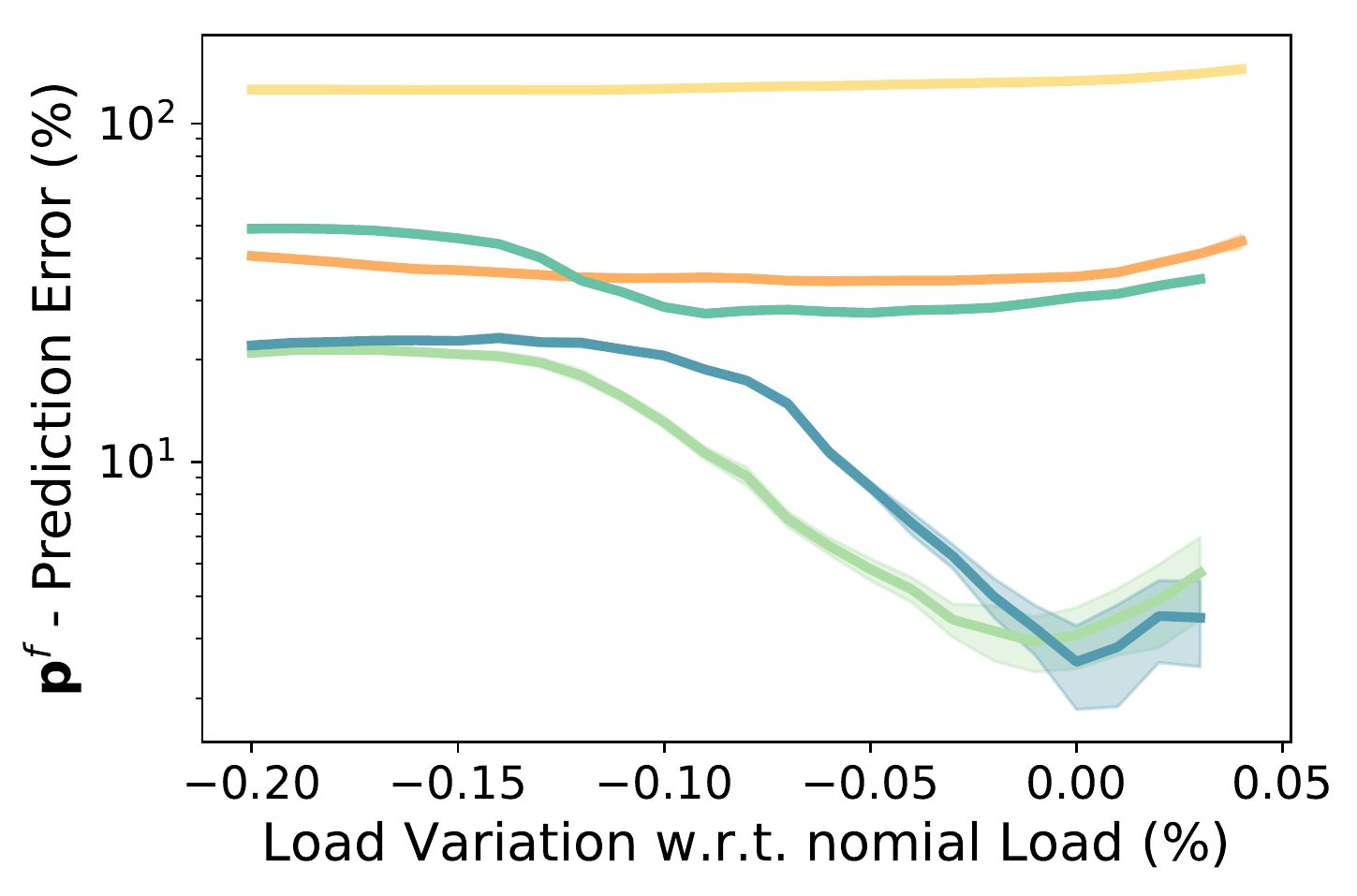}

\caption{Prediction Errors (\%) at the varying of the nominal network loads.}
\label{fig:prediction_errors2}
\end{figure*}

\smallskip
Figure \ref{fig:prediction_errors1} and \ref{fig:prediction_errors2} 
further illustrate the importance of modeling the problem constraints 
and exploiting a hot-start state.  The figures illustrate the prediction 
errors on the operational parameters $\bm{v}$ and $\bm{p}^g$, as well as on 
the angle magnitude $\bm{\theta}$ and the power flows $\bm{p}^f$, at the 
varying of the load demands in the power networks (from -20\% to 20\% of the 
aggregated nominal load values). The reason for the differences in the 
$x$-axis range in the various networks, is due to that, the increased load values 
may produce congested scenarios that cannot be accommodated. 
The plots are in log-10 scale and clearly indicate that the models exploiting 
the problem structure better generalize to the different network settings.

\begin{table*}[!h]
\centering
\small
\resizebox{0.99\linewidth}{!}
  {
  \begin{tabular}{l l|rrrrrrr|rrrrrrrr}
  \toprule
   \multicolumn{2}{l}{\textbf{Test case}}  
                  & {DC}&  ${\cal M}_\B$&  ${\cal M}_\C$& ${\cal M}_\CD$& ${\cal M}_\CL$&  ${\cal M}_\CLL$& ${\cal M}_\CLD$
                  & {DC}&  {$\LFL$} & ${\cal M}_\B$&  ${\cal M}_\C$& ${\cal M}_\CD$& ${\cal M}_\CL$&  ${\cal M}_\CLL$& ${\cal M}_\CLD$\\
  \midrule

     \multirow{2}{*}{\textbf{14\_ieee}}
                        & $\bm{p}^g$ & 2.4020 & 5.7048 & 6.0474 &5.5056 & 0.2131 & 0.2052 & \b{0.0233}&     0.1359 &4.3518 & 1.1061 & 0.8154 &0.8314 & 0.2649 & 0.2571 & \b{0.0003}\\
                        & $\bm{v}$   & 1.8352 & 0.9174 & 0.8636 &0.3169 & 0.0937 & 0.0944 & \b{0.0017}&     3.0365 &0.3450 & 0.3075 & 0.7808 &0.6916 & 0.1516 & 0.2170 & \b{0.0018}\\
     \multirow{2}{*}{\textbf{30\_ieee}}                                                                                                                          
                        & $\bm{p}^g$ & 2.6972 & 2.0793 & 1.9688 &1.7344 & 0.1815 & 0.1320 & \b{0.0007}&     0.1907 &13.504 & 2.1353 & 1.8268 &1.5523 & 0.2735 & 0.2853 & \b{0.0058}\\
                        & $\bm{v}$   & 1.2929 & 83.138 & 0.4309 &0.2869 & 0.0944 & 0.0728 & \b{0.0037}&     3.4931 &0.4829 & 6.2996 & 2.7458 &0.2270 & 0.4299 & 0.4168 & \b{0.0086}\\
     \multirow{2}{*}{\textbf{39\_epri}}                                                                                                                          
                        & $\bm{p}^g$ & 0.0731 & 0.2067 & 0.1700 &0.0857 & 0.0516 & 0.0488 & \b{0.0005}&     0.2163 &2.1260 & 0.1350 & 0.1467 &0.1160 & 0.0140 & 0.0155 & \b{0.0023}\\
                        & $\bm{v}$   & 1.1086 & 95.944 & 4.2008 &22.921 & 0.3273 & 0.3131 & \b{0.0222}&     1.7251 &0.8573 & 6.8089 & 3.1999 &6.7880 & 2.3860 & 2.3476 & \b{0.0313}\\
     \multirow{2}{*}{\textbf{57\_ieee}}                                                                                                                        
                        & $\bm{p}^g$ & 0.3354 & 0.9733 & 0.7507 &0.8837 & 0.1166 & 0.1165 & \b{0.0076} & \b{0.0112}&4.7722 & 1.2882 & 1.2378 &1.4518 & 0.1710 & 0.1752 & 0.0206\\
                        & $\bm{v}$   & 0.9091 & 4.0504 & 2.5770 &0.4038 & 0.1771 & 0.1650 & \b{0.0384} &    0.2825 &1.2243 & 1.5682 & 1.1176 &0.4724 & 0.2772 & 0.2626 & \b{0.0482}\\
 \multirow{2}{*}{\textbf{73\_ieee\_rts}}                                                                                                                        
                        & $\bm{p}^g$ & 0.0204 & 0.2078 & 0.0482 &0.0214 & 0.0054 & \b{0.0053} & \b{0.0053}& 0.0304 &5.2103 & 0.1144 & 0.0543 &0.0239 & 0.0081 & 0.0082 & \b{0.0077}\\
                        & $\bm{v}$   & 0.1528 & 14.074 & 0.1995 &0.4470 & 0.0611 & 0.0599 & \b{0.0337} &    4.1537 &0.1496 & 4.8500 & 2.8951 &0.2525 & 0.2488 & 0.3433 & \b{0.0516}\\
   \multirow{2}{*}{\textbf{89\_pegase}}                                                                                                                        
                        & $\bm{p}^g$ & 0.1641 & 0.1440 & 0.2228 &0.2166 & \b{0.0420} & 0.0443 & 0.0486 &    3.0662 &7.3622 & 0.1834 & 0.2524 &0.2623 & 0.0900 & 0.1035 & \b{0.0827}\\
                        & $\bm{v}$   & 3.8584 & 86.795 & 73.506 &56.519 & 5.6660 & 6.4734 & \b{1.2492} &\b{1.1776} &0.6913  &4.1315 & 4.0437 &4.0373 & 3.9890 & 3.9410 & 1.2610\\
    \multirow{2}{*}{\textbf{118\_ieee}}                                                                                                                        
                        & $\bm{p}^g$ & 0.2011 & 0.1071 & 0.0359 &0.0216 & 0.0043 & 0.0056 & \b{0.0038} &    0.5865 &3.8034 & 0.1353 & 0.1557 &0.1050 & 0.0372 & 0.0391 & \b{0.0368}\\
                        & $\bm{v}$   & 1.9971 & 3.4391 & 0.8995 &0.0791 & 0.0956 & 0.0920 & \b{0.0866} &    2.2780 &0.9772 & 4.5972 & 6.0326 &0.4303 & 0.1599 & 0.1768 & \b{0.1335}\\
\multirow{2}{*}{\textbf{162\_ieee\_dtc}}                                                                                                                         
                        & $\bm{p}^g$ & 0.6727 & 0.1054 & 0.0673 &0.1442 & 0.0587 & 0.0571 & \b{0.0558} &    0.6917 &17.873 & 0.1648 & 0.2389 &0.1575 & 0.0977 & 0.0981 & \b{0.0954}\\
                        & $\bm{v}$   & 3.7718 & 3.6372 & 6.7930 &4.2568 & 0.3276 & 0.3202 & \b{0.2565} &    0.5820 &1.4595  &0.4378 & 0.6922 &0.5824 & \b{0.2846} & 0.2906 & 0.2921\\
    \multirow{2}{*}{\textbf{189\_edin}}                                                                                                                         
                        & $\bm{p}^g$ & 1.0514 & 0.2694 & 0.3742 &0.1915 & 0.0951 & 0.0966 & \b{0.0438} &    0.9891 &3.9627 & 0.3669 & 0.5026 &0.3999 & 0.1194 & 0.1202 & \b{0.0869}\\
                        & $\bm{v}$   & 5.5054 & 39.188 & 3.5797 &10.661 & 1.6986 & 2.4568 & \b{0.1041} &    0.4561 &0.4525 & 1.8800 & 1.3474 &1.6144 & 0.3469 & 0.4621 & \b{0.0882}\\
    \multirow{2}{*}{\textbf{300\_ieee}}                                                                                                                         
                        & $\bm{p}^g$ & 0.1336 & 0.0447 & 0.0339 &0.0241 & 0.0091 & 0.0096 & \b{0.0084} &    0.1717 &14.813 & 0.0644 & 0.0766 &0.0476 & 0.0204 & 0.0205 & \b{0.0175}\\
                        & $\bm{v}$   & 3.8526 & 31.698 & 10.292 &4.0253 & 0.2383 & 2.2161 & \b{0.1994} &    0.6854 &1.5737 & 2.9985 & 2.1296 &2.3136 & 1.1553 & 0.2539 & \b{0.2196}\\
   
    \cmidrule(r){1-9} 
    \cmidrule(l){10-16} 
    \multirow{2}{*}{\textbf{Total Avg. (\%)}}                       
  
                       & $\bm{p}^g$ & 0.7751 & 0.9843 & 0.9719 & 0.8829 & 0.0777 & 0.0721 & \b{0.0198} &      0.6090 &0.8214 & 0.5694 & 0.5307 & 0.4948 & 0.1096 & 0.1123 & \b{0.0356} \\
                       & $\bm{v}$   & 2.4284 & 36.2881 & 10.3342 & 9.9916 & 0.8780 & 1.2264 & \b{0.1995} &    1.7870 &0.9818 & 3.3879 & 2.4985 & 1.7409 & 0.9429 & 0.8712 & \b{0.2136} \\
   \bottomrule
  \end{tabular}
  }
  \caption{Average distances (\%) for the active power (top rows) 
  		   and voltage magnitude (bottom rows) of the Load Flow solutions 
  		   w.r.t. the corresponding predictions (left table) and w.r.t.~the 
  		   AC-OPF solutions (right table).}
  \label{tbl:load_flow} 
\end{table*}

\subsection{Load Flow Analysis}

Having assessed the predictive capabilities of OPF-DNN, the next
results focus on evaluating its practicality by simulating the
prediction results in an operational environment. The idea is to
measure how much the predictions need to be adjusted in order to
satisfy the operational and physical constraints.  The experiments
perform a load flow (Model \ref{model:load_flow}) on the predicted
$\hat{\bm{p}}^g$ and $\hat{\bm{v}}$ values. In addition to comparing
the DNN model variants, the results also report the deviations of the
linear DC model from an AC-feasible solution. The DC model is widely
used in power system industry. The results also reports the performance of a baseline load flow model $\LFL$ that finds a
feasible solution using the hot-start state $\bm{s}_0$ as reference
point in its objective function. These results highlight the value of
learning in OPF-DNN: The reference point alone is not sufficient to
find high quality solutions.

The results are tabulated in Table \ref{tbl:load_flow}. The left table
reports the L1 distances, in percentage, of the predictions
$\hat{\bm{p}}^g$ and $\hat{\bm{v}}$ to the solutions $\bm{p}^g$ and
$\bm{v}$ of the load flows. Trends similar to the previous section are
observed, with ${\cal M}_\CLD$ being substantially more accurate than
all other DNN versions. The table also shows that ${\cal M}_\CLD$ is
up to two orders of magnitude more precise than the DC model.  The
right table reports the L1 distances of the load flow solutions to the
optimal AC-OPF solutions. The results follow a similar trend, with the
OPF-DNN model (${\cal M}_\CLD$) being at least one order of magnitude
more precise than the DC model and the baseline $\LFL$ model.  The
bottom rows of the table show the average results over all the power
network adopted in the experimental analysis.  Note that the very high
accuracy of OPF-DNN may render the use of a load flow optimization, to restore feasibility, unnecessary.  \emph{These results are
significant: They suggest that OPF-DNN has the potential to replace
the DC model as an AC-OPF approximation and deliver generator
setpoints with greater fidelity.}

\begin{table}[!t]
\centering
\small
\resizebox{\linewidth}{!}
  {
  \begin{tabular}{l|rrrrrr}
  \toprule
     \multicolumn{1}{l}{\textbf{Test Case}} &    \textbf{DC}&  
     $\LFL$ & \textbf{${\cal M}_\B$}&  \textbf{${\cal M}_\C$}&  \textbf{${\cal M}_\CL$}& \textbf{${\cal M}_\CLD$}\\
     \midrule
     \textbf{14\_ieee}  &5.1792&4.5246&0.7562&0.6290&0.2614&\textbf{0.0007}\\
     \textbf{30\_ieee}  &7.9894&8.2411&2.9447&2.1316&0.5433&\textbf{0.0180}\\
     \textbf{39\_epri}  &0.9094&2.2869&0.1901&0.0752&0.0537&\textbf{0.0003}\\
     \textbf{57\_ieee}  &1.7758&3.8445&1.1115&1.0609&0.2025&\textbf{0.0527}\\
	\textbf{73\_ieee\_rts}  &2.6846&1.4581&9.4364&3.2399&0.5143&\textbf{0.4586}\\
   \textbf{89\_pegase}  &1.5089&2.6287&0.3284&0.3274&0.3347&\textbf{0.1494}\\
    \textbf{118\_ieee}  &4.7455&1.0389&1.0973&1.1897&\textbf{0.5300}&0.5408\\
	\textbf{162\_ieee\_dtc} &6.2090&4.2094&0.5021&0.8360&0.3162&\textbf{0.2845}\\
    \textbf{189\_edin}  &9.9803&7.5561&5.3851&2.7770&0.7135&\textbf{0.3177}\\
    \textbf{300\_ieee}  &4.7508&6.6394&1.9543&1.1576&0.3233&\textbf{0.3011}\\
    \cmidrule(r){1-7} 
    {\textbf{Total Avg. (\%)}} 
                       & 4.5733 & 4.2428 & 2.3706 & 1.3424 & 0.3793 & \textbf{0.2124}\\
   \bottomrule
  \end{tabular}
  }
  \caption{Load Flow vs.~AC-OPF cost distances (\%).}
  \label{tbl:opf_cost} 
\end{table}

\subsection{Solution Quality and Runtime}

The next results compare the accuracy and runtime of the
proposed DNN models, the DC approximation, and the load flow baseline 
$\LFL$, against the \emph{optimal} AC-OPF solutions.  The solution quality is
measured by first finding the closest AC feasible solution to the
predictions returned by the DC or by the DNN models.  Then, the cost
of the dispatches are compared to the original ones.
Table \ref{tbl:opf_cost} reports the average L1-distances of the
dispatch costs.  The last row reports the average distances across all
the test cases.  The analysis of the DNN variants exhibits the same
trends as before, with the networks progressively improving the
results as they exploit the problem constraints (${\cal M}_\C$), a
hot-start state (${\cal M}_\CL$), and use the Lagrangian dual (${\cal
M}_\CLD$).

\begin{table}[!t]
\centering
\small
\resizebox{0.85\linewidth}{!}
  {
  \begin{tabular}{l|rrrr}
    \toprule
     \multicolumn{1}{l}{\textbf{Test Case}}&    \textbf{AC}& {$\LFL$} &   \textbf{DC}&   \textbf{OPF-DNN}\\
     \midrule
     \textbf{14\_ieee}	&0.0332 &0.0430&0.0075&0.0000\\
     \textbf{30\_ieee}	&0.1023 &0.0755&0.0148&0.0000\\
     \textbf{39\_epri}	&0.2169 &0.0968&0.0232&0.0000\\
     \textbf{57\_ieee}	&0.3288 &0.1394&0.0359&0.0000\\
 \textbf{73\_ieee\_rts}	&0.3081 &0.2979&0.0496&0.0000\\
   \textbf{89\_pegase}	&1.4503 &0.6014&0.0601&0.0000\\
    \textbf{118\_ieee}	&0.4207 &0.7819&0.0785&0.0001\\
\textbf{162\_ieee\_dtc}	&1.8909 &0.7393&0.2016&0.0000\\
    \textbf{189\_edin}	&4.0081 &0.4490&0.0865&0.0001\\
    \textbf{300\_ieee}	&8.0645 &1.4850&0.2662&0.0001\\
    \hline
	\textbf{Avg speedup}& $1$x  & $2.76$x & $15.2$x& $>\!10^4$x\\
   \bottomrule
  \end{tabular}
  }
  \caption{Average runtime in seconds.}
  \label{tbl:runtime} 
\end{table}

Table \ref{tbl:runtime} illustrates the average time required
to find an AC OPF solution, the AC load flow with a reference
solution, a linear DC approximation, and a prediction using OPF-DNN
(${\cal M}_\CLD$) on the test dataset. Recall that the dataset
adopted uses a load stress value of up to 20\% of the nominal loads
and hence the test cases are often much more challenging than their
original counterparts. The last row of the table reports the average
speedup of the models compared to the AC OPF. \emph{Observe that
OPF-DNN finds dispatches whose costs are at least one order of
magnitude closer to the AC solution than those returned by the DC
approximation, while being several order of magnitude faster}.

\subsection{Hot-Start Robustness Analysis}

Finally, the last results analyze the robustness of the DNN-OPF model 
when trained using test cases whose hot-start states differ from the 
input state by 1\%, 2\%, and 3\% in the total active loads. 

Table \ref{tbl:SA_pred_errs} reports the average L1 distances 
(in percentage) between the predicted generator power 
$\hat{\bm{p}}^g$, voltage magnitude $\hat{\bm{v}}$, voltage 
angles $\hat{\bm{\theta}}$ and the original quantities. It also 
reports the errors of the predicted flows $\tilde{\bm{p}}^f$ 
which use the generator power and voltage predictions. 
Table \ref{tbl:SA_load_flow} illustrates the load flow results. 
The left table reports the L1 distances, in percentage, of the 
predictions $\hat{\bm{p}}^g$,  and $\hat{\bm{v}}$, to the 
solutions $\bm{p}^g$, and $\bm{v}$ of the load flows. 
The right table reports the L1 distances of the load flow solutions 
to the optimal AC-OPF solutions. 
Finally, Table \ref{tbl:SA_opf_cost} compares the accuracy of the 
OPF-DNN models and the DC approximation against the optimal AC OPF 
solutions.

Observe that DNN-OPF is insensitive, in general, to the 
different hot-start datasets adopted during its training. 
\emph{These results are significant as they indicate that the 
DNN-OPF predictions may be robust to different hot-start range 
accuracies, such as those that may arise in networks with high 
penetration of renewable energy sources.}

\begin{table*}[!t]
\centering
\small
\resizebox{0.8\linewidth}{!}
  {
  \begin{tabular}{llrrrr|llrrrr}
  \toprule
  \textbf{Test case} & \textbf{Dataset} & $\hat{\bm{p}}^g$ & $\hat{\bm{v}}$& $\hat{\bm{\theta}}$& $\tilde{\bm{p}}^f$  &
  \textbf{Test case} & \textbf{Dataset} & $\hat{\bm{p}}^g$ & $\hat{\bm{v}}$& $\hat{\bm{\theta}}$& $\tilde{\bm{p}}^f$  \\
  \midrule
   \multirow{3}{*}{\textbf{14\_ieee}}   
     &     {$\Delta_{1\%}~p^d$} &0.0234&0.0050 &0.0070 &0.0530 & \multirow{3}{*}{\textbf{89\_pegase}}& {$\Delta_{1\%}~p^d$} &0.0797&0.0862&0.0530&5.0160\\
     &     {$\Delta_{2\%}~p^d$} &0.0530&0.0090 &0.0160 &0.0800 && 					                 {$\Delta_{2\%}~p^d$} &0.1380&0.1330&0.1630&5.7420\\
     &     {$\Delta_{3\%}~p^d$} &0.0760&0.0070 &0.0200 &0.0880 && 					                 {$\Delta_{3\%}~p^d$} &0.0690&0.1140&0.2480&5.3680\\
    \hline 
    \multirow{3}{*}{\textbf{30\_ieee}}
     &    {$\Delta_{1\%}~p^d$} &0.0055&0.0070&0.0041&0.0620   &\multirow{3}{*}{\textbf{118\_ieee}}& {$\Delta_{1\%}~p^d$} &0.0340&0.0290&0.2070&0.4550\\
     &    {$\Delta_{2\%}~p^d$} &0.0480&0.0140&0.0170&0.0990	  &&                   					{$\Delta_{2\%}~p^d$} &0.0360&0.1450&0.1120&0.4390\\
     &    {$\Delta_{3\%}~p^d$} &0.0030&0.0160 &0.0080 &0.2120 && 					                {$\Delta_{3\%}~p^d$} &0.0070&0.0210&0.0590&0.3030\\
    \hline
    \multirow{3}{*}{\textbf{39\_epri}}     
     &    {$\Delta_{1\%}~p^d$} &0.0024&0.0280&0.0100&1.2660	&\multirow{3}{*}{\textbf{162\_ieee}}&   {$\Delta_{1\%}~p^d$} &0.0710&0.0770&0.3660&0.4920\\
     &    {$\Delta_{2\%}~p^d$} &0.0140&0.0110&0.0130&0.9400	&&                                      {$\Delta_{2\%}~p^d$} &0.0700&0.2040&0.2640&0.6610\\
     &    {$\Delta_{3\%}~p^d$} &0.0140&0.0130&0.0160&1.5100 && 					                    {$\Delta_{3\%}~p^d$} &0.0680&0.3660&0.2400&0.6500\\
    \hline
    \multirow{3}{*}{\textbf{57\_ieee}}    
    &    {$\Delta_{1\%}~p^d$} &0.0170&0.0150&0.0080&0.1520    &\multirow{3}{*}{\textbf{189\_edin}}& {$\Delta_{1\%}~p^d$} &0.0907&0.0982&0.3330&1.6520\\
    &    {$\Delta_{2\%}~p^d$} &0.0001&0.0150&0.0080&0.3870	&&                                      {$\Delta_{2\%}~p^d$} &0.0150&0.2960&0.0690&2.6160\\
    &    {$\Delta_{3\%}~p^d$} &0.0410&0.0290&0.0090&0.1890 && 					                    {$\Delta_{3\%}~p^d$} &0.0780&0.4040&0.1480&1.9180\\
    \hline
    \multirow{3}{*}{\textbf{73\_ieee}}    
    &    {$\Delta_{1\%}~p^d$} &0.0050&0.0235&0.1180&0.3300 &\multirow{3}{*}{\textbf{300\_ieee}} &  {$\Delta_{1\%}~p^d$} &0.0126&0.0610&2.5670&1.1360\\
    &    {$\Delta_{2\%}~p^d$} &0.0050&0.0235&0.1180&0.3300 &&                                      {$\Delta_{2\%}~p^d$} &0.0220&0.1810&0.8110&1.6890\\
    &    {$\Delta_{3\%}~p^d$} &0.0010&0.0150&0.0170&0.2670 && 					                   {$\Delta_{3\%}~p^d$} &0.0260&0.2270&0.9980&1.9300\\

   \bottomrule
  \end{tabular}
  }
  \caption{OPF-DNN hot-start robustness analysis: Prediction errors (\%).}
  \label{tbl:SA_pred_errs} 
\end{table*}

\begin{table*}[!h]
\centering
\small
\resizebox{0.7\linewidth}{!}
  {
  \begin{tabular}{l l|rrrr|rrrr}
  \toprule
   \multicolumn{2}{l}{\textbf{Test case}}  & \multicolumn{1}{c}{DC}&  \multicolumn{3}{c|}{${\cal M}_\CLD$} 
                                           & \multicolumn{1}{c}{DC}&  \multicolumn{3}{c}{${\cal M}_\CLD$}\\
   \cmidrule(r){4-6}
   \cmidrule(r){8-10}
   \multicolumn{2}{l}{\textbf{Dataset}}  
                  & &  {$\Delta_{1\%}~p^d$}  & {$\Delta_{2\%}~p^d$}  & {$\Delta_{3\%}~p^d$} 
                  & &  {$\Delta_{1\%}~p^d$}  & {$\Delta_{2\%}~p^d$}  & {$\Delta_{3\%}~p^d$} \\
  \midrule     
\multirow{2}{*}{\textbf{14\_ieee}}
                        & $\bm{p}^g$ & 2.4020 & 0.0233 &0.0534 & 0.0764&  0.1359 & 0.0003 & 0.0000 & 0.0000 \\
                        & $\bm{v}$   & 1.8352 & 0.0017 &0.0113 & 0.0034&  3.0365 & 0.0018 & 0.0005 & 0.0005 \\     
\multirow{2}{*}{\textbf{30\_ieee}} 
                        & $\bm{p}^g$ & 2.6972 & 0.0007 &0.0461 & 0.0019&  0.1907 & 0.0058 & 0.0060 & 0.0030 \\
                        & $\bm{v}$   & 1.2929 & 0.0037 &0.0055 & 0.0019&  3.4931 & 0.0086 & 0.0037 & 0.0126 \\     
\multirow{2}{*}{\textbf{39\_epri}} 
                        & $\bm{p}^g$ & 0.0731 & 0.0005 &0.0140 & 0.0105&  0.2163 & 0.0023 & 0.0039 & 0.0089 \\
                        & $\bm{v}$   & 1.1086 & 0.0222 &0.0100 & 0.0686&  1.7251 & 0.0313 & 0.0039 & 0.0776 \\     
\multirow{2}{*}{\textbf{57\_ieee}}
                        & $\bm{p}^g$ & 0.3354 & 0.0076 &0.0136 & 0.0410&  0.0112 & 0.0206 & 0.0000 & 0.0000 \\
                        & $\bm{v}$   & 0.9091 & 0.0384 &0.0031 & 0.0222&  0.2825 & 0.0482 & 0.0145 & 0.0219 \\ 
\multirow{2}{*}{\textbf{73\_ieee\_rts}}
                        & $\bm{p}^g$ & 0.0204 & 0.0053 &0.0003 & 0.0010&  0.0304 & 0.0077 & 0.0005 & 0.0014 \\
                        & $\bm{v}$   & 0.1528 & 0.0337 &0.0024 & 0.0063&  4.1537 & 0.0516 & 0.0169 & 0.0208 \\   
\multirow{2}{*}{\textbf{89\_pegase}}
                        & $\bm{p}^g$ & 0.1641 & 0.0486 &0.1275 & 0.0551&  3.0662 & 0.0827 & 0.1820 & 0.0870 \\
                        & $\bm{v}$   & 3.8584 & 1.2492 &0.9275 & 0.2549&  1.1776 & 1.2610 & 1.0323 & 0.1774 \\    
\multirow{2}{*}{\textbf{118\_ieee}}
                        & $\bm{p}^g$ & 0.2011 & 0.0038 &0.0189 & 0.0010&  0.5865 & 0.0368 & 0.0393 & 0.0068 \\
                        & $\bm{v}$   & 1.9971 & 0.0866 &0.0642 & 0.0050&  2.2780 & 0.1335 & 0.1637 & 0.0200 \\
\multirow{2}{*}{\textbf{162\_ieee\_dtc}}
                        & $\bm{p}^g$ & 0.6727 & 0.0558 &0.0661 & 0.0630&  0.6917 & 0.0954 & 0.0682 & 0.0479 \\
                        & $\bm{v}$   & 3.7718 & 0.2565 &0.4336 & 0.6034&  0.5820 & 0.2921 & 0.3724 & 0.4342 \\    
\multirow{2}{*}{\textbf{189\_edin}}
                        & $\bm{p}^g$ & 1.0514 & 0.0438 &0.0107 & 0.0183&  0.9891 & 0.0869 & 0.0206 & 0.0887 \\
                        & $\bm{v}$   & 5.5054 & 0.1041 &0.1623 & 0.1337&  0.4561 & 0.0882 & 0.1699 & 0.3075 \\    
\multirow{2}{*}{\textbf{300\_ieee}}
                        & $\bm{p}^g$ & 0.1336 & 0.0084 &0.0121 & 0.0124&  0.1717 & 0.0175 & 0.0209 & 0.0247 \\
                        & $\bm{v}$   & 3.8526 & 0.1994 &0.1623 & 0.2507&  0.6854 & 0.2196 & 0.0991 & 0.1575 \\   
    \cmidrule(r){1-6} 
    \cmidrule(l){7-10} 
    \multirow{2}{*}{\textbf{Total Avg. (\%)}}                    
    					& $\bm{p}^g$ & 0.7751 & 0.0198 & 0.0363 & 0.0281 & 0.6090 & 0.0356 & 0.0341 & 0.0268\\
						  & $\bm{v}$   & 2.4284 & 0.1996 & 0.1782 & 0.1350 & 1.7870 & 0.2136 & 0.1877 & 0.1230\\
   \bottomrule
  \end{tabular}
  }
  \caption{Sensitivity analysis of the average errors for the active power (top rows) and voltage magnitude 
  		  (bottom rows) of the load flow solutions w.r.t. the corresponding DC solution or DNN predictions (left table)
           and w.r.t. the AC-OPF solutions (right table), at varying of the 
           distance between the loads in the previous state $\b{s}_0$ and the current load observation. }
  \label{tbl:SA_load_flow} 
\end{table*}

\begin{table}[!h]
\centering
\resizebox{0.8\linewidth}{!}
  {
    \begin{tabular}{l|rrrr}
   \toprule
   \multicolumn{1}{l}{\textbf{Test case}}  & {DC}&  \multicolumn{3}{c}{${\cal M}_\CLD$} \\
   \cmidrule(r){2-5} 
   \multicolumn{1}{l}{\textbf{Dataset}} &  &  {$\Delta_{1\%}~p^d$}   & {$\Delta_{2\%}~p^d$}   & {$\Delta_{3\%}~p^d$}  \\
   \midrule     
     \textbf{14\_ieee}  &5.1792&0.0007&0.0001&0.0001 \\
     \textbf{30\_ieee}  &7.9894&0.0180&0.0028&0.0078 \\
     \textbf{39\_epri}  &0.9094&0.0003&0.0000&0.0027 \\
     \textbf{57\_ieee}  &1.7758&0.0527&0.0000&0.0001 \\
\textbf{73\_ieee\_rts}  &2.6846&0.4586&0.0663&0.0356 \\
   \textbf{89\_pegase}  &1.5089&0.1494&0.1273&0.0237 \\
    \textbf{118\_ieee}  &4.7455&0.5408&0.3913&0.1620 \\
\textbf{162\_ieee\_dtc} &6.2090&0.2845&0.2704&0.1535 \\
    \textbf{189\_edin}  &9.9803&0.3177&0.1064&0.3500 \\
    \textbf{300\_ieee}  &4.7508&0.3011&0.6430&0.6226 \\
    \cmidrule(r){1-5} 
    {\textbf{Total Avg. (\%)}} 
                       & 4.5733 & 0.2124 & 0.1608 &0.1358 \\
   \bottomrule
  \end{tabular}
  }
  \caption{Sensitivity analysis of the LoadFlow OPF solution costs distances from optimal AC-OPF cost (in percentage)
  at varying of the 
           distance between the loads in the previous state $\b{s}_0$ and the current load observation. }
  \label{tbl:SA_opf_cost} 
\end{table}

\section{Related Work}
Within the energy research landscape, DNN architectures 
have mainly been adopted to predict exogenous factors 
affecting renewable resources, such  as solar or wind. 
For instance, Anwar et al.~\citeyear{Anwar:16} uses a 
DNN-based system to  predict wind speed and adopt the 
predictions to schedule generation units ahead of the 
trading period, and Boukelia et al.~\citeyear{Boukelia:17} 
studied a DDN framework to predict the electricity costs 
of solar power plants coupled with a fuel backup system 
and energy storage. \citep{Chatziagorakis:16} studied the 
control of hybrid renewable energy systems, using recurrent 
neural networks  to forecast weather conditions. 

Another power system area in which DNNs have been adopted 
is that of \emph{security assessment}: \citep{Ince:16} proposed 
a convolutional neural network (CNN) model for real-time power
system fault classification to  detect faulted power system 
voltage signals. \citep{Arteaga:19} proposed a convolutional 
neural network to identify safe vs.~unsafe operating points to 
reduce the risks of a blackout. \citep{donnot:hal-02268886} use 
a ResNet architecture to predict the effect of interventions that 
reconnect disconnected transmission lines in a power network.

In terms of OPF prediction, the literature is much sparser.  The most
relevant work uses a DNN architecture to learn the set of active
constraints (e.g., those that, if removed, would improve the value of
the objective function) at optimality in the linear DC
model \cite{ng2018statistical,deka:2019}.  Once the set of relevant
active constraints are identified, exploiting the fact that the DC OPF
is a linear program, one can run an exhaustive search to find a
solution that satisfies the active constraints.  While this strategy
is efficient when the number of active constraints is small, its
computational efficiently decreases drastically when its number
increases due to the combinatoric nature of the problem. Additionally,
this strategy applies only to the linear DC approximation.

This work departs from these proposals and predicts the optimal
setpoints for the network generators and bus voltages in the AC-OPF
setting. Crucially, the presented model actively exploits the OPF
constraints during training, 
producing reliable results that significantly outperform classical
model approximations (e.g., DC-OPF). This work also provides a
compelling alternative to real-time OPF
tracking~\cite{tang17realtime,liu18coordinate}: OPF-DNN always
converges instantly with very high accuracy and can be applied to a
wider class of applications.

\section{Conclusions}

The paper studied a DNN approach for predicting the generators
setpoint in optimal power flows. The AC-OPF problem is a non-convex
non-linear optimization problem that is subject to a set of
constraints dictated by the physics of power networks and engineering
practices. The proposed OPF-DNN model exploits the problem constraints
using a Lagrangian dual method as well as a related hot-start state.
The resulting model was tested on several power network test cases of
varying sizes in terms of prediction accuracy, operational
feasibility, and solution quality. The computational results show that
the proposed OPF-DNN model can find solutions that are up to several
order of magnitude more precise and faster than existing approximation
methods (e.g., the commonly adopted linear DC model). These results
may open a new avenue in approximating the AC-OPF problem, a key
building block in many power system applications, including expansion
planning and security assessment studies which typically requires a
huge number of multi-year simulations based on the linear DC model.
Current work aims at improving the (currently naive) implementation to 
test the approach on very large networks whose entire data sets are 
significantly larger than the GPU memory.

\smallskip\noindent\textbf{Acknowledgments} This research is partly supported by NSF Grant 1709094.

\fontsize{9.0pt}{10.0pt} \selectfont 
\bibliographystyle{aaai}
\bibliography{dl_opf}

\appendix
\section{Appendix} 

\subsection{Network Architectures}
\label{app:networks}
This section includes additional details on the DNN models architecture.
Throughout the text, it considers a power system represented 
by the network graph $({\cal N}, {\cal E})$, 
and use $n$ to denote the number of buses ${\cal N}$ and $e$ to 
denote the number of directed transmission lines ${\cal E}$. 
We also use $l$ to denote the number of buses serving a network load, 
and $g$ to denote the number of buses serving a generator.

\subsubsection{Model ${\cal M}_\B$}
\begin{figure}[!htb]
\centering\includegraphics[width=\linewidth]{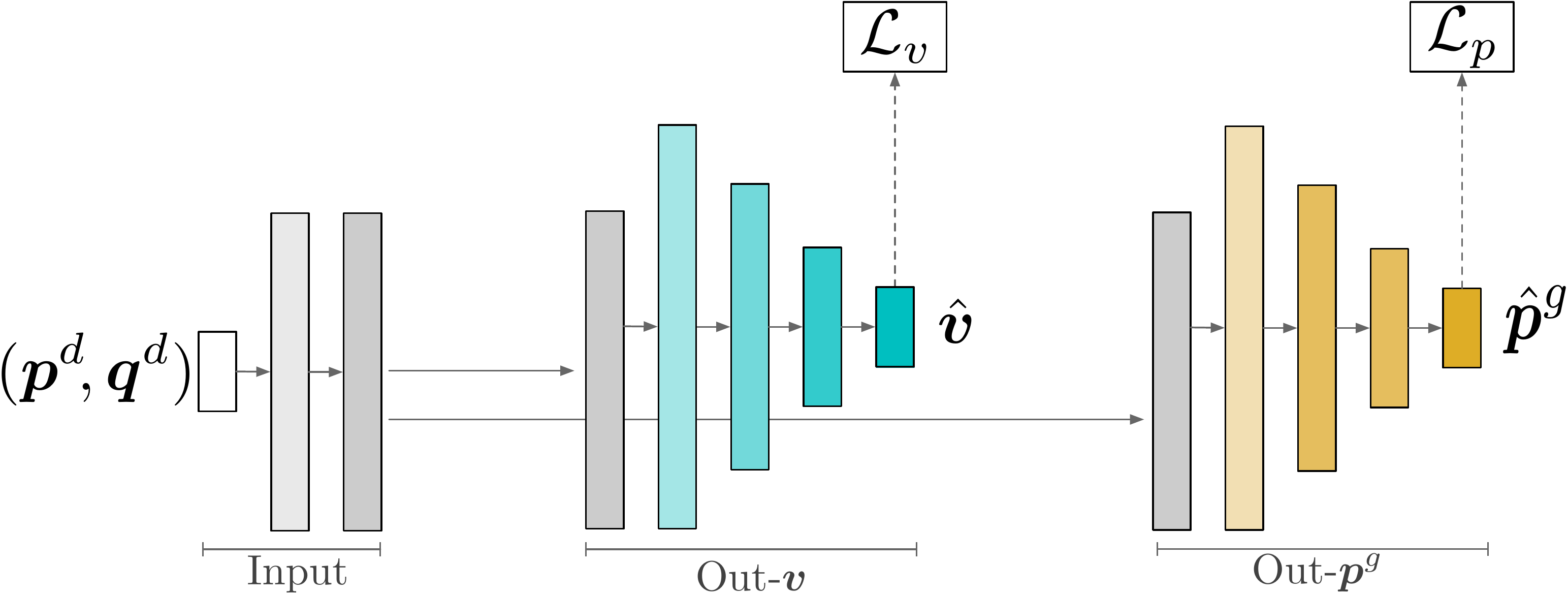}
\caption{\label{fig:dnn1} A representation of the AC-OPF Neural Network Model ${\cal M}_\B$.
    Each layer is fully connected with ReLU activation. 
    White boxes correspond to input tensors, dark, colored, boxes correspond to output 
    layers. Loss components are shown in the rectangles with black borders.}
\end{figure}

It refers to the baseline model that minimizes the 
following loss: 
\[
  {\cal L}_o(\bm{y}, \hat{\bm{y}}) = 
  \| \bm{v} - \hat{\bm{v}}\| ^2 + 
  \| \bm{p}^g - \hat{\bm{p}}^g\| ^2
\]
\noindent
The associated DNN architecture is summarized in the following table.
\begin{center}
\resizebox{0.7\linewidth}{!}
{
\begin{tabular}{l | r r r r}
  \toprule
  \textbf{Alias} & \textbf{Layer} & \textbf{size in} & \textbf{size out} & \text{AF}\\
  \midrule
  Input & FC & $2l$ & $4l$ & ReLU \\
        & FC & $4l$ & $4l$ & ReLU \\
  \hline
  Out-$\bm{v}$ 
        & FC & $4l$ & $8l$ & ReLU \\
        & FC & $8l$ & $4l$ & ReLU \\
        & FC & $4l$ & $2g$ & ReLU \\
        & FC & $2g$ & $g$  &      \\
  \hline
  Out-$\bm{p}^g$ 
        & FC & $4l$ & $8l$ & ReLU \\
        & FC & $8l$ & $4l$ & ReLU \\
        & FC & $4l$ & $2g$ & ReLU \\
        & FC & $2g$ & $g$  &      \\
  \bottomrule
\end{tabular}
}
\end{center}
Therein, the first column identifies the name given to the 
associated group of layers, as used in the illustration
in Figure \ref{fig:dnn1}; FC denotes a fully connected 
layer, size in and size out describe the input and output
dimensions of each layer, and, finally, AF describes the 
activation function adopted at each layer.
The architecture is illustrated in Figure \ref{fig:dnn1}.

\subsubsection{Model ${\cal M}_\C$}
This model exploits the OPF problem constraints using the 
Lagrangian framework under violation degrees. The model minimizes 
the following loss:
\begin{align*}
  {\cal L}_o(\bm{y}, \hat{\bm{y}}) &= 
     \| \bm{v} - \hat{\bm{v}}\|^2 + \| \bm{\theta} - \hat{\bm{\theta}}\|^2 \\
  &+ \| \bm{p}^g - \hat{\bm{p}}^g\| ^2 + \| \bm{q}^g - \hat{\bm{q}}^g\| ^2 \\
  &+  \sum_{c \in {\cal C}} \lambda_c \nu_c(\bm{\hat{y}})
\end{align*}
with ${\cal C}$ being the set of the OPF constraints as defined in Model 1, 
and $\nu_c(\bm{\hat{y}})$ represent the constraint penalty 
associated to constraint $c \in {\cal C}$. All weights $\lambda_c$ are set to $1$.

Its architecture is summarized in the following table:
\begin{center}
\resizebox{0.7\linewidth}{!}
{
\begin{tabular}{l | r r r r}
  \toprule
  \textbf{Alias} & \textbf{Layer} & \textbf{size in} & \textbf{size out} & \text{AF}\\
  \midrule
  Input & FC & $2l$ & $4l$ & ReLU \\
        & FC & $4l$ & $4l$ & ReLU \\
  \hline
  Out-$\bm{v}$ 
        & FC & $4l$ & $8l$ & ReLU \\
        & FC & $8l$ & $4l$ & ReLU \\
        & FC & $4l$ & $2n$ & ReLU \\
        & FC & $2n$ & $n$  &      \\
  \hline
  Out-$\bm{\theta}$ 
        & FC & $4l$ & $8l$ & ReLU \\
        & FC & $8l$ & $4l$ & ReLU \\
        & FC & $4l$ & $2n$ & ReLU \\
        & FC & $2n$ & $n$  &      \\
  \hline
  Out-$\bm{p}^g$ 
        & FC & $4l$ & $8l$ & ReLU \\
        & FC & $8l$ & $4l$ & ReLU \\
        & FC & $4l$ & $2g$ & ReLU \\
        & FC & $2g$ & $g$  &      \\
  \hline
  Out-$\bm{q}^g$ 
        & FC & $4l$ & $8l$ & ReLU \\
        & FC & $8l$ & $4l$ & ReLU \\
        & FC & $4l$ & $2g$ & ReLU \\
        & FC & $2g$ & $g$  &      \\
  \bottomrule
\end{tabular}
}
\end{center}

\begin{figure*}[!h]
\centering\includegraphics[width=0.9\linewidth]{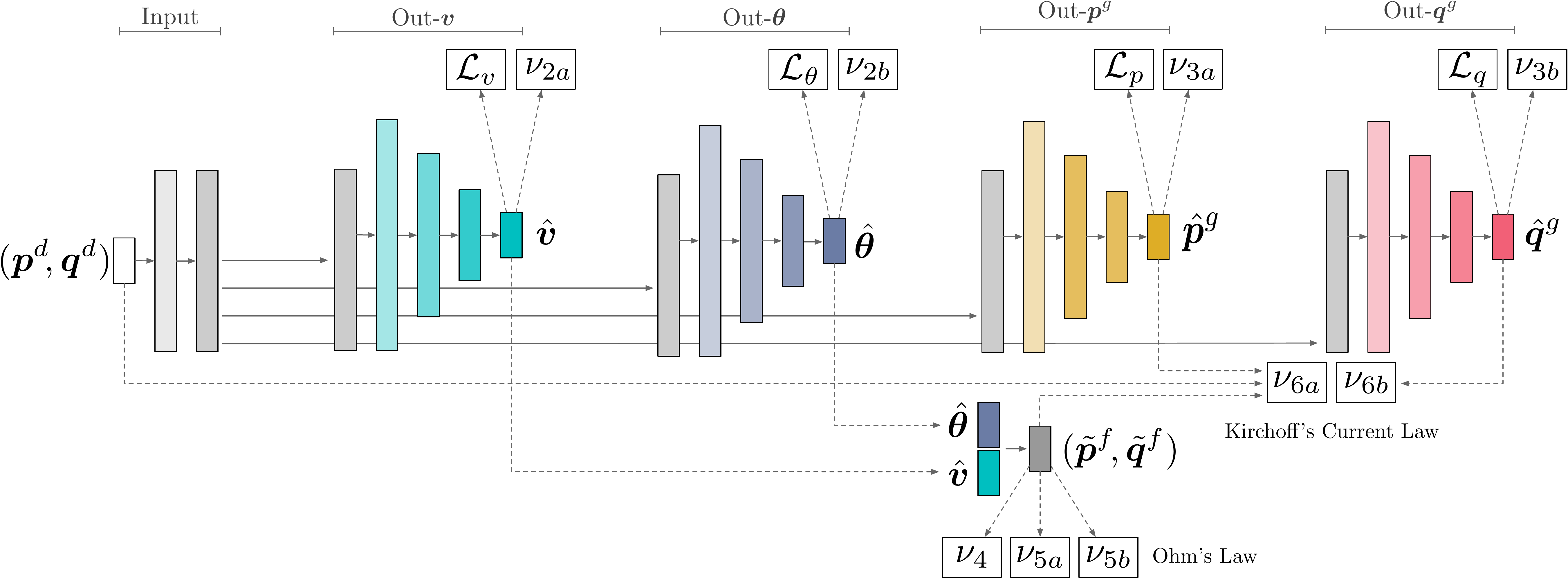}
\caption{\label{fig:dnn2} 
    A representation of the AC-OPF DNN model ${\cal M}_\C$.
    White boxes correspond to input tensors, dark, colored, boxes correspond to output 
    layers. Loss components and violation degrees are shown in the rectangles with black borders.}
\end{figure*}

An illustration of the above architecture is provided in Figure \ref{fig:dnn2}.
The input layers on the left process the tensor of loads 
$(\bm{p}^d, \bm{q}^d)$.  The network has four basic units, each following 
a decoder-encoder structure and composed by a number of layers 
as outlined in the table above. Each subnetwork predicts a target variable: 
voltage magnitudes $\hat{\bm{v}}$, phase angles $\hat{\bm{\theta}}$, 
active power generations $\hat{\bm{p}}^g$, and reactive power generations
$\hat{\bm{q}}^g$.  Each sub-network takes as input the last hidden layer 
of its input subnetwork, that processes the load tensors.
Each of the four prediction outputs is used to compute the penalties 
associated to the quantity bounds ($\nu_{2a}$ for the voltage 
magnitudes, $\nu_{2b}$ for voltage angles, $\nu_{3a}$ for the active 
generator power, and $\nu_{3b}$ for the reactive generator power).
The predictions for the voltage magnitude $\hat{\bm{v}}$ and angle
$\hat{\bm{\theta}}$ are used to compute the load flows
$(\tilde{\bm{p}}^f, \tilde{\bm{q}}^f)$, as illustrated on the bottom
of the Figure and produce penalties $\nu_4, \nu_{5a},$ and $\nu_{5b}$.
Finally, the resulting flows $(\tilde{\bm{p}}^f, \tilde{\bm{q}}^f)$ and 
the predictions for active $\hat{\bm{p}}^g$ and reactive $\hat{\bm{q}}^g$ 
generator power are used to compute the penalties associated to the 
Kirchhoff's Current Law ($\nu_{6a}$ and $\nu_{6b}$).

\subsubsection{Model ${\cal M}_\CD$}
This model extends ${\cal M}_\C$ by estimating the Lagrangian weights 
$\lambda_c$ using the iterative Lagrangian dual scheme described 
in Algorithm 1. 
Its loss function and architecture are analogous to those of 
model {${\cal M}_\C$}.

\subsubsection{Model ${\cal M}_\CL$}
This model extend ${\cal M}_\C$ by exploiting both the problem 
constraints and the previous power system state; 
It uses the same loss function as that used by ${\cal M}_\C$ 
but it adopts the architecture outlined in Figure \ref{fig:dlopf}, 
that uses the information related to the previous power 
network state, as input to each of the four output subnetwork,
in addition to  the last hidden layer of the input subnetwork 
that processes the load tensors.

Its architecture is summarized in the following table:
\begin{center}
\resizebox{0.75\linewidth}{!}
{
\begin{tabular}{l | r r r r}
  \toprule
  \textbf{Alias} & \textbf{Layer} & \textbf{size in} & \textbf{size out} & \text{AF}\\
  \midrule
  Input & FC & $4l$ & $8l$ & ReLU \\
        & FC & $8l$ & $8l$ & ReLU \\
  \hline
  Out-$\bm{v}$ 
        & FC & $8l+n$ & $16l+2n$ & ReLU \\
        & FC & $16l+2n$ & $8l+n$ & ReLU \\
        & FC & $8l+n$ & $4n$ & ReLU \\
        & FC & $4n$ & $2n$   & ReLU \\
        & FC & $2n$ & $n$    &  \\
  \hline
  Out-$\bm{\theta}$ 
        & FC & $8l+n$ & $16l+2n$ & ReLU \\
        & FC & $16l+2n$ & $8l+n$ & ReLU \\
        & FC & $8l+n$ & $4n$ & ReLU \\
        & FC & $4n$ & $2n$   & ReLU \\
        & FC & $2n$ & $n$    &  \\
  \hline
  Out-$\bm{p}^g$ 
        & FC & $8l+g$ & $16l+2g$ & ReLU \\
        & FC & $16l+2g$ & $8l+g$ & ReLU \\
        & FC & $8l+g$ & $4g$ & ReLU \\
        & FC & $4g$ & $2g$   & ReLU \\
        & FC & $2g$ & $g$    &  \\
  \hline
  Out-$\bm{q}^g$ 
        & FC & $8l+g$ & $16l+2g$ & ReLU \\
        & FC & $16l+2g$ & $8l+g$ & ReLU \\
        & FC & $8l+g$ & $4g$ & ReLU \\
        & FC & $4g$ & $2g$   & ReLU \\
        & FC & $2g$ & $g$    &  \\
  \bottomrule
\end{tabular}
}
\end{center}

\subsubsection{Model ${\cal M}_\CLL$}
This model extends ${\cal M}_\CL$ by using trainable Lagrangian multipliers 
$\lambda_c$ associated to each constraint penalty $\nu_c$, for 
$c \in {\cal C}$, whose value is learned during the training 
cycle. Its loss function is thus:
\begin{align*}
  {\cal L}_o(\bm{y}, \hat{\bm{y}}) &= 
     \| \bm{v} - \hat{\bm{v}}\|^2 + \| \bm{\theta} - \hat{\bm{\theta}}\|^2 \\
  &+ \| \bm{p}^g - \hat{\bm{p}}^g\| ^2 + \| \bm{q}^g - \hat{\bm{q}}^g\| ^2 \\
  &+  \sum_{c \in {\cal C}} \lambda_c \nu_c(\bm{\hat{y}})
\end{align*}

${\cal M}_\CLL$ has the same network architecture that the one adopted by 
${\cal M}_\CL$.

\subsubsection{Model ${\cal M}_\CLD$}
Finally, ${\cal M}_\CLD$, (aka OPF-DNN) uses a different approach to estimate the Lagrangian 
weights $\lambda_c$: It does so by using the iterative Lagrangian dual scheme
described in Algorithm 1. 
Its loss function and architecture are analogous to those of model {${\cal M}_\CLL$}.

\end{document}